\documentclass[preprint,times]{qjrmsC}

\usepackage{commath}

\usepackage[makeroom]{cancel}

\renewcommand{\P}{\mathbf{P}}

\newcommand{\C}{\mathbf{C}}

\newcommand{\U}{\mathbf{U}}

\newcommand{\Cc}{\boldsymbol{\mathcal{C}}}

\newcommand{\Pe}{\mathbf{P}^e}

\newcommand{\Pf}{\mathbf{P}^f}
\newcommand{\Pc}{\mathbf{P}^c}

\newcommand{\Ph}{\mathbf{P}^h}

\newcommand{\I}{\mathbf{I}}

\newcommand{\B}{\mathbf{B}}
\newcommand{\A}{\mathbf{A}}

\newcommand{\Z}{\mathbf{Z}}

\newcommand{\Sm}{\mathbf{S}_{m}}

\newcommand{\Eb}{\mathbf{E}}
\newcommand{\Fb}{\mathbf{F}}
\newcommand{\Lambdam}{\boldsymbol{\Lambda}_m}
\newcommand{\V}{\mathbf{V}}
\newcommand{\Deltab}{\boldsymbol{\Delta}}
\newcommand{\F}{\mathbf{F}}
\renewcommand{\E}{\mathbf{E}} 

\newcommand{\co}{\mathbf{c}}
\newcommand{\z}{\mathbf{z}}
\newcommand{\w}{\mathbf{w}}
\renewcommand{\a}{\mathbf{a}}
\newcommand{\phim}{\boldsymbol{\varphi}_m}
\newcommand{\Phim}{\boldsymbol{\Phi}_m}
\newcommand{\phimp}{\boldsymbol{\varphi}_{m^\prime}}
\newcommand{\Phimp}{\boldsymbol{\Phi}_{m^\prime}}

\newcommand{\fmnr}{\Check{\mathbf{f}}_{n,m,r}}
\newcommand{\hr}{\mathbf{h}_r}
\newcommand{\unm}{\Check{\boldsymbol{\upsilon}}_{n,m}}

\newcommand{\Up}{\boldsymbol{\Upsilon}}

\newcommand{\x}{\mathbf{x}}

\newcommand{\ef}{\boldsymbol{\epsilon}^{f}}

\newcommand{\e}{\boldsymbol{\epsilon}}

\newcommand{\xs}{\tilde{\mathbf{x}}}

\newcommand{\efs}{\tilde{\boldsymbol{\epsilon}}^{f}}

\newcommand{\Pes}{\tilde{\mathbf{P}}^e}

\newcommand{\Pcs}{\tilde{\mathbf{P}}^c}
\newcommand{\Phs}{\tilde{\mathbf{P}}^h}

\newcommand{\ns}{\tilde{n}}

 \renewcommand{\tp}{^\intercal} 
 
 \renewcommand{\I}{\mathbf{I}}

 \usepackage{xspace}

\newif\ifYellowBoxOn

\YellowBoxOnfalse

 \DeclareFontFamily{U}{mathb}{\hyphenchar\font45}
\DeclareFontShape{U}{mathb}{m}{n}{
      <5> <6> <7> <8> <9> <10> gen * mathb
      <10.95> mathb10 <12> <14.4> <17.28> <20.74> <24.88> mathb12
      }{}
\DeclareSymbolFont{mathb}{U}{mathb}{m}{n}
 \DeclareMathSymbol{\smalltriangleup}{2}{mathb}{"98}

\newif \ifShowColorBox
\newcommand{\colorboxToBeHidden}[2]{\ifShowColorBox \colorbox{#1}{#2} \fi}
\ShowColorBoxfalse

\renewcommand{\Pr}{\Check{\P}}
\newcommand{\xr}{\Check{\x}}

\newif\ifBookmarkOn
\BookmarkOnfalse
\definecolor{orcidlogocol}{HTML}{A6CE39}
\definecolor{orcidlinkcol}{HTML}{0C4160}
\definecolor{BlueFromUniMelbWebsite}{HTML}{083973}
\definecolor{LightBlueFromUniMelbWebsite}{HTML}{8087A2}
\definecolor{DarkBlueFromUniMelbWebsite}{HTML}{000F46}
\definecolor{VeryLightBlueFromUniMelbWebsite}{HTML}{BFC3D1}
\definecolor{GreenStone}{HTML}{8087A2}
\definecolor{Celeste}{HTML}{406D80}
\definecolor{PaleOrange}{HTML}{F1F5D9}
\definecolor{PaleYellow}{HTML}{FDFD96}
\definecolor{AbstractColour}{HTML}{FDFD96}
\definecolor{AbstractColourBorder}{gray}{0.8}
\definecolor{AbstractTitleColour}{gray}{0.65}
\definecolor{GreenForEpigraphRule}{HTML}{8AD24D}
\definecolor{BlueForLinks}{HTML}{0025ff}
\definecolor{PcColor}{HTML}{BBD64A}
\definecolor{PhCurrentColor}{HTML}{D64A75}
\definecolor{PhTensorHIColor}{HTML}{654AD6}
\definecolor{PhTensorColor}{HTML}{4AD6AB}
\definecolor{PtColor}{HTML}{AEAEAE}
\definecolor{PeColor}{HTML}{FFA000}
\definecolor{PeLocColor}{HTML}{005FFF}
\definecolor{GraphicalAbstractBackgroundColor}{HTML}{F1FFC0}
\definecolor{AxisBackgroundColor}{HTML}{F0F6D7}
\definecolor{PcClimatologyColor}{HTML}{664BD6}
\definecolor{ClimatologyColor1}{rgb}{0.9010, 0.5593, 0.4990}
\definecolor{ClimatologyColor2}{rgb}{0.8981,  0.9010,  0.4990}
\definecolor{ClimatologyColor3}{rgb}{0.5536 ,   0.9010 ,   0.4990}
\definecolor{ClimatologyColor4}{rgb}{0.4990  ,  0.9010  ,  0.7890}
\definecolor{ClimatologyColor5}{rgb}{ 0.4990  ,  0.6684  ,  0.9010}
\definecolor{ClimatologyColor6}{rgb}{0.6742  ,  0.4990  ,  0.9010}
\definecolor{ClimatologyColor7}{rgb}{ 0.9010    0.4990    0.7833}
\definecolor{PtClimatologyColor}{HTML}{999999}
\definecolor{TuningDotsColor}{HTML}{A291E6}
\definecolor{TuningFitLineColor}{HTML}{5233D1}
\definecolor{ResultsBoxesBorderColour}{gray}{0.8}
\definecolor{ResultsBoxesColour}{HTML}{F1F5D9}
\definecolor{HighlightColour}{HTML}{E0F65D}

\definecolor{PsColor}{HTML}{BEE31C}
\definecolor{PColor}{HTML}{1CBEE3}

\usepackage{mathdots}
\usepackage{nicematrix}
\usepackage{scalerel}
\usepackage{mathtools}
\usepackage{url}
\usepackage{etoolbox}
\usepackage[toc]{appendix}
\usepackage{tikz}
\usetikzlibrary{positioning, arrows.meta}
\newcommand{\here}[2]{\tikz[remember picture]{\node[inner sep=0](#2){#1}}}

\renewbibmacro*{doi+eprint+url}{%
  \printfield{doi}%
  \newunit\newblock
  \iffieldundef{doi}{%
    \usebibmacro{eprint}%
    \newunit\newblock
    \usebibmacro{url+urldate}%
  }{}%
}

\usepackage[colorlinks,bookmarksopen,bookmarksnumbered,citecolor=BlueForLinks,urlcolor=BlueForLinks,linkcolor=BlueForLinks]{hyperref}
\usepackage{enumitem}

\newcommand\BibTeX{{\rmfamily B\kern-.05em \textsc{i\kern-.025em b}\kern-.08em
T\kern-.1667em\lower.7ex\hbox{E}\kern-.125emX}}

\usepackage{moreverb}

\def\volumeyear{2013}

\begin{document}

\runningheads{F.~Sardelli and C.~H.~Bishop}{An algorithm for a left inverse of a square root of a horizontally localized ensemble covariance matrix}

\title{A tool for new Hybrid models: a swift algorithm to calculate the left inverse of a square root of a horizontally localized climatological ensemble covariance matrix}

\author{Francesco~Sardelli\affil{a}\corrauth and Craig H.~Bishop\affil{a,b}}

\address{\affilnum{a}School of Geography, Earth and Atmospheric Sciences, University of Melbourne, Australia\\
\affilnum{b}Bureau of Meteorology, Australia}

\corraddr{F.~Sardelli, School of Geography, Earth and Atmospheric Sciences, McCoy Building, Corner Swanston and Elgin Streets, The University of Melbourne, Parkville, VIC, Australia, 3010 \\ E-mail: \url{fsardelli@student.unimelb.edu.au}}

\abstract{Computing matrix inverses (or, e.g., left inverses) using known algorithms is in general computationally prohibitive in very high dimension, such as in an operational weather forecasting context. In this article, a swift, scalable algorithm to calculate the left inverse (through singular values decompositions techniques) of a square root of a horizontally localized ensemble covariance matrix is introduced. In this algorithm, calculations involving model variables located in different grid vertical columns can be performed separately. This is the aspect that makes this algorithm scalable by allowing for parallel computing. The algorithm is developed in the case when the number of ensemble members equals the number of model variables in each grid vertical column (this number is of order $10^3$ for current high-resolution models). The algorithm is tested in a synthetic experiment on a three-dimensional grid, and it is found to have very good accuracy. Additionally, a version of the algorithm outputting the left-multiplication of a generic vector by that above-mentioned left inverse is also introduced. For operational applications, if executed on a single processor, that version of the algorithm is already $9$ orders of magnitude faster than a traditional algorithm, and, with multiple, parallel processors, it  scales until the number of processors  reaches the order of $10^6$. In addition, a relevant application of the algorithm is pointed out: performing the transform from model variables to an alternative set of variables, defined herein as standardized variables, being climatologically uncorrelated and each having unit climatological error variance. Furthermore, it is pointed out that the standardized variables allow constructing a new Hybrid covariance model, which is positive definite as well as climatologically unbiased and it features a geographically dependent climatological error covariance matrix.}

\keywords{parallelizable algorithm; inverse square root; standardized variables; control variables; horizontal localization; localized ensemble covariance matrix; Hybrid covariance matrix; climatological ensemble of forecast error proxies }

\maketitle



\section{Introduction} \label{sec:introduction}

How do we forecast the weather? If we give the state of the atmosphere at the current time as an input to an \emph{atmospheric model}, this will calculate the state at a future time, i.e. the forecast. But, due to chaos theory \parencite{chaos1993}, even a small uncertainty in the current atmospheric state results in a bigger uncertainty in the predicted future state. Determining the current state (i.e. the model input) as accurately as possible is crucial for \emph{weather forecasting} accuracy. To estimate the current state, one combines two pieces of information: \begin{enumerate*}[label=\textit{(\roman*)}]
    \item an (imperfect) forecast produced in the past for the current atmospheric state;
    \item incomplete, spacially sparse, imperfect measurements (called \emph{observations}), in a recent time window, by instruments (e.g. radiosondes, weather stations, satellites, \dots) all around the atmosphere of the Earth. 
\end{enumerate*}
For the task of optimally combining these two pieces of information, a set of mathematical techniques are available, collectively referred to as \emph{(atmospheric) data assimilation} \parencite{kalnay_2002,daley1993atmospheric,Reich2015,OptimalFiltering}. To perform that task, in those techniques, a paramount step is to first determine the inaccuracy of each of those two pieces of information. In particular, one needs to estimate the uncertainty in the forecast produced in the past for the current time. This uncertainty is rigorously described by a mathematical concept called \emph{true forecast error covariance matrix}, here denoted as $\Pf$, also referred to as \emph{background error covariance matrix}. 
This matrix is unknown and it has to be estimated. Any method to estimate $\Pf$ is called a \emph{forecast covariance model}. A widely used forecast covariance model is the \emph{Hybrid covariance model} \parencite{Hamill.Snyder.2000,Bannister2017}. This estimates  $\Pf$ as a matrix $\Ph$ given by a linear combination between: \begin{enumerate*}[label=\textit{(\roman*)}]
    \item the \emph{climatological error covariance matrix} $\Pc$ \parencite{Bannister2008} (which can be defined as the time average of $\Pf$ over a long period, namely, its \emph{climatological mean}) and
    \item the \emph{localized ensemble covariance matrix} $\hat{\P}^e$ \parencite{FurrerBengtsson2007}.
\end{enumerate*}  
The matrix $\hat{\P}^e$, in turn, is given by the \emph{ensemble covariance matrix} $\Pe$ (describing the uncertainty of the forecast performed probabilistically with Monte Carlo techniques known as \emph{ensemble data assimilation} \parencite{Evensen,ReviewoftheEnKF}) element-wise multiplied by a matrix $\C$ with the aim of curing the low rank of $\Pe$ and the sample noise in $\Pe$ \parencite{EvensenChapterOnLocalization}. The matrix $\C$ is referred to as \emph{localization matrix}.\par
Although the Hybrid covariance model was successful in operational weather forecasting centres, \cite{https://doi.org/10.1002/qj.4008} provided empirical evidence suggesting that it should be modified. The experimental work in \cite{https://doi.org/10.1002/qj.4008} concerns the fact that the Hybrid $\Ph$ does not have the correct climatological mean. In other words, the Hybrid $\Ph$ systematically misspecifies the statistical structure of the true forecast errors \parencite{Sardelli2026PhDThesis}. To tackle this problem, \cite{https://doi.org/10.1002/qj.4008} suggested a new Hybrid covariance model (see section \ref{sec:standardizedVariables}), with the correct climatological mean. However, that new Hybrid suffers from a critical flaw. Namely, in some cases, its spectrum might include negative eigenvalues. This flaw can degrade the performance of a weather forecasting system, should one use that new Hybrid.   In the present article, we will show (in section \ref{sec:standardizedVariables}) that it is possible, as a variant of the Hybrid suggested by \cite{https://doi.org/10.1002/qj.4008}, to construct a further novel Hybrid, but without that flaw. Namely, this novel Hybrid is always positive definite. Moreover, it is climatological unbiased (unlike the current Hybrid). To construct this novel Hybrid, we will use a set of variables (to describe the state of the atmosphere) which will be referred to as the \emph{standardized variables}. These are non-dimensional variables with the property that the climatological error covariance matrix expressed in them is just the identity matrix. 
To use standardized variables for that novel Hybrid, we need to be able to compute them  starting from the native model variables, i.e. we need to perform the \emph{standardized variables transform}. This is a linear transform involving a left inverse of a square root of the climatological error covariance matrix. But, calculating that left inverse in the high dimensional weather forecasting operational contexts is computationally unfeasible with known, brute force algorithms. The main focus of the present article is to address this computational problem, by introducing a new, swift, parallelizable (and thus scalable) algorithm to calculate the left inverse of a square root of the climatological error covariance matrix $\Pc$, in the case when $\Pc$ is a horizontally localized sample covariance matrix of an historical ensemble of forecast error proxies. This new algorithm is explained in section \ref{sec:swiftAlgorithm}, with its technical core described in section \ref{app:SwiftAlgorithmTechnical} of the supporting information. Whereas, in section \ref{sec:swiftAlgorithmTest}, we present a synthetic experiment to illustrate and test this algorithm on a 3D grid using spherical harmonics techniques. Finally, in section \ref{sec:conclusions}, we summarize our results and point out possible future research directions and applications.\par 

\section{Motivation} \label{sec:standardizedVariables}

Before diving into the central focus of the present article (starting from section \ref{sec:swiftAlgorithm}), in this section, we will expand on the motivation for our work. More importantly, the present section will act as a useful junction between this paper and future papers in which new Hybrid covariance models \parencite{Sardelli2026PhDThesis} will be introduced.\par
Now, as a starting point of our reasoning, let us consider the Hybrid covariance model which is currently used in operational centres worldwide. This Hybrid is defined by:
\begin{equation} \label{eq:currentHybrid}
    \Ph = \beta_c \,\Pc + \beta_e \, \C \odot \Pe
\end{equation}
where $\beta_c$ and $\beta_e$ are positive real numbers, typically between $0$ and $1$, and the symbol $\odot$ denotes the element-wise product between matrices, also referred to as \emph{Hadamard product} or \emph{Schur product} \parencite{RaoRao}. \cite{https://doi.org/10.1002/qj.4008} observed that the climatological mean $\langle \Ph \rangle_{\text{clim}}$ does not equal $\Pc$, as it should. Namely, the current Hybrid is \emph{climatologically biased}. To address this issue, and based on empirical evidence, \cite{https://doi.org/10.1002/qj.4008} suggested to modify the current Hybrid formula as follows:
\begin{equation} \label{eq:CBKHybrid}
    \Ph =  (\U - \beta_e \C) \odot \Pc + \beta_e \,\C \odot \Pe
\end{equation}
where $\U$ denotes the $n \times n$ matrix whose elements all equal $1$. They showed that the new Hybrid \eqref{eq:CBKHybrid} is \emph{climatologically unbiased}. However, the spectrum of a Hybrid $\Ph$ of the form \eqref{eq:CBKHybrid} may include negative eigenvalues. This, in turn, may seriously degrade the forecasting performance of a data assimilation system which utilizes such a $\Ph$. Thus, we now ask a question. How can we construct a Hybrid covariance model that is both climatologically unbiased and positive definite?\par
We will achieve this goal in a moment. But, for that, we need to first define a particular set of variables, other than the native model variables, to describe the atmospheric state. (And, prior to giving this definition, we will need to first establish some notation.) 
\subsection{The standardized variables transform}
First of all, let us list the native model variables in a $n$-dimensional column vector $\x$. i.e. the state vector. The variables that we are about to define will be referred to as \emph{standardized variables}. In general, given any quantity, vector or matrix in model variables, by adding a $\tilde{\phantom{\P}}$ on top of its symbol, we will obtain the notation for its counterpart in the new variables. For example, $\ns$ will denote the number of standardized variables, and the $\ns$-dimensional column vector listing the standardized variables will be denoted by $\xs$.\par
Then, let the $n \times \ns$ matrix $(\Pc)^{\frac{1}{2}}$ be a square root of the climatological error covariance matrix $\Pc$, i.e. $(\Pc)^{\frac{1}{2}}\, ((\Pc)^{\frac{1}{2}})\tp = \Pc$. And, let the $\ns \times n$ matrix $(\Pc)^{-\frac{1}{2}}$ be a left inverse of $(\Pc)^{\frac{1}{2}}$, that is, $(\Pc)^{-\frac{1}{2}} \, (\Pc)^{\frac{1}{2}} = \I_{\ns}$, where $\I_{\ns}$ is the $\ns \times \ns$ identity matrix. We are now ready to define the standardized variables , through 
 \begin{equation} \label{eq:svTransform}
     \tilde{\x} = (\Pc)^{-\frac{1}{2}} \, \x
 \end{equation}
 We will refer to \eqref{eq:svTransform} as the \emph{standardized variables transform}.\par
A moment of reflection will convince the reader that the standardized variables $\xs$ are non-dimensional. Moreover, they have a remarkable property. To see this, let us first define the standardized variables counterpart of $\Pc$ as the matrix $\Pcs \doteq \langle \efs  (\efs)\tp\rangle_{\text{clim}}$, where $\efs$ is the true forecast error vector in standardized variables. The just-mentioned remarkable property is that $\Pcs$ is the $\ns \times \ns$ identity matrix $\I_{\ns}$:
\begin{equation} \label{eq:PcsvIsTheIdentity}
\begin{split}
    \Pcs & \doteq \langle \efs  (\efs)\tp\rangle_{\text{clim}} =  \langle [(\Pc)^{-\frac{1}{2}}\ef ] \,[(\Pc)^{-\frac{1}{2}}\ef]\tp\rangle_{\text{clim}} \\ & = (\Pc)^{-\frac{1}{2}} \; \langle \ef  (\ef)\tp\rangle_{\text{clim}} \; [(\Pc)^{-\frac{1}{2}}]\tp \\ &= (\Pc)^{-\frac{1}{2}}\Pc [(\Pc)^{-\frac{1}{2}}]\tp = \I_{\ns}
    \end{split}
\end{equation}
In words, in standardized variables, the climatological mean of every forecast error variance is $1$. Whereas, the climatological mean of every covariance between forecast errors of two distinct variables is zero, namely, the standardized variables are \emph{climatologically uncorrelated}. 
Now, having defined and expanded on the standardized variables, let us briefly go back to the Hybrid covariance models.
\subsection{Suggesting an application of the standardized variables to a new Hybrid} \label{sec:SuggestingApplication}
Let us ask a question. What happens if we formulate the Hybrid covariance model \eqref{eq:CBKHybrid} in standardized variables? Let us explore this question in the following.\par 
As $\Pcs=\I_{\ns}$, the standardized variables formulation of \eqref{eq:CBKHybrid} is:
\begin{equation} \label{eq:svCBKHybridBeforeSimplification}
     \Phs =  (\U - \tilde{\beta}_e \, \tilde{\C}) \odot \I_{\ns} + \tilde{\beta}_e \,\tilde{\C} \odot \Pes
\end{equation}
where $0 < \tilde{\beta}_e < 1$, $\tilde{\C}$ is a localization matrix and $\Pes=(\Pc)^{-\frac{1}{2}} \: \Pe \, ((\Pc)^{-\frac{1}{2}} )\tp$ is the standardized variables ensemble covariance matrix. Now, as the off-diagonal elements of the identity $\I_{\ns}$ are zero and each diagonal element of the localization matrix $\tilde{\C}$ equals $1$, the first term on the right-hand side of \eqref{eq:svCBKHybridBeforeSimplification} is simplified as follows, as one can easily verify:
\begin{equation} \label{eq:svCBKHybrid}
     \Phs =  (1- \tilde{\beta}_e ) \, \I_{\ns} + \tilde{\beta}_e \,\tilde{\C} \odot \Pes
\end{equation}
where we have recalled the definition of $\U$ as the $n \times n$ matrix whose elements all equal $1$. A thorough study and experimental tests for the Hybrid model \eqref{eq:svCBKHybrid} are beyond the scope of the present article. Moreover, as touched on above, novel Hybrid models (involving standardized variables), more advanced than \eqref{eq:svCBKHybrid} \parencite{Sardelli2026PhDThesis}, will be introduced and tested in separate articles. Here, let us limit ourselves to mentioning that formula \eqref{eq:svCBKHybrid} produces a matrix $\Phs$ which is always positive definite (using a positive semi-definite $\tilde{\C}$ and provided that $\tilde{\beta}_e$ as well as $\tilde{\beta}_c \doteq 1- \tilde{\beta}_e $ are positive). Furthermore, provided we use an ensemble data assimilation scheme with a climatologically unbiased $\Pe$\footnote{A good ensemble data assimilation scheme should be designed so that the resulting $\Pe$ is climatologically unbiased, i.e. $\langle \Pe\rangle_{\text{clim}}=\Pc$. It is not difficult to prove (but we will not do this for concision) that $\langle \Pe\rangle_{\text{clim}}=\Pc$ implies $\langle \Pes\rangle_{\text{clim}}=\Pcs=\I_{\ns}$, which, in turn, can be used to show that $\langle \Phs\rangle_{\text{clim}}=\Pcs=\I_{\ns}$.}, the Hybrid $\Phs$ is climatologically unbiased!\par
Constructing $\Phs$ was made possible by the use of standardized variables. Being able to build the novel Hybrid $\Phs$ (and even more advanced new Hybrids \parencite{Sardelli2026PhDThesis} to be introduced in future papers) is the main motivation and application of the research work of the present article. In turn, a natural application for $\Phs$ would be utilizing this Hybrid as a forecast covariance model within a variational data assimilation framework \parencite{Sardelli2026PhDThesis}, i.e. constructing a Hybrid ensemble-variational scheme using $\Phs$. Another possibility would be to utilize the Hybrid $\Phs$ as a forecast covariance model within an ensemble data assimilation framework, e.g. similarly to how \cite{Kretschmer01122015}, \cite{KotsukiBishop2022} or \cite{FrolovWhitakerDraper2022} included traditional Hybrid forecast error covariances (i.e. constructed as per formula \eqref{eq:currentHybrid}) in an ensemble scheme.\par
Along the route towards the operational implementability of the Hybrid $\Phs$, next steps include testing its accuracy as a forecast covariance model; and, in the context of variational methods in particular, constructing a control variable theory for $\Phs$, and studying the efficiency of the minimization for a cost function which uses $\Phs$, namely thoroughly studying the Hessian of such a cost function \parencite{HABENEtAl2011,TabeartEtAl2022}. These future steps are beyond the scope of this article. However, here, especially for readers familiar with the work of \cite{LorencBowlerEtal2015}, let us mention that one can obtain a control variable theory suitable for the Hybrid $\Phs$ essentially as a special case of the \emph{alpha control variables} approach of \cite{LorencBowlerEtal2015}. More specifically and precisely, one can obtain a control variables theory for $\Phs$ (when back-transformed to model variables) by choosing the standardized variables (instead of, e.g., spectral variables or model variables) to represent the ensemble members perturbations in the alpha control variables approach (see equation (15) in \cite{LorencBowlerEtal2015}). Moreover, that control variables theory for $\Phs$ results in a cost function (for the control variables) of the form given by equation (8) in \cite{LorencBowlerEtal2015} (having a positive definite Hessian). As touched on above, a full, thorough explanation of these aspects is left to future papers. Here, let us just remark that, in that control variables theory for $\Phs$, the standardized variables transform (and its inverse) plays a crucial role.
\par
Having pointed out possible applications of the standardized variables transform and emphasized its relevance, let us now come to the focus of the present article, namely, tackling the problem of making this transform computationally feasible. In particular, for this aim, a difficulty needed to be overcome. With known algorithms, calculating the left inverse $(\Pc)^{-\frac{1}{2}}$ is not computationally tractable in the context of operational, and thus high dimensional, applications. %
\colorboxToBeHidden{PaleYellow}{What is the computational cost of computing a left inverse? and an inverse? }
We will tackle this issue by introducing a novel swift algorithm (section \ref{sec:swiftAlgorithm}) to compute $(\Pc)^{-\frac{1}{2}}$ in the case when $\Pc$ takes the following form\footnote{Let us remark that the localization matrix $\Cc$ appearing in formula \eqref{eq:HLocalizedPc} for $\Pc$ should not be confused with the localization matrix $\C$ appearing in some Hybrid covariance models, such as \eqref{eq:currentHybrid} and \eqref{eq:CBKHybrid}. The two localization matrices $\boldsymbol{\mathcal{C}}$ and $\C$ are distinct objects, and, for this reason, we denote the former by a bold curl letter, i.e. $\boldsymbol{\mathcal{C}}$, and the latter by just a bold letter, i.e. $\C$.}:
\begin{equation} \label{eq:HLocalizedPc}
    \Pc = \Cc \odot \P^s
\end{equation}
where
\begin{itemize}
    \item $\P^s$ is the sample covariance matrix of a climatological ensemble of forecast error proxies (as we will see in section \ref{sec:AssumptionAboutK}, we will make an assumption about the number of ensemble members $K$; as a consequence of that assumption, we will consider a $K$ of order $10^3$);
    \item $\boldsymbol{\mathcal{C}}$ is a \emph{horizontal localization} matrix (we will give a definition of horizontal localization matrix in section \ref{sec:HorizontalLocalization}).
\end{itemize}
We will refer to a climatological error covariance matrix $\Pc$ of the form \eqref{eq:HLocalizedPc} as a \emph{horizontally localized climatological error sample covariance matrix}. Incidentally, let us notice that such a $\Pc$ features geographically dependent climatological error covariances, constructed from the (empirical) above-mentioned climatological ensemble of error proxies. Let us just touch on the fact that, when working in standardized variables with the novel Hybrid $\Phs$ and back-transforming to model variables through a mapping of the form $\x= (\Pc)^{\frac{1}{2}}\, \xs$, interestingly, the use of $(\Pc)^{\frac{1}{2}}$ injects geographically dependent climatological covariance information into the data assimilation system \parencite{Sardelli2026PhDThesis}. Delving into this aspect is beyond the scope of the present article, and it is left to future ones.\par
Now, let us construct and explain our swift algorithm in section \ref{sec:swiftAlgorithm}. Then, in section \ref{sec:swiftAlgorithmTest}, we will test it in a synthetic experiment. 


\section{A swift algorithm for $(\Pc)^{-\frac{1}{2}}$} \label{sec:swiftAlgorithm}

\subsection{Overview of this section} \label{sec:Overview}

Before diving into a description of the concepts leading to the construction of our novel algorithm, here it may be helpful to get a panoramic view of this section. The aim of our algorithm is to swiftly build the matrix $(\Pc)^{-\frac{1}{2}}$ through parallel computing. This is possible because, as we will explain, a part of the calculations to build $(\Pc)^{-\frac{1}{2}}$ involving variables located at different grid vertical columns can be executed independently (i.e. in parallel) for each grid vertical column. Thus, the working of our novel algorithm is strictly related to the vertical (as well as the horizontal) structure of the grid (which is outlined in section \ref{sec:gridTheorySection}). Because of this, vectors involved in our algorithm should be have their elements listed according to an order that reflects the vertical as well as the horizontal structure of the grid (this listing order will be explained in section \ref{sec:segmentStructure}). That way of organizing elements into vectors will lead to a particular block structure of matrices involved in our algorithm (this block structure is described in section \ref{sec:blockMatrixStructure}). Regarding the main ingredients for our algorithm, they include the ensemble and the sample covariance matrix $\P^s$ (section \ref{sec:PsTheorySection}), the localization matrix $\Cc$ (section \ref{sec:HorizontalLocalization}), and the square roots $\Cc^{\frac{1}{2}}$ of $\Cc$ and $(\Pc)^{\frac{1}{2}}$ of $\Pc = \Cc \odot \P^s$ (section \ref{sec:SquareRoots}). A technical assumption which we made in the derivation of our algorithm is discussed in section \ref{sec:AssumptionAboutK}. After these needed preliminary steps, the description of the swift algorithm to build $(\Pc)^{-\frac{1}{2}}$ is given in section \ref{sec:AlgorithmDescription}. For operational applications, we are particularly interested in left-multiplying a vector $\x$ by $(\Pc)^{-\frac{1}{2}}$ to implement the standardized variables transform \eqref{eq:svTransform}. A swift, operationally advantageous algorithm to perform this left-multiplication without having to explicitly form the whole matrix $(\Pc)^{-\frac{1}{2}}$ is provided in section \ref{sec:ActionOnVector}. In its development, the algorithm of section \ref{sec:ActionOnVector} is heavily based on (and, thus, in this sense, can be seen as a different version of) the algorithm of section \ref{sec:AlgorithmDescription} for $(\Pc)^{-\frac{1}{2}}$. Finally, the practical efficacy and feasibility of the algorithm of section \ref{sec:ActionOnVector} in terms of computational time and required memory space is explained in section \ref{sec:ComputationalAspects}. 

\subsection{The grid} \label{sec:gridTheorySection}

Before diving into the construction of our novel algorithm, we need to take a few preliminary steps. First of all, let us briefly fix our notation about aspects of the grid on which our atmospheric model $\mathcal{M}$ is defined. Let us suppose that the grid points are arranged on horizontal layers. Let $n_{v}$ be the number of horizontal layers. Let $n_{h}$ denote the number of grid points on each horizontal layer. Moreover, let $n_{G}$ be the number of variables located at each grid point. Thus, the total number $n$ of model variables is given by
\begin{equation} \label{eq:NumberOfModelVariables}
    n= n_{G} \, n_{h} \, n_{v}
\end{equation}

\subsection{The climatological ensemble and the sample covariance matrix $\P^{s}$} \label{sec:PsTheorySection}
Let us suppose that we are working with model variables. Let us consider a climatological ensemble $\{ \e_k\}_{k=1}^K$ of $K$ forecast error proxies. Let us then define the vectors $\z_k$  (they will be useful later) as
\begin{equation} \label{eq:zkDefinition}
    \z_k \doteq  \frac{1}{\sqrt{K-1}} \, (\e_k - \bar{\e})   \qquad \text{with} \quad 1 \le k \le K
\end{equation}
where $\bar{\e}$ denotes the mean of our ensemble, i.e. $\bar{\e} = (\sum_{k=1}^K \e_k) / K$.\par
The sample covariance matrix $\P^s$ is given by:
\begin{equation} \label{eq:PsAndItsSquareRoot}
    \P^{s} =\frac{1}{K-1} \, \sum_{k=1}^K (\e_k - \bar{\e}) (\e_k - \bar{\e})\tp = \sum_{k=1}^K \z_k \z_k\tp = \Z \Z\tp
\end{equation}
where $\Z$ is the $n \times K$ matrix whose columns are $\z_k$ ($1 \le k \le K$). From \eqref{eq:PsAndItsSquareRoot}, we see that $\Z$ is a square root of the sample covariance matrix $\P^s$.

 \par
Now, before proceeding to describe our novel algorithm, we will need a further preliminary step. Namely, we need to define a convenient order to arrange the vector elements along all column vectors $\e_k$ and $\z_k$ ($1 \le k \le K$).
\subsection{A segment structure for vectors} \label{sec:segmentStructure}
First of all, let us remind ourselves that the $i$-th vector element $(\e_k)_i$ is the value of the error on the $i$-th model variable within the $k$-th error proxy $\e_k$ in our ensemble. Thus, essentially, the vector element $(\e_k)_i$ pertains to the $i$-th model variable. Therefore, defining an order to arrange the $(\e_k)_i$'s along the column vector $\e_k$ is simply about choosing an order to list the model variables.\par
Let us also remind ourselves that the model variables at each grid point are of different kinds, such as temperature, pressure, relative humidity, the three components of the wind, and so on. We have $n_{G}$ kinds of variables per grid point\footnote{The algorithm that we are introducing can be easily generalized to the case when the number of variables per grid point is \emph{not} necessarily the same on all horizontal layers. But, for expounding simplicity, throughout this article, we will assume that that number, denoted as $n_G$, is the same for all grid points, irrespective of the horizontal layer.}.  Let us now first consider the vectors elements of $\e_k$ pertaining to the temperature in the lowest grid horizontal layer. These vector elements are $n_h$. Let us list them in a $n_h$-dimensional vector, which we will denote by $\e_{1,k}$. The vector $\e_{1,k}$ will be useful in a moment. But before, we need to define other $n_h$-dimensional vectors in an analogous way. Namely, let us consider the vectors elements of $\e_k$ pertaining to the pressure in lowest grid horizontal layer. Let us list these vectors elements in a $n_h$-dimensional vector $\e_{2,k}$. Let us proceed analogously and define the vectors $\e_{2,k}$, \dots, $\e_{n_G,k}$ for all variable kinds in the lowest horizontal layer. Let us then move on to the second lowest horizontal layer. For that layer (as done for the lowest one), let us list the vector elements of $\e_k$ pertaining to the temperature, pressure, and so on (for all variables kinds) in the $n_h$-dimensional vectors $\e_{n_G+1,k}$, $\e_{n_G+2,k}$, \dots, $\e_{2n_G,k}$, respectively. Let us keep going with our definitions, for all horizontal layers, up to the highest one, whose corresponding vectors are $\e_{n_G(n_v -1)+1,k}$, $\e_{n_G(n_v -1)+2,k}$, \dots, $\e_{n_G n_v,k}$. Thus, in total, the number of $n_h$-dimensional vectors that we have defined equals the number of variables $n_w$ in each grid vertical column, that is, $n_{w} = n_v \, n_{G}$ (let us recall that $n_v$ denotes the number of horizontal layers in our grid). At this point, utilizing the just-defined vectors, it is straightforward to describe how we arrange the vector elements in each ensemble member $\e_k$. Namely, we arrange those according to the following vector equation:
\begin{align} \label{eq:MemberSegments}
    \e_k &= \begin{bmatrix}
                \e_{1,k} \\[10pt]
                \e_{2,k} \\[10pt]
                \vdots   \\[10pt]
                \e_{l,k} \\[10pt]
                \vdots   \\[10pt]
                \e_{n_w,k}
            \end{bmatrix}      \qquad \text{for all} \; k \; \text{with} \; 1 \le k \le K
\end{align}

In other words, we are listing the vector elements in $\e_k$ in successive ``segments'' $\e_{l,k}$, with each segment listing the vector elements pertaining to all variables at a given horizontal layer and given kind. \par
Our definitions, given above, of the $n_h$-dimensional vectors $\e_{l,k}$ (i.e. the segments) are incomplete because we have not specified a vector element order within each segment yet. Let us then specify this in the following. For that, first of all, let us choose an order for the set of vertical columns in our grid. Then, let us consider the $m$-th vertical grid column according to this chosen order, with $1 \le m \le n_h$. Let $(\e_{l,k})_{m} $ be the $m$-th element of the segment $\e_{l,k}$, with $1 \le m \le n_{h}$. Let us emphasize that the element  $(\e_{l,k})_{m} $ is a quantity pertaining to a single model variable. Now, let us choose a vector element order within each segment $\e_{l,k}$ so that $(\e_{l,k})_{m} $ pertains to a variable which belongs to the $m$-th grid vertical column, for all $m$, and for all $l$ and $k$ ($1 \le m \le n_{h}$, $1 \le j \le n_w$ and $1 \le k \le K$). This convention ensures that the vector element order within each segment $\e_{l,k}$ is consistent and matching across all segments. We will see below that this will simplify the derivation of our algorithm. Furthermore, for each given $m$, one can consider the set of all quantities $(\e_{l,k})_{m} $ placed in the $m$-th position across all segments $\e_{l,k}$ for all $l$ and $k$ ($1 \le l \le n_{w}$ and $1 \le k \le K$). This is the set of vector elements pertaining to all variables belonging to the $m$-th grid vertical column. Let us refer to this set as the \emph{quantities of the $m$-th grid vertical column}. Let us keep this concept in mind as we will make extensive use of it.\par
Now, let us define the segments
\begin{equation} \label{eq:ZSegmentDef}
    \z_{l,k}= \frac{1}{\sqrt{K-1}} \e_{l,k} \qquad \begin{array}{l}
1 \le l \le n_w \\
 1 \le k \le K
\end{array}  
\end{equation}
Thanks to this definition, the vectors $\z_k$ inherit a segment structure of the kind \eqref{eq:MemberSegments} from the vectors $\e_k$:
\begin{align} \label{eq:ZSegments}
    \z_k &= \begin{bmatrix}
                \z_{1,k} \\[10pt]
                \z_{2,k} \\[10pt]
                \vdots   \\[10pt]
                \z_{l,k} \\[10pt]
                \vdots   \\[10pt]
                \z_{n_w,k}
            \end{bmatrix}      \qquad \text{for all} \; k \; \text{with} \; 1 \le k \le K
\end{align}
Now, analogously to what we did for the segments $\e_{k,l}$, let us denote the $m$-th element of each vector $\z_{k,l}$ by $(\z_{k,l})_m$ ($1 \le m \le n_h$, $1 \le k \le K$ and $1 \le l \le n_w$).  When dealing with the vectors $\z_k$, we will also refer to the set of quantities $(\z_{k,l})_m$ for a given, fixed $m$, for all $l$ and $k$ as the \emph{quantities of the $m$-th grid vertical column}. In the remainder of this chapter, for simplicity, when using the expression \emph{quantities of the $m$-th grid column}, we will not explicitly specify whether we are referring to the vectors $\e_k$ of $\z_k$ if that is clear from the context.\par
At this stage, having defined the concept of \emph{quantities of the $m$-th grid vertical column}, we can already give a glimpse into the key aspect of the functioning of our novel swift algorithm that we will introduce in this article. As we will see, in calculating $(\Pc)^{-\frac{1}{2}}$, the elements of the vectors $\z_k$ are involved. More specifically, it turns out that the calculations on the quantities of the $m$-th grid vertical column are independent from the calculations on the quantities of the $m^\prime$-th grid vertical column for all $m^\prime$ different from $m$. This will allow for parallel computing, and thus, for a scalable, swift algorithm. We will provide a full, complete explanation of these aspects in section \ref{sec:AlgorithmDescription}, where the algorithm will be described, and, even more in depth, in the technical section \ref{app:SwiftAlgorithmTechnical} (in the supporting information of this article).
%
%
%
%

\subsection{A block structure for matrices} \label{sec:blockMatrixStructure}
In the previous section \ref{sec:segmentStructure}, we defined a segment structure for the vectors $\e_k$ and $\z_k$. This will naturally lead to a block structure for some matrices. For example, a moment of reflection will convince the reader that the sample covariance matrix $\P^s$, given by \eqref{eq:PsAndItsSquareRoot}, can be written as:
\begin{align} \label{eq:blockStructurePs}
    \P^s &= \begin{bmatrix}
        \B_{11} & \B_{12} & \dots & \B_{1l^\prime} & \dots & \B_{1n_{w}}\\[10pt]
        \B_{21} & \B_{22} & \dots & \B_{2l^\prime} & \dots & \B_{2n_{w}}\\[10pt]
        \vdots & \vdots & \ddots & \vdots & \iddots & \vdots\\[10pt]
        \B_{l1} & \B_{l2} & \dots & \B_{ll^\prime} & \dots & \B_{ln_{w}}\\[10pt]
        \vdots & \vdots & \iddots & \vdots & \ddots & \vdots\\[10pt]
        \B_{n_{w}1} & \B_{n_{w}2} & \dots & \B_{n_{w}l^\prime} & \dots & \B_{n_{w}n_{w}}\\[10pt]
    \end{bmatrix}
\end{align}
where each block $\B_{l l^\prime}$ is the $n_h \times n_h$ matrix
\begin{equation}
    \B_{l l^\prime}= \sum_{k=1}^K \z_{l,k} \; (\z_{l^\prime,k}) \tp \qquad \begin{array}{l}
1 \le \,l \,\le n_w\\
 1 \le \, l^\prime \le n_w
\end{array}  
\end{equation}
where $\z_{l,k}$ are the segments (defined in \eqref{eq:ZSegmentDef}) of the vectors $\z_k$. Thus, the matrix $\P^s$ is made of $n_w^2$ blocks $\B_{l l^\prime}$ with $1 \le l \le n_w$ and $1 \le l^\prime \le n_w$.

Now, the localization matrix $\Cc$ can be thought of as having a block structure matching the one \eqref{eq:blockStructurePs} of $\P^s$, that is:

\begin{align} \label{eq:blockStructureC}
    \Cc &= \begin{bmatrix}
        \Cc_{11} & \Cc_{12} & \dots & \Cc_{1l^\prime} & \dots & \Cc_{1n_{w}}\\[10pt]
        \Cc_{21} & \Cc_{22} & \dots & \Cc_{2l^\prime} & \dots & \Cc_{2n_{w}}\\[10pt]
        \vdots & \vdots & \ddots & \vdots & \iddots & \vdots\\[10pt]
        \Cc_{l1} & \Cc_{l2} & \dots & \Cc_{ll^\prime} & \dots & \Cc_{ln_{w}}\\[10pt]
        \vdots & \vdots & \iddots & \vdots & \ddots & \vdots\\[10pt]
        \Cc_{n_{w}1} & \Cc_{n_{w}2} & \dots & \Cc_{n_{w}l^\prime} & \dots & \Cc_{n_{w}n_{w}}\\[10pt]
    \end{bmatrix}
\end{align}

Namely, the localization matrix $\Cc$ is made of $n_w^2$ blocks $\Cc_{l l^\prime}$ with $1 \le l \le n_w$ and $1 \le l^\prime \le n_w$. Each block $\Cc_{l l^\prime}$ is a $n_h \times n_h$ matrix and corresponds to the block $\B_{l l^\prime}$ of $\P^s$. Consequently, the matrix $\Pc = \Cc \odot \P^s $ will inherit a block structure from $\P^s$ and $\Cc$, namely
 \makeatletter 
 \if@submission \makeatother
\else \begin{adjustwidth}{-1.8cm}{-1.8cm} \fi
   \begin{align} \label{eq:blockStructurePHat} 
    \Cc \odot \P^s&= \begin{bmatrix}
        \Cc_{11} \odot \B_{11} & \Cc_{12}\odot \B_{12} & \dots & \Cc_{1l^\prime} \odot \B_{1l^\prime}& \dots & \Cc_{1n_{w}} \odot \B_{1n_{w}}\\[10pt]
        \Cc_{21} \odot \B_{21}& \Cc_{22}\odot \B_{22} & \dots & \Cc_{2l^\prime} \odot \B_{2l^\prime}& \dots & \Cc_{2n_{w}}\odot \B_{2n_{w}}\\[10pt]
        \vdots & \vdots & \ddots & \vdots & \iddots & \vdots\\[10pt]
        \Cc_{l1} \odot \B_{l1}& \Cc_{l2} \odot \B_{l2}& \dots & \Cc_{ll^\prime} \odot \B_{ll^\prime} & \dots & \Cc_{ln_{w}} \odot \B_{ln_{w}}\\[10pt]
        \vdots & \vdots & \iddots & \vdots & \ddots & \vdots\\[10pt]
        \Cc_{n_{w}1} \odot \B_{n_{w}1}& \Cc_{n_{w}2} \odot \B_{n_{w}2}& \dots & \Cc_{n_{w}l^\prime} \odot \B_{n_{w}l^\prime}& \dots & \Cc_{n_{w}n_{w}}\odot \B_{n_{w}n_{w}}\\[10pt]
    \end{bmatrix} 
\end{align}
\makeatletter 
 \if@submission \makeatother
\else \end{adjustwidth} \fi

\subsection{Definition of horizontal localization matrix} \label{sec:HorizontalLocalization}
As mentioned in section \ref{sec:introduction}, the swift algorithm for $(\Pc)^{-\frac{1}{2}}$ that we are introducing in this article, is defined in the case when $\Cc$ is a horizontal localization matrix. What do we mean by horizontal localization matrix? To give a definition of this concept, let us first remind ourselves that each model variable is identified by the horizontal and vertical coordinates of its grid point location and its variable kind (i.e. whether it is a temperature, a pressure, a relative humidity, \dots). Let us also remind ourselves that, in the element-wise product $\Cc \odot \P^s$, each element of $\Cc$ multiples a sample covariance between two model variables. In this sense, each element of $\Cc$ corresponds to a pair of model variables. In general, each element of $\Cc$ will be a function of the grid point coordinates and of the variables kinds of its two corresponding model variables. Now, let us define a \emph{horizontal localization} matrix $\Cc$ as a localization matrix whose elements do not depend on the vertical coordinates of their corresponding pair of model variables. Moreover, in the present article, we will restrict our study to a horizontal localization matrix $\Cc$ whose elements do not depend on the model variable kinds, either. A generalization of our research work to cases when the elements of $\Cc$ \emph{do} depend on the model variables vertical coordinates and/or on their kinds is left for a future investigation.\par
Now, our efforts to choose a convenient order for the elements of the vectors $\e_k$ and $\z_k$ will bring their fruits. Namely, as we have seen, that order implies a block structure for some matrices, such as structure \eqref{eq:blockStructureC} for $\Cc$. And, \eqref{eq:blockStructureC} is very handy now because it significantly simplifies in the case of the $\Cc$ considered in this article (i.e. horizontal localization matrix; independent of the variables kind). More specifically, a moment of reflection will convince the reader that all blocks $\Cc_{ll^\prime}$ ($1 \le l \le n_w$ and $1 \le l^\prime \le n_w$) in $\Cc$ equal each other. Let us denote all these equal blocks $\Cc_{ll^\prime}$ by simply $\Cc_{\textbf{b}}$, where the subscript $\phantom{i}_{\textbf{b}}$ stands for \emph{block}. Namely, $\Cc$ takes the simple form:
\begin{align} \label{eq:blockStructureCEqualBlocks}
    \Cc &= \begin{bmatrix}
        \Cc_{\textbf{b}} & \Cc_{\textbf{b}} & \dots & \Cc_{\textbf{b}} & \dots & \Cc_{\textbf{b}}\\[10pt]
        \Cc_{\textbf{b}} & \Cc_{\textbf{b}} & \dots & \Cc_{\textbf{b}} & \dots & \Cc_{\textbf{b}}\\[10pt]
        \vdots & \vdots & \ddots & \vdots & \iddots & \vdots\\[10pt]
        \Cc_{\textbf{b}} & \Cc_{\textbf{b}} & \dots & \Cc_{\textbf{b}} & \dots & \Cc_{\textbf{b}}\\[10pt]
        \vdots & \vdots & \iddots & \vdots & \ddots & \vdots\\[10pt]
        \Cc_{\textbf{b}} & \Cc_{\textbf{b}} & \dots & \Cc_{\textbf{b}} & \dots & \Cc_{\textbf{b}}\\[10pt]
    \end{bmatrix}
\end{align}
\subsection{Matrix square roots} \label{sec:SquareRoots}
Let $\Cc_{\textbf{b}}^{\frac{1}{2}}$ be a $n_h \times L$ square root of the matrix block $\Cc_{\textbf{b}}$, i.e. $ \Cc_{\textbf{b}}^{\frac{1}{2}} (\Cc_{\textbf{b}}^{\frac{1}{2}})\tp$. Hence, as the reader can easily verify, a square root $\Cc^{\frac{1}{2}}$ of the localization matrix $\Cc$ is given by

\begin{align} \label{eq:SqrtOfC}
    \Cc^{\frac{1}{2}} &= \begin{bmatrix}
        \Cc_{\textbf{b}}^{\frac{1}{2}} \\[10pt]
        \Cc_{\textbf{b}}^{\frac{1}{2}} \\[10pt]
        \vdots\\[10pt]
        \Cc_{\textbf{b}}^{\frac{1}{2}}\\[10pt]
        \vdots \\[10pt]
        \Cc_{\textbf{b}}^{\frac{1}{2}} \\[10pt]
    \end{bmatrix}
\end{align}

As it will be useful in a moment, let us denote the columns of the square root $\Cc^{\frac{1}{2}}$ of the localization matrix $\Cc$ by $\co_l$ with $1 \le l \le L$.\par
Regarding the sample covariance matrix $\P^s$, a square root of it is the matrix $\Z$, as we showed in \eqref{eq:PsAndItsSquareRoot}.  Now, having square roots of both $\Cc$ and $\P^s$, we can utilize a technique described in \cite{Bishop_Hodyss_th_sqrt_loc} to construct a square root of the element-wise product $\Pc =\Cc \odot \P^s$. Namely, let $\Cc^{\frac{1}{2}} \smalltriangleup \Z$ be the matrix whose columns are all possible element-wise products $\co_l \odot \z_k$ between a column $\co_l$ of $\Cc^{\frac{1}{2}}$ and a column $\z_k$ of $\Z$ for all $l$ and $k$ with $1 \le l \le L $ and $1 \le k \le K$. Thus, the so-constructed matrix $\Cc^{\frac{1}{2}} \smalltriangleup \Z$ is $n \times KL$. As shown in \cite{Bishop_Hodyss_th_sqrt_loc}, $\Cc^{\frac{1}{2}} \smalltriangleup \Z$ is a square root of $\Cc \odot \P^s$, i.e.
\begin{equation}
    (\Cc^{\frac{1}{2}} \smalltriangleup \Z) \, (\Cc^{\frac{1}{2}} \smalltriangleup \Z)\tp = \Cc \odot \P^s=  \Pc
\end{equation}
We will now express the matrix $ (\Pc)^{\frac{1}{2}}\doteq \Cc^{\frac{1}{2}} \smalltriangleup \Z  $ through a formula involving basic matrix operations, such as the usual matrix product. Before doing that, let us establish some notation. Given any positive integer $d$, and any $d$-dimensional column vector $\a$, we will denote the $d \times d$ diagonal matrix whose diagonal elements are the elements of $\a$ by $\diag (\a)$. For this, we will use the (natural) convention that the order of the elements in the diagonal of the matrix $\diag (\a)$ is the same as the order in which they are listed in the column vector $\a$. With this notation, the matrix $ (\Pc)^{\frac{1}{2}}\doteq \Cc^{\frac{1}{2}} \smalltriangleup \Z  $ can be written as
\begin{equation} \label{eq:SQRTofPCModulation}
   (\Pc)^{\frac{1}{2}}\doteq \Cc^{\frac{1}{2}} \smalltriangleup \Z =\left[ \diag(\z_1) \Cc^{\frac{1}{2}} , \, \diag(\z_2) \Cc^{\frac{1}{2}} , \, \dots, \, \diag(\z_K) \Cc^{\frac{1}{2}} \right]
\end{equation}
where the notation with square brackets and commas denotes the concatenation of the matrices $\diag(\z_k) \Cc^{\frac{1}{2}}$ ($1 \le k \le K$).\par

Then, remembering the block structure \eqref{eq:SqrtOfC} of the square root $(\Cc)^{\frac{1}{2}}$ of the localization matrix $\Cc$, we find:

\begin{align} \label{eq:DiagZC}
    \diag(\z_k) \Cc^{\frac{1}{2}} &= \begin{bmatrix}
                \diag(\z_{1,k}) \\[10pt]
                \diag(\z_{2,k}) \\[10pt]
                \vdots   \\[10pt]
                \diag(\z_{l,k}) \\[10pt]
                \vdots   \\[10pt]
                \diag(\z_{n_w,k})
            \end{bmatrix}  \Cc_{\textbf{b}}^{\frac{1}{2}}    \qquad \text{for all} \; k \; \text{with} \; 1 \le k \le K
\end{align}
where $\z_{k,l}$ ($1 \le k \le K$ and $1 \le l \le n_w$) are the segments (defined in \eqref{eq:ZSegmentDef}) of the vectors $\z_k$.\par
Using \eqref{eq:DiagZC}, equation \eqref{eq:SQRTofPCModulation} for the matrix $(\Pc)^{\frac{1}{2}}= \Cc^{\frac{1}{2}} \smalltriangleup \Z$ can be rewritten as
 \NiceMatrixOptions{xdots/shorten=2em,xdots/radius=0.65pt}

 \begin{adjustwidth}{-1.3cm}{-1.3cm}
\begin{align} \label{eq:SQRTofPCSegmentsBlockC}
   (\Pc)^{\frac{1}{2}}  &= \begin{bmatrix}
                \diag(\z_{1,1}) & \diag(\z_{1,2}) & \cdots & \diag(\z_{1,K})\\[10pt]
                \diag(\z_{2,1}) & \diag(\z_{2,2}) & \cdots & \diag(\z_{2,K})\\[10pt]
                \vdots  & \vdots & \cdots& \vdots\\[10pt]
                \diag(\z_{l,1}) & \diag(\z_{l,2}) & \cdots & \diag(\z_{l,K})\\[10pt]
                \vdots & \vdots& \cdots& \vdots  \\[10pt]
                \diag(\z_{n_w,1})& \diag(\z_{n_w,2}) & \cdots & \diag(\z_{n_w,K})
            \end{bmatrix}
            \begin{bNiceMatrix}
                \Cc_{\textbf{b}}^{\frac{1}{2}}  &  &  &  \Block{3-3}<\Huge>{0}& & \\[10pt]
                 & \Cc_{\textbf{b}}^{\frac{1}{2}}  &  &  & & \\[10pt]
                 &  &\Ddots &   & & \\[10pt]
                \Block{3-3}<\Huge>{0}  & &  & \Cc_{\textbf{b}}^{\frac{1}{2}} & & \\[10pt]
                 & & &  &\Ddots &  \\[10pt]
                &  &  & & & \Cc_{\textbf{b}}^{\frac{1}{2}}
             \end{bNiceMatrix}
\end{align}
\end{adjustwidth}

For notational simplicity, let us denote the first matrix of the right-hand side of equation \eqref{eq:SQRTofPCSegmentsBlockC} by $\A$, i.e.

\begin{align} \label{eq:ADef}
  \A &\doteq \begin{bmatrix}
                 \diag(\z_{1,1}) & \diag(\z_{1,2}) & \cdots & \diag(\z_{1,K})\\[10pt]
                \diag(\z_{2,1}) & \diag(\z_{2,2}) & \cdots & \diag(\z_{2,K})\\[10pt]
                \vdots  & \vdots & \cdots& \vdots\\[10pt]
                \diag(\z_{l,1}) & \diag(\z_{l,2}) & \cdots & \diag(\z_{l,K})\\[10pt]
                \vdots & \vdots& \cdots& \vdots  \\[10pt]
                \diag(\z_{n_w,1})& \diag(\z_{n_w,2}) & \cdots & \diag(\z_{n_w,K})
            \end{bmatrix}
\end{align}


\subsection{Assumption about the number of ensemble members} \label{sec:AssumptionAboutK}
Now, before proceeding further with our derivation, we will make an assumption. Namely, we will assume that the number $n_w$ of model variables in a grid vertical column equals the number $K$ of ensemble members, i.e.
\begin{equation} \label{eq:nwEqualsK}
    n_w=K
\end{equation}
We will use this assumption below. Thus, in this article, our swift algorithm to compute a left inverse of $(\Pc)^{\frac{1}{2}}$ will be developed in the special case when $n_w=K$.  Now, before resuming our derivation, let us look at assumption \eqref{eq:nwEqualsK} within the context of operational weather forecasting. For today's high resolution atmospheric models, the number of variables variables per grid point is $n_{G} \sim 10$ and the number of horizontal layers is $n_v \sim 10^2$. Thus, the number of model variables per grid vertical column is $n_w= n_G \, n_v \sim 10^3$. Hence, assumption \eqref{eq:nwEqualsK} would imply a number of ensemble members $K \sim 10^3$. Let us recall here that the intended, direct application of the research work in this article is performing the standardized variables transform $\xs = (\Pc)^{\frac{1}{2}} \x$ with $\Pc$ constructed as $\Pc= \Cc \odot \P^s$. In particular, $\P^s$ is the sample covariance matrix of a climatological ensemble of forecast error proxies. In turn, these error proxies are drawn from historical archives. In this context, to fulfil assumption \eqref{eq:nwEqualsK}, we simply need to choose the number $K$ of climatological error proxies (drawn from those archives) to equal $n_w$, which, as mentioned, is $n_w \sim 10^3$. In different contexts, such as for operational forecast ensembles at a given data assimilation cycle, there may be less flexibility in the choice of $K$. Envisaging applications of our work to such contexts, and, for those applications, generalizing our swift algorithm to the case when $n_w$ does not necessarily equal $K$ could be an interesting research route, which we leave to future research work.
\par
Now, let us go back to our derivation. By assuming that $n_w = K$, we turned the $n \times n_h K$ matrix $\A$ into a square matrix, more precisely, a $n \times n$ matrix because $n_h K = n_h n_w = n$. The core aspect of our swift algorithm is that, as we will see, in the case when $n_w = K$, the singular value decomposition of $\A$ is a parallelizable problem!\par

\subsection{Description of the algorithm to construct $(\Pc)^{-\frac{1}{2}}$} \label{sec:AlgorithmDescription}
In a moment, we will give equation \eqref{eq:sdvOfA}, which is a formula for the singular value decomposition of $\A$. A proof of \eqref{eq:sdvOfA} can be found in section \ref{app:SwiftAlgorithmTechnical} (in the supporting information of this article), which is the kernel, technical part of the derivation of our swift algorithm.\par
To write formula \eqref{eq:sdvOfA}, we first need to construct its ingredients. To do that, let us start by recalling the structure of the matrix $\A$ given by \eqref{eq:ADef}, in which the segments $\z_{l,k}$ appear.  Let us also recall the concept, introduced in section \ref{sec:segmentStructure}, of \emph{quantities of the $m$-th grid vertical column} for the elements of the segments $\z_{l,k}$. Namely, these quantities are, for a \textbf{given fixed} $m$, the collection of $m$-th vector elements $(\z_{l,k})_m$ across all segments $\z_{l,k}$.  Now, for any given integer $m$ with $1 \le m \le n_h$, let us define the $n_w \times n_w$ matrix $\A_m$ as the output of the following simple procedure:
\begin{itemize}
    \item[$\rhd$] Consider the matrix $\A$ and delete, from it, all the rows and columns \emph{apart from} those rows and columns containing quantities of the $m$-th grid vertical column.
\end{itemize}
By carefully inspecting the structure of the matrix $\A$ in equation \eqref{eq:ADef}, one realizes that it is possible to give a different, but equivalent definition of the matrices $\A_m$ with respect to the one just given above. Namely, for any fixed integer $m$ with $1 \le m \le n_h$, we can define $\A_m$ as the matrix that we obtain from $\Z$ by deleting all  rows of $\Z$ \emph{apart from} those rows containing quantities of the $m$-th grid vertical column (let us recall that $\Z$ is the square root of the sample covariance matrix $\P^s$ as shown by equation \eqref{eq:PsAndItsSquareRoot}).\par
Having defined the matrices $\A_m$ (for every $m$ with $1 \le m \le n_h$), let us then consider a singular value decomposition of each of them
\begin{equation} \label{eq:svdOfAm}
  \A_m =   \Eb_m \Lambdam \Fb_m\tp 
\end{equation}
where $\Lambdam$ is a $n_w \times n_w$ diagonal matrix and $\Eb_m$ and $\Fb_m$ are $n_w \times n_w$ orthogonal matrices.\par 
Now, for any given integer $m$ with $1 \le m \le n_h$, let us define $\Check{\Eb}_m$ as the matrix that we obtain starting from $\Eb_m$ and performing the following operations:
\begin{enumerate}
    \item \label{expansionStep:middle} Between each pair of consecutive rows of $\Eb_m$, insert $n_h-1$ rows of zeros;
    \item \label{expansionStep:top}Insert $m-1$ rows of zeros on top of the matrix obtained at the previous step;
    \item \label{expansionStep:bottom}Insert $n_h-m$ rows of zeros to the bottom of the matrix obtained at the previous step.
\end{enumerate}
In other words, the matrix $\Check{\Eb}_m$ has the following block structure:
\begin{align} \label{eq:EmCheck}
    \Check{\Eb}_m &= \begin{bmatrix}
        \Check{\Eb}_{m,1} \\[10pt]
        \Check{\Eb}_{m,2} \\[10pt]
        \vdots\\[10pt]
        \Check{\Eb}_{m,l}\\[10pt]
        \vdots \\[10pt]
         \Check{\Eb}_{m,n_w} \\[10pt]
    \end{bmatrix}
\end{align}
where each block $\Check{\Eb}_{m,l}$ ($1 \le l \le n_w$) is $n_h \times n_w$ with all rows filled with zeros except its $m$-th row, which equals the $l$-th row of the matrix $\Eb_m$. Namely,
\begin{align} \label{eq:EmCheckBlock}
  \Check{\Eb}_{m,l} &\doteq \begin{bmatrix}
                 0 & 0 & \cdots & 0\\[10pt]
                 0 & 0 & \cdots & 0\\[10pt]
                \vdots  & \vdots & \cdots& \vdots\\[10pt]
                 0 & 0 & \cdots & 0\\[10pt]
              (\Eb_m)_{l,1} & (\Eb_m)_{l,2} & \cdots & (\Eb_m)_{l,n_w}\here{$\phantom{i}$}{fromhereraw}\\[10pt]
                 0 & 0 & \cdots & 0\\[10pt]
                \vdots & \vdots& \cdots& \vdots  \\[10pt]
               0& 0 & \cdots & 0
            \end{bmatrix}
\end{align}
\begin{tikzpicture}[remember picture, overlay]
\node[font=\small,  right=45pt of fromhereraw] (tohere) {
\begin{minipage}{0.25\textwidth}
    $m$-th row of $\Check{\Eb}_{m,l}$ \\which equals\\
the $l$-th row of $\Eb_m$
\end{minipage}};
\node [font=\small,  right=-10pt of fromhereraw] (fromhere) {};
\draw[Latex-, orcidlogocol, line width=2.3pt] (fromhere) edge ([shift={(-2.35cm, 0.5cm)}]tohere);
\end{tikzpicture}
\newline \vspace{0em}%
where $[ (\Eb_m)_{l,1}\; (\Eb_m)_{l,2} \; \cdots \; (\Eb_m)_{l,n_w} ]$ is the $l$-th row of the matrix $\Eb_m$ placed as the $m$-th row of $ \Check{\Eb}_{m,l}$.\par
Now, for any given integer $m$ with $1 \le m \le n_h$, let us define $\Check{\Fb}_m$ as the matrix that we obtain by performing the same operations in the three steps above, but starting with $\Fb_m$ instead of $\Eb_m$ at step \ref{expansionStep:middle}. Regarding their dimensions, it is easy to verify that $\Check{\Eb}_m$ and $\Check{\Fb}_m$ are $n \times n_w$ matrices (see appendix \ref{app:SwiftAlgorithmTechnical} in the supporting information of this article).\par
We now have all needed ingredients to write the following formula giving a singular value decomposition of $\A$ 
\begin{equation} \label{eq:sdvOfA}
    \A  =  \Eb \:\boldsymbol{\Lambda} \: \Fb\tp
\end{equation}
where $\boldsymbol{\Lambda}$ is the diagonal matrix

\begin{equation} \label{eq:LambdaBlocks}
   \boldsymbol{\Lambda}= \begin{bNiceMatrix}
               \boldsymbol{\Lambda}_1 &  &  &  \Block{3-3}<\Huge>{0}& & \\[10pt]
                 & \boldsymbol{\Lambda}_2  &  &  & & \\[10pt]
                 &  &\Ddots &   & & \\[10pt]
                \Block{3-3}<\Huge>{0}  & &  & \Lambdam & & \\[10pt]
                 & & &  &\Ddots &  \\[10pt]
                &  &  & & & \boldsymbol{\Lambda}_{n_h}
    \end{bNiceMatrix} 
\end{equation}

and $\Eb$ and $\Fb$ have the following block structure:
\begin{equation} \label{eq:EBlocks}
   \Eb= \begin{bNiceMatrix}
             \Check{\Eb}_1 & \Check{\Eb}_2 & \dots &\Check{\Eb}_m & \dots &  \Check{\Eb}_{n_h}
    \end{bNiceMatrix} 
\end{equation}
and
\begin{equation} \label{eq:FBlocks}
   \Fb= \begin{bNiceMatrix}
             \Check{\Fb}_1 & \Check{\Fb}_2 & \dots &\Check{\Fb}_m & \dots &  \Check{\Fb}_{n_h}
    \end{bNiceMatrix} 
\end{equation}
As mentioned above, equation \eqref{eq:sdvOfA} will be proven in section \ref{app:SwiftAlgorithmTechnical} in the supporting information of this article. Moreover, in that section, we will show that the matrices $\Eb$ and $\Fb$ are orthogonal and, therefore, formula \eqref{eq:sdvOfA} is indeed a singular value decomposition of $\A$.\par
Thus, with \eqref{eq:LambdaBlocks}, \eqref{eq:EBlocks} and \eqref{eq:FBlocks}, we have expressed left and right singular vectors and singular values of $\A$ in terms of left and right singular vectors and singular values of the matrices $ \A_m = \Eb_m \Lambdam \Fb_m\tp$ ($1 \le m \le n_h$). In other words, by performing the singular value decompositions  $\A_m = \Eb_m \Lambdam \Fb_m\tp$ in parallel, we get the singular value decomposition $\A = \Eb \:\boldsymbol{\Lambda} \: \Eb\tp$. Namely, we have parallelized the singular value decomposition of $\A$. Each parallel branch of this algorithm consists in performing a singular value decomposition of a matrix $\A_m$ for a given $m$ ($1 \le m \le n_h$). %
A moment of reflection about the definition of $\A_m$ given above will convince the reader that $\A_m$ is a $n_w \times n_w$ matrix, for any given $m$ with $1 \le m \le n_h$. As mentioned earlier, in today's operational weather forecasting, $n_w \sim 10^3$. Thus, for each matrix $\A_m$, a singular value decomposition can be swiftly performed with today's computers. \colorboxToBeHidden{PaleYellow}{How many nodes does a supercomputer need to have for this? Or cores \dots?} \par
Now, let us recall that in this section we are aiming at a formula for a left inverse of the square root $\Cc^{\frac{1}{2}} \smalltriangleup \Z$ of the matrix $\Pc =\Cc \odot \P^s$. To proceed further towards such a formula, we will assume that the square matrix $\A$ is full rank, and, thus, invertible. In operational applications, to construct an $\A$ fulfilling this property, a twofold strategy can be adopted. The first component of this strategy is a procedure which randomly discards very few members from the ensemble. This approach will be explained in section \ref{sec:GeneratingEnsemble} and it will be utilized in our experiment, showing that this will allow us to perform our swift algorithm with very good accuracy (section \ref{sec:Accuracy}). As a second component of the above-mentioned strategy, one can, if needed, exclude the balanced part of the model variables involved in linear balance equations from the atmospheric state vector \parencite{Bannister2008}. Both components of this strategy aim at eliminating causes of rank deficiency for the matrix $\A$. An explanation of these technical aspects, on which we have just touched here, can be found in section \ref{sec:RankOfA} (in the supporting information of this article), to which the interested reader is referred. A different route to address these aspects would be to generalize of our swift algorithm to the case when $\A$ is \emph{not} full rank. We leave the development of this research avenue to future work. Now, let us go back to our derivation.\par
As we assumed $\A$ to be invertible, we can now compute its inverse. For this, we can utilize the singular value decomposition of $\A$ obtained through the above-explained swift parallel procedure, and we can thus get the inverse of $\A$ through the following formula
\begin{equation} \label{eq:svdInverseOfA}
    \A^{-1}= \Fb \boldsymbol{\Lambda}^{-1} \Eb\tp
\end{equation}
Let us emphasize that $\boldsymbol{\Lambda}$ is diagonal, so computing its inverse $\boldsymbol{\Lambda}^{-1}$ is computationally feasible even in an operational context.\par
Let us now recall that, in our notation, $\Cc_{\textbf{b}}^{\frac{1}{2}}$ denotes a $n_h \times L$ square root of the matrix block $\Cc_{\textbf{b}}$, which in turn appears in the localization matrix $\Cc$. The block $\Cc_{\textbf{b}}$ is a $n_h \times n_h$ matrix. Incidentally, let us emphasize that, often, in operational applications, $\Cc_{\textbf{b}}$ will not be full rank. Now, let $\Cc_{\textbf{b}} = \tilde{\V}_{\textbf{b}}  \, \tilde{\Deltab}_{\textbf{b}}  \tilde{\V}_{\textbf{b}} \tp$ be an eigendecomposition of $\Cc_{\textbf{b}}$. \colorboxToBeHidden{PaleYellow}{Is this decomposition swift?} Then, let $\Deltab_{\textbf{b}}$ be the diagonal matrix of the non-zero eigenvalues of $\Cc_{\textbf{b}}$.  More precisely, let $\Deltab_{\textbf{b}}$ the matrix that we get from $\tilde{\Deltab}_{\textbf{b}} $ by deleting its rows and columns containing a zero eigenvalue. Moreover, let us define $\V_{\textbf{b}}$ as the matrix that we get from $\tilde{\V}_{\textbf{b}}$ by deleting the columns which are eigenvectors corresponding to a zero eigenvalue. Thus, we can write the block $\Cc_{\textbf{b}}$ as $\Cc_{\textbf{b}} = \V_{\textbf{b}}  \, \Deltab_{\textbf{b}}  \V_{\textbf{b}} \tp$ (let us refer to this decomposition as an \emph{economical eigendecomposition} of $\Cc_{\textbf{b}}$). Let us recall that 
$ \tilde{\V}_{\textbf{b}}$ is an orthogonal matrix. Hence, in particular, $ \tilde{\V}_{\textbf{b}}\tp  \tilde{\V}_{\textbf{b}} = \I_{n_h}$. From this, we get
\begin{equation} \label{eq:OrthogonalityDeltaBlock}
   \V_{\textbf{b}}\tp \, \V_{\textbf{b}} = \I_L 
\end{equation}
where $L$ is the number of non-zero eigenvalues of $\Cc_{\textbf{b}}$ and $\I_L$ is the $L \times L$ identity matrix.\par
Let us now specify our choice for the square root $\Cc_{\textbf{b}}^{\frac{1}{2}}$ as
\begin{equation}
    \Cc_{\textbf{b}}^{\frac{1}{2}} = \V_{\textbf{b}}  \, \Deltab_{\textbf{b}}^{\frac{1}{2}}
\end{equation}
It is now convenient to define the block diagonal matrices

\begin{equation} \label{eq:VBlocks}
   \V \doteq \begin{bNiceMatrix}
              \V_{\textbf{b}} &  &  &  \Block{3-3}<\Huge>{0}& & \\[10pt]
                 & \V_{\textbf{b}}  &  &  & & \\[10pt]
                 &  &\Ddots &   & & \\[10pt]
                \Block{3-3}<\Huge>{0}  & &  &\V_{\textbf{b}} & & \\[10pt]
                 & & &  &\Ddots &  \\[10pt]
                &  &  & & & \V_{\textbf{b}}
    \end{bNiceMatrix} 
\end{equation}

and

\begin{equation} \label{eq:DeltaBlocks}
   \Deltab \doteq \begin{bNiceMatrix}
               \Deltab_{\textbf{b}} &  &  &  \Block{3-3}<\Huge>{0}& & \\[10pt]
                 &  \Deltab_{\textbf{b}}  &  &  & & \\[10pt]
                 &  &\Ddots &   & & \\[10pt]
                \Block{3-3}<\Huge>{0}  & &  &  \Deltab_{\textbf{b}} & & \\[10pt]
                 & & &  &\Ddots &  \\[10pt]
                &  &  & & &  \Deltab_{\textbf{b}}
    \end{bNiceMatrix} 
\end{equation}

where both $\V$ and $\Deltab$ have $n_w$ blocks along the diagonal. Let us emphasize that, as the blocks $\Deltab_{\textbf{b}}$ are diagonal, the matrix $\Deltab$ is itself diagonal. Thus, finding the square root $\Deltab^{\frac{1}{2}}$ and its inverse $\Deltab^{-\frac{1}{2}}$ is computationally feasible even in an operational context. With the above definitions for $\V$ and $\Deltab$, expression \eqref{eq:SQRTofPCSegmentsBlockC} for the square root $(\Pc)^{\frac{1}{2}}$ of the $\Pc =\Cc \odot \P^s$ can be rewritten as
\begin{equation}
    (\Pc)^{\frac{1}{2}}= \A \V \Deltab^{\frac{1}{2}}
\end{equation}
Moreover, from \eqref{eq:OrthogonalityDeltaBlock}, we get
\begin{equation}
    \V\tp \V = \I_{\ns}
\end{equation}
where $\ns\doteq L \,n_w$ and  $\I_{\ns}$ is the $\ns \times \ns$ identity matrix.\par
Thus, a left inverse $(\Pc)^{-\frac{1}{2}}$ of $(\Pc)^{\frac{1}{2}}$ is 
\begin{equation} \label{eq:InverseSquareRootAlmost}
    (\Pc)^{-\frac{1}{2}}=   \Deltab^{-\frac{1}{2}} \V\tp \A^{-1}
\end{equation}
Substituting expression \eqref{eq:svdInverseOfA} for $\A^{-1}$ into \eqref{eq:InverseSquareRootAlmost} yields
\begin{equation} \label{eq:InverseSquareRootSwift}
    (\Pc)^{-\frac{1}{2}}=   \Deltab^{-\frac{1}{2}} \V\tp \Fb \boldsymbol{\Lambda}^{-1} \Eb\tp
\end{equation}
where every matrix on the right-hand side is obtained in a computationally swift way, as described above. This concludes our derivation of the rapid algorithm to construct the left inverse $(\Pc)^{-\frac{1}{2}}$ of the square root $(\Pc)^{\frac{1}{2}}$ of $\Pc =\Cc \odot \P^s$. Let us now make an essential remark. In operational applications, forming and, thus, storing in memory very large matrices, such as $(\Pc)^{-\frac{1}{2}}$, is not possible due to memory space limits of supercomputers. In this sense, for operational applications, rather than constructing $(\Pc)^{-\frac{1}{2}}$, we are interested in being able to compute products of the form $(\Pc)^{-\frac{1}{2}} \, \x$, where $\x$ is a generic $n$-dimensional vector, in a computationally feasible way without forming the whole matrix $(\Pc)^{-\frac{1}{2}}$. In the next section \ref{sec:ActionOnVector}, utilizing the theory developed in this section \ref{sec:AlgorithmDescription}, we will introduce an algorithm which carries out this task.

\subsection{Swift algorithm for the product between $(\Pc)^{-\frac{1}{2}}$ and a vector} \label{sec:ActionOnVector}

For operational applications, we are particularly interested in having a swift algorithm to left-multiply a generic ($n$-dimensional) atmospheric state vector $\x$ by the matrix $(\Pc)^{-\frac{1}{2}}$, namely, for the product $\xs = (\Pc)^{-\frac{1}{2}}\, \x$, i.e. for the standardized variables transform (equation \eqref{eq:svTransform}). In this section, we will introduce and describe this algorithm. Whereas, a technical derivation of it can be found in section \ref{sec:ActionOnVectorTechnical} in the supporting information of this article. In the next section, we will explain why this algorithm is swift, efficient and convenient, with a particular emphasis on the context of operational weather forecasting applications.\par
Before describing this algorithm, we need to establish some notation concerning the generic state vector $\x$. Let the model variables in $\x$ be listed in the same order as the one used for the error proxy vectors $\e_k$ in our ensemble (see section \ref{sec:segmentStructure}). Thus, the vector $\x$ has a fully analogous segment structure as the error proxy vectors $\e_k$:
\begin{align} \label{eq:xSegments}
    \x &= \begin{bmatrix}
                \x_{1} \\[10pt]
                \x_{2} \\[10pt]
                \vdots   \\[10pt]
                \x_{l} \\[10pt]
                \vdots   \\[10pt]
                \x_{n_w}
            \end{bmatrix}      
\end{align}
where each ``segment'' $\x_{l}$ ($1 \le l \le n_w$) lists all variables of a given kind at a given horizontal grid layer. Let $(\x_{l})_m$ denote the $m$-th element of the ``segment'' $\x_{l}$ ($1 \le m \le n_h$ and $1 \le l \le n_w$). Moreover, analogously to how we proceeded for the $\e_k$'s, for each given $m$ ($1 \le m \le n_h$), let us refer to the collection of variables $(\x_{l})_m$ for all $l$ with $1 \le l \le n_w$, but with $m$ fixed, as the \emph{quantities of the $m$-th grid vertical column}. Then, let us define $\x^{(m)}$ ($1 \le m \le n_h$) as the $n_w$-dimensional vector that we obtain from $\x$ by deleting all its elements apart from quantities of the $m$-th grid vertical column. In other words, the vector $\x^{(m)}$ lists all variables in $\x$ located on the $m$-th vertical grid column (as this will be crucial in what follows, please note the different definitions of the $n_w$-dimensional vectors $\x^{(m)}$ and of the $n_h$-dimensional segments $\x_{l}$). As a last preliminary step, let us mention that, in the algorithm, a $n$-dimensional vector $\w$ will appear. Let this vector $\w$ have the same segment structure of $\x$ (equation \eqref{eq:xSegments}) with segments denoted by $\w_{l}$ ($1 \le l \le n_w$). Furthermore, let the quantities of the $m$-th grid vertical column and the vectors $\w^{(m)}$ be defined for the vector $\w$ as the analogous concepts just defined for $\x$. Now, we have all needed ingredients, and we are thus are ready to describe the algorithm:
\def\RegularTheEnumi{\theenumi}
\renewcommand{\theenumi}{V\arabic{enumi}}
\begin{enumerate}
    \item For each $m$ ($1 \le m \le n_h$), compute $\w^{(m)} \doteq\F_m \Lambdam^{-1} \E_m\tp \, \x^{(m)}$, where $\A_m = \F_m \Lambdam \E_m\tp$ is a singular vector decomposition of the $n_w \times n_w$ matrices $\A_m$ (see equation \eqref{eq:svdOfAm}); \label{V:wm}
    \item Arrange all elements of all vectors $\w^{(m)}$ ($1 \le m \le n_h$) into the $n$-dimensional vector $\w$; \label{V:w}
    \item Consider the $n_h$-dimensional ``segments'' $\w_l$ ($1 \le l \le n_w$) of the vector $\w$; \label{V:wl}
    \item For each $l$ ($1 \le l \le n_w$), compute $\xs_l \doteq  \Deltab_{\textbf{b}}^{-\frac{1}{2}}\V_{\textbf{b}} \,\w_l$. As $\Deltab_{\textbf{b}}^{-\frac{1}{2}}$ is a $L \times L$ matrix (see section \ref{sec:AlgorithmDescription}), the just-defined vectors $\xs_l$ are $L$-dimensional; \label{V:xsl}
    \item Concatenate the ``segments'' $\xs_l$ ($1 \le l \le n_w$) into a $\ns$-dimensional vector $\xs$, where $\ns= L n_w$. \label{V:xs}
\end{enumerate}
\renewcommand{\theenumi}{\RegularTheEnumi}
The vector $\ns$ is the output of the just-described algorithm. It is possible to prove that this vector is the standardized variables transform of the $n$-dimensional vector $\x$, i.e. $\xs = (\Pc)^{-\frac{1}{2}}\, \x$. A proof of this is provided in section \ref{sec:ActionOnVectorTechnical}.\par
To conclude our description of the algorithm \ref{V:wm} to \ref{V:xs}, let us explain its parallelization structure. Step \ref{V:wm} consists of $n_h$ independent procedures, one for each integer $m$ (i.e. each grid vertical column) with $1 \le m \le n_h$. Thus, these procedures can be performed in parallel on different supercomputer processors. Step \ref{V:w} requires communication between processors in that, during this step, the output of the independent procedures of the previous step, i.e. the vectors $\w^{(m)}$, is utilized to build a single vector $\w$. Then, during step \ref{V:wl}, the ``segments'' $\w_l$ of the vector $\w$ are sent to different processors, each of which can independently perform one of the $n_w$ procedures of step \ref{V:xsl} (one procedure for each integer $l$ with $1 \le l \le n_w$). Finally, during step \ref{V:xs}, the output of step \ref{V:xsl} from the different processors is joined into the vector $\xs$.

\subsection{Computational aspects of the swift algorithm} \label{sec:ComputationalAspects}

In this section, we will discuss the computational aspects of the algorithm \ref{V:wm} to \ref{V:xs} (described in the previous section \ref{sec:ActionOnVector}) which performs the standardized variables transform. In particular, focusing on the context of operational weather forecasting applications, we will study: \begin{enumerate*}[label=\textcolor{gray!70}{(\roman*)}]
     \item the \emph{time complexity} of algorithm \ref{V:wm} to \ref{V:xs}, i.e. the order of magnitude of its computing time in relation to the size of the input data;
     \item the \emph{space complexity} of this algorithm, that is, the memory space that this algorithm requires to run as a function of the size of the input.
 \end{enumerate*}\par
Let us start by the time complexity. Let us suppose that our algorithm \ref{V:wm} to \ref{V:xs} has been implemented on a supercomputer. In the following, let $N$ denote the number of parallel processors of this supercomputer. A variety of types of operations are involved in our algorithm, including deleting rows and columns of a matrix, singular value decompositions, computing the inverse of a diagonal matrix, left-multiplying vectors by a matrix, arranging quantities as elements of a vector in a specific order, splitting a vector into smaller vectors, concatenating vectors into a bigger vector. Let us be more specific and list the operations involved in steps \ref{V:wm} to \ref{V:xs} in table \ref{tab:TimeComplexity}, giving the time complexity of each operation using the big $\mathcal{O}$ notation. Let us note that, for the time complexity of an operation that is executed in parallel, we have to distinguish two cases:
\begin{itemize}
    \item The number of processors $N$ is less than or equal to the total number of required independent runs of that operation. In such a case, the time complexity of that operation is given by the time complexity of each of its parallel runs (in big $\mathcal{O}$ notation in table \ref{tab:TimeComplexity}) multiplied by the total number of parallel runs and divided by the number of processors $N$. For example, the second operation given in table \ref{tab:TimeComplexity}, i.e. the singular value decomposition of the $\A_m$'s, consists of $n_h$ independent runs (one for each $m$ with $1 \le m \le n_h$). Thus, in the case when $N \le n_h$, the time complexity of this operation is given by $\mathcal{O}(n_w^3 ) \cdot \frac{n_h}{N}$, where $\mathcal{O}(n_w^3 )$ is the time complexity of a singular value decomposition of one $\A_m$ (for each given $m$).
    \item The number of processors $N$ is greater than the total number of required independent runs of that operation. In such a case, the time complexity of that operation is simply given by the time complexity of one of its independent runs. For example, the second-to-last operation in table \ref{tab:TimeComplexity}) consists of $n_w$ independent runs, in each of which a vector $\xs_l \doteq  \Deltab_{\textbf{b}}^{-\frac{1}{2}}\V_{\textbf{b}} \,\w_l$  is computed for a given $l$ ($1 \le l \le n_w$). 
    Thus, if we consider the case when $N > n_w $, the time complexity for that operation is simply given by the time complexity of one of its independent runs, i.e. $\mathcal{O}(L n_h )  +\mathcal{O}(L^2 ) $. 
\end{itemize}

\begin{figure}[h!]
    \centering
    \setlength{\unitlength}{1\textwidth}
     \begin{picture}(1,0.51)(0,0)
      \put(0.19,0.03){ \includegraphics[width=0.65\textwidth]{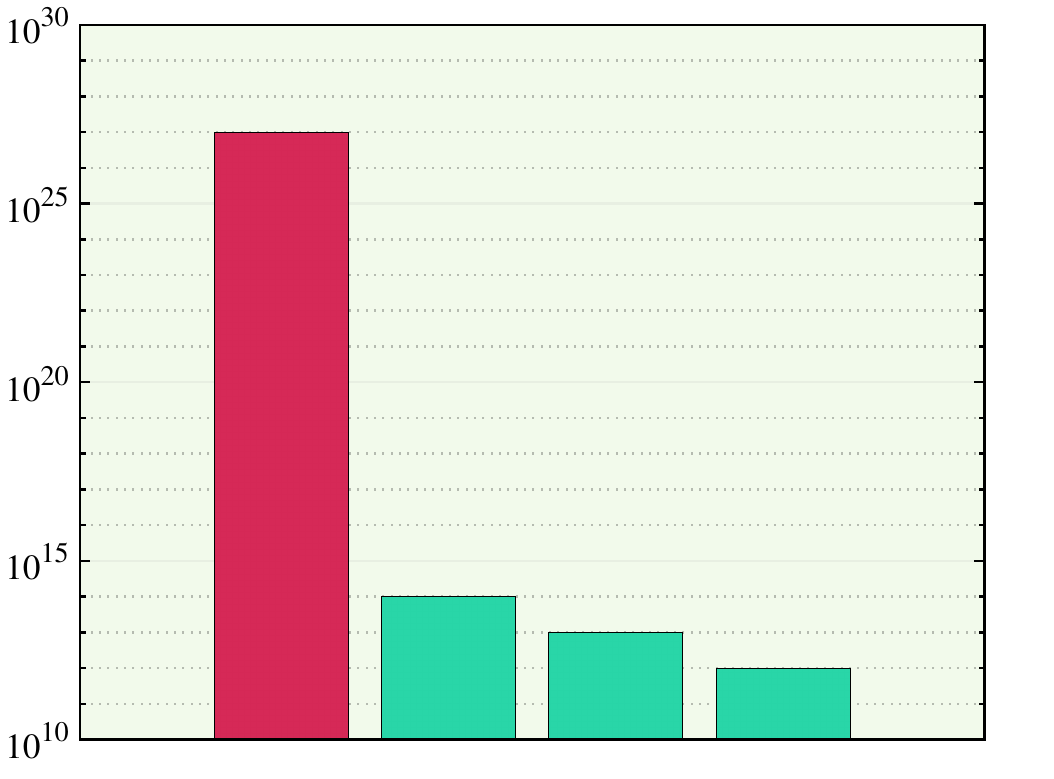}}
        \put(0.163,0.18){\rotatebox{90}{ \large \textit{Computational time} }}
         \put(0.735,0.18){\rotatebox{90}{\Huge \color{black!15!white}\textit{Log scale} }}
         \put(0.425,0.175){\begin{minipage}{0.1\textwidth}
 \vspace{0.5em}
     \centering  $10^4$ \\ \textit{processors}
     \vspace{0.3em}
   \end{minipage}  }
    \put(0.525,0.15){\begin{minipage}{0.1\textwidth}
 \vspace{0.5em}
     \centering  $10^5$ \\ \textit{processors}
     \vspace{0.3em}
   \end{minipage}  }
    \put(0.627,0.13){\begin{minipage}{0.1\textwidth}
 \vspace{0.5em}
     \centering  $10^6$ \\ \textit{processors}
     \vspace{0.3em}
   \end{minipage}  }
   \put(0.313, 0.01) { \fbox{\begin{minipage}{0.1\textwidth}
   \vspace{0.5em}
      \centering \textit{Traditional} \\ \textit{algorithm}
      \vspace{0.3em}
   \end{minipage}}}
 \put(0.524, 0.01) {\fbox{\begin{minipage}{0.1\textwidth}
 \vspace{0.5em}
     \centering  \textit{Swift} \\ \textit{algorithm}
     \vspace{0.3em}
   \end{minipage}}}
    \put(0.37,0.063){\linethickness{0.7mm} \color{lightgray} \line(0,-1){0.028}}
       \put(0.475,0.06){\linethickness{0.7mm} \color{lightgray} \line(0,-1){0.042}}
       \put(0.575,0.06){\linethickness{0.7mm} \color{lightgray}\line(0,-1){0.025}}
       \put(0.68,0.06){\linethickness{0.7mm} \color{lightgray} \line(0,-1){0.042}}
        \put(0.68,0.02){\linethickness{0.7mm} \color{lightgray} \line(-1,0){0.062}}
         \put(0.475,0.02){\linethickness{0.7mm} \color{lightgray} \line(1,0){0.056}}
\end{picture}
\vspace{1em}
    \caption[Time complexities]{In this bar chart, we compare the computational time to perform the standardized variables transform $\xs = (\Pc)^{-\frac{1}{2}}\, \x$ utilizing a traditional algorithm (red bar) or the swift algorithm of section \ref{sec:ActionOnVector} with $10^4$, $10^5$ or $10^6$ parallel processors (mint colour bars). The bar chart is in logarithmic scale. Thus, when running on $10^4$, $10^5$, or $10^6$ parallel processors, the algorithm introduced in section \ref{sec:ActionOnVector} is, respectively, $13$, $14$, or $15$ orders of magnitude faster than a traditional algorithm. In this comparison, both the traditional and the novel algorithm of section \ref{sec:ActionOnVector} start with the following input: \begin{enumerate*}[label=\textcolor{gray!60}{\checkmark}]
     \item a $n$-dimensional vector $\x$, with $n \sim 10^9$;
     \item a $n \times K$ matrix $\Z$  of climatological forecast error proxies, with $K \sim10^3$;
     \item the matrices $ \V_{\textbf{b}} $ ($n_h \times L$) and  $\Deltab_{\textbf{b}}$ ($L \times L$), with $n_h \sim L \sim 10^6$ ($ \V_{\textbf{b}} $ and  $\Deltab_{\textbf{b}}$ are constructed, prior to the algorithms compared in figure \ref{fig:CompareTimeComplexities}, through analytical techniques, such as spherical harmonics; these two matrices define the $n_h \times n_h$ horizontal localization matrix block $\Cc_{\textbf{b}} = \V_{\textbf{b}}  \, \Deltab_{\textbf{b}}  \V_{\textbf{b}} \tp$ for the climatological error covariance matrix $\Pc$; in turn, $\Pc$ is defined from $\Z$ and $\Cc_{\textbf{b}}$ through \eqref{eq:blockStructureCEqualBlocks}, \eqref{eq:PsAndItsSquareRoot} and \eqref{eq:HLocalizedPc}).\end{enumerate*}}
    \label{fig:CompareTimeComplexities}
\end{figure}
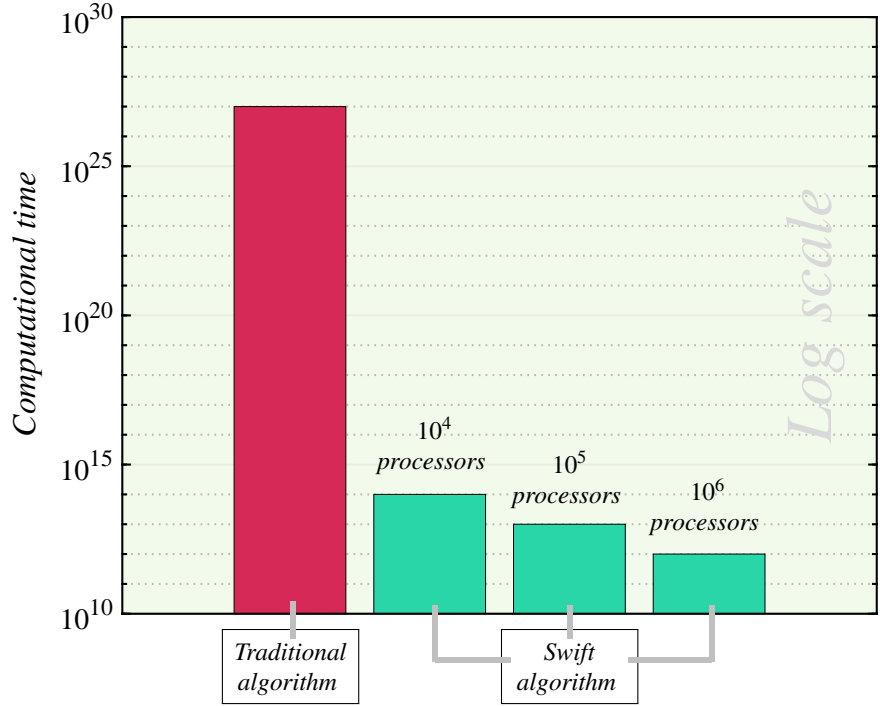

\begin{table}[h!] 
    \centering \renewcommand{\arraystretch}{1.5} 
        \begin{tabular}{cll} 
        \toprule
        \textbf{Step} & \textbf{Algorithmic operation}  & \textbf{Time complexity} \\
        \midrule
        \ref{V:wm} & Forming the $\A_m$'s given $\Z$   & $\mathcal{O}(n K ) \cdot \frac{n_h}{N}$ \hspace{0.3em} if $N \le n_h$ \hspace{0.7em} or \hspace{0.7em} $\mathcal{O}(n K ) $ \hspace{0.3em} if $N > n_h $ \\
        \ref{V:wm} & Singular value decomposition of the $\A_m$'s & $\mathcal{O}(n_w^3 ) \cdot \frac{n_h}{N}$ \hspace{0.3em} if $N \le n_h$ \hspace{0.7em} or \hspace{0.7em} $\mathcal{O}(n_w^3 )$ \hspace{0.3em} if $N > n_h $\\
        \ref{V:wm} & Computing $\Lambdam^{-1}$ given $\Lambdam$ & $\mathcal{O}(n_w ) \cdot \frac{n_h}{N}$ \hspace{0.3em} if $N \le n_h$ \hspace{0.7em} or \hspace{0.7em} $\mathcal{O}(n_w^3 )$ \hspace{0.3em} if $N > n_h $ \\
        \ref{V:wm} & Computing the vectors $\w^{(m)} \doteq\F_m \Lambdam^{-1} \E_m\tp \, \x^{(m)}$    & $\mathcal{O}(n_w^2 ) \cdot \frac{n_h}{N}$ \hspace{0.3em} if $N \le n_h$ \hspace{0.7em} or \hspace{0.7em} $\mathcal{O}(n_w^2 )$ \hspace{0.3em} if $N > n_h $\\
        \ref{V:w} & Arranging the vectors $\w^{(m)}$ into $\w$    & $\mathcal{O}(n)$\\
        \ref{V:wl} & Splitting the vector $\w$ into segments $\w_l$    & $\mathcal{O}(n)$\\
        \ref{V:xsl} & Computing $\Deltab_{\textbf{b}}^{-\frac{1}{2}}$ given $\Deltab_{\textbf{b}}$ & $\mathcal{O}(n_h ) $ \\
        \ref{V:xsl} & Computing the vectors $\xs_l \doteq  \Deltab_{\textbf{b}}^{-\frac{1}{2}}\V_{\textbf{b}} \,\w_l$    & $\left[ \mathcal{O}(L n_h )  +\mathcal{O}(L^2 ) \right]\cdot \frac{n_w}{N}$  \hspace{0.3em} if $N \le n_w $ \hspace{0.7em} or \hspace{0.7em}  $\mathcal{O}(L n_h )  +\mathcal{O}(L^2 ) $  \hspace{0.3em} if $N > n_w $\\
        \ref{V:xs} & Concatenate the segments $\xs_l$ into $\xs$   & $\mathcal{O}(\ns)$\\
        \bottomrule
    \end{tabular}
     \caption{Algorithmic operations involved in steps \ref{V:wm} to \ref{V:xs} to perform the standardized variables transform. For each operation, its time complexity is given using the big $\mathcal{O}$ notation. The number of parallel processors of the supercomputer is denoted by $N$.}
     \label{tab:TimeComplexity}
\end{table}
The time complexity $\Delta t_{\text{swift}}$ of the algorithm \ref{V:wm} to \ref{V:xs} is thus given by the sum (or simply by the dominant terms in the sum) of the time complexities of the operations listed in the rows of table \ref{tab:TimeComplexity}. We are now particularly interested in comparing our algorithm \ref{V:wm} to \ref{V:xs} with a traditional algorithm to perform the standardized variables transform $\xs = (\Pc)^{-\frac{1}{2}}\, \x$. For example, a traditional algorithm would first compute the left inverse $(\Pc)^{-\frac{1}{2}}$ through a (full) singular value decomposition of $(\Pc)^{\frac{1}{2}}$, with time complexity $\mathcal{O} (n \ns^2)$, and, then, by left-multiplying $\x$ by $(\Pc)^{-\frac{1}{2}}$, with time complexity $\mathcal{O} (n \ns)$. Essentially, the total time complexity $\Delta t_{\text{traditional}}$ of a traditional algorithm would be  $\Delta t_{\text{traditional}} \sim \mathcal{O} (n \ns^2) + \mathcal{O} (n \ns) \sim \mathcal{O} (n \ns^2)$. Let us now consider a typical weather forecasting situation, with $n \sim 10^9$, $K=n_w \sim 10^3$, $n_h \sim 10^6$, $L \sim n_h \sim 10^6$, $\ns \sim n \sim 10^9$. As one can easily verify, in this situation, the sum giving $\Delta t_{\text{swift}} $ is dominated by the term $\mathcal{O}(n K ) \cdot \frac{n_h}{N}$ (first row of table \ref{tab:TimeComplexity}) in the case when $N$ is on the order of or less than $10^5$. Whereas, in the case when $N$ is on the order of $10^6$, the sum giving $\Delta t_{\text{swift}} $ is dominated by the terms $\mathcal{O}(n K ) \cdot \frac{n_h}{N}+\mathcal{O}(L n_h )  +\mathcal{O}(L^2 ) $ (first and penultimate rows of table \ref{tab:TimeComplexity}). Hence, as one can straightforwardly verify, in both cases just considered, namely, in other words, when $N$ is on the order of or less than $10^6$, the ratio $\Delta t_{\text{swift}} / \Delta t_{\text{traditional}}$ can be expressed by:
\begin{equation} \label{eq:ratioTimeComplexity}
    \frac{\Delta t_{\text{swift}}}{\Delta t_{\text{traditional}}} \sim 
    \frac{10^{9} \cdot 10^3 \cdot 10^{6} \cdot N^{-1} + 10^{6 \cdot 2} }{10^{9 \cdot 3}}
    \sim \frac{10^{18}}{10^{27}N} + \frac{10^{12}}{10^{27}} = \frac{10^{-9}}{N} +10^{-15}
\end{equation}
From \eqref{eq:ratioTimeComplexity}, we find that, if we run the algorithm \ref{V:wm} to \ref{V:xs} on a single processor only, i.e. for $N=1$, that algorithm is $9$ orders of magnitude faster than a traditional algorithm! Moreover, from \eqref{eq:ratioTimeComplexity}, one finds that, for example, if $N \sim 10^4$ or $N \sim 10^5$ or $N \sim 10^6$, the executional time for the algorithm \ref{V:wm} to \ref{V:xs} is, respectively, $10^{13}$ $10^{14}$ and $10^{15}$ times faster than the traditional algorithm! %
These just-mentioned results about comparing the computational times of the swift algorithm of section \ref{sec:ActionOnVector} and of a traditional algorithm are visually represented in figure \ref{fig:CompareTimeComplexities}. Now, let us consider the case when $N$ is on the order of or greater than $10^6$. It is straightforward to verify that, in this case, the sum giving $\Delta t_{\text{swift}}$ is dominated by the terms $\mathcal{O}(n K ) +\mathcal{O}(L n_h )  +\mathcal{O}(L^2 ) $ (first and penultimate rows of table \ref{tab:TimeComplexity}). Thus, in this case, the computational time does not depend on $N$. Therefore, we can summarize as follows: the algorithm \ref{V:wm} to \ref{V:xs} scales until the number of processors $N$ reaches the order of $10^6$.\par
\par
Having studied the time complexity of algorithm \ref{V:wm} to \ref{V:xs}, let us now discuss its space complexity in the context of operational weather forecasting. To do this, let us start by considering the vectors involved in this algorithm. As one can easily check, the total memory space to store those vectors is equivalent to the memory space required for a few state vectors. This is perfectly feasible in operational weather forecasting. Now, let us focus our attention on the matrices involved in the algorithm \ref{V:wm} to \ref{V:xs}. Thus, let us consider the $\A_m$'s and the matrices involved in their singular value decomposition, i.e. the $\F_m$'s, the $\Lambdam$'s and the $\E_m$'s. Moreover, let us also consider each inverse $\Lambdam^{-1}$ of $\Lambdam$. All these matrices are $n_w \times n_w$. Namely, the space complexity to store each of them is $\mathcal{O}(n_w^2)$ (to be more precise, regarding each $\Lambdam$ and each $\Lambdam^{-1}$, as these are diagonal matrices, the space complexity for each of them is just $\mathcal{O}(n_w)$). As $n_w \sim 10^3$ and, among those matrices, those with different values of the integer $m$ do not need to be stored at the same time, forming these matrices in our algorithm is easily feasible. Finally, let us discuss the space complexity for the $L \times L$ matrices $\Deltab_{\textbf{b}}$, $\Deltab_{\textbf{b}}^{-\frac{1}{2}}$ and the $L \times n_h$ matrix $\V_{\textbf{b}}$. Let us recall that $L$ is at most $n_h \sim 10^6$. Hence, the space complexity for the matrices $\Deltab_{\textbf{b}}$ and $\Deltab_{\textbf{b}}^{-\frac{1}{2}}$, which are diagonal, is at most $\mathcal{O} (n_h)$. Regarding $\V_{\textbf{b}}$, let us recall that this appears in the economical eigendecomposition $\Cc_{\textbf{b}} = \V_{\textbf{b}}  \, \Deltab_{\textbf{b}}  \V_{\textbf{b}} \tp$ of the block $\Cc_{\textbf{b}}$ of the localization matrix $\Cc$. A way to construct $\Cc_{\textbf{b}}$ is to build $\V_{\textbf{b}}$ and $\Deltab_{\textbf{b}}$ first through analytical techniques, such as spherical harmonics techniques (actually, the matrix $\Cc_{\textbf{b}}$ does not even need to be formed explicitly because our algorithm \ref{V:wm} to \ref{V:xs} only requires $\V_{\textbf{b}}$ and $\Deltab_{\textbf{b}}$, and not $\Cc_{\textbf{b}}$, as part of its input). In particular, analytical techniques allow us to compute the elements of $\V_{\textbf{b}}$ when needed thus making storing the entire matrix $\V_{\textbf{b}}$ unnecessary. This strategy makes the algorithmic operations involving $\V_{\textbf{b}}$ operationally viable. This concludes this section \ref{sec:ComputationalAspects}, in which we explained why the algorithm \ref{V:wm} to \ref{V:xs} to perform the standardized variables transform is swift, convenient and practical in the context of operational weather forecasting.

\subsection{The algorithm is applicable to a more general case} 
Along the derivation of our novel algorithm to calculate $(\Pc)^{-\frac{1}{2}}$ (section \ref{sec:AlgorithmDescription}) as well as of our novel algorithm for $(\Pc)^{-\frac{1}{2}}\, \x$ (section \ref{sec:ActionOnVector}), the nature of the ensemble $\e_k$ ($1 \le k \le K$) was fully irrelevant. Namely, these algorithms work (to calculate a left inverse of a square root of $\Cc \odot \P^s$ as well as to calculate the result of the left-multiplication of a vector by that left inverse) irrespective of whether $\{ \e_k \}_{k=1}^K$ is a climatological ensemble of error proxies or, e.g., a forecast ensemble or any other ensemble. Thus, we can be more general, and to this aim, let us modify a few aspects of our notation. More specifically, starting from the next section, unless specified otherwise, $\{ \e_k \}_{k=1}^K$ will be a generic ensemble (not necessarily a climatological ensemble of forecasts error proxies). As before $\P^s$ will be the sample covariance of the ensemble $\{ \e_k \}_{k=1}^K$. The symbol $\Cc$ will always denote a horizontal localization matrix, independent of the variable kinds (as explained in section \ref{sec:HorizontalLocalization}). We will refer to $\Cc \odot \P^s$ as a \emph{horizontally localized sample covariance matrix}, and we will denote it by the more general notation $\hat{\P}^s$ (instead of $\Pc$), i.e. $\hat{\P}^s \doteq \Cc \odot \P^s$. Its square root, given by \eqref{eq:SQRTofPCSegmentsBlockC}, we will denoted as $\hat{\Z}$ (instead of $(\Pc)^{\frac{1}{2}}$). The left inverse of $\hat{\Z}$, given by \eqref{eq:InverseSquareRootSwift}, and computed by our algorithm of section \ref{sec:AlgorithmDescription}, will be denoted by $\hat{\Z}^{-1}$. In this more general setting, in the next section, we shall illustrate and test our novel algorithm of section \ref{sec:AlgorithmDescription} in a synthetic computer experiment.

\section{Experimental test for the swift algorithm} \label{sec:swiftAlgorithmTest}

Let us start by describing the grid we constructed for this synthetic computer experiment.

\subsection{The grid} \label{sec:grid}

We considered a three dimensional grid for the atmosphere of a spherical planet. The grid has $n_v= 64$ horizontal layers. There is only $n_{G}=1$ model variable per grid point. Thus, the number of model variables in each vertical column is $n_w = n_G  n_v =64$. Moreover, each horizontal layer of our grid has $n_h = \num{200}$ grid points, with a longitudinal grid resolution $\Delta\theta =\ang{18}$ and a latitudinal grid resolution $\Delta\varphi =\ang{16}$. The total number $n$ of grid points in our experiment is thus $n= n_w n_h=\num{12800}$. Regarding the grid point coordinates, let $\theta_i$, (with $1 \le i \le n_\theta$ and $n_\theta=20$), $\varphi_j$ (with $1 \le j \le n_\varphi$ and $n_\varphi=10$), and $h_l$ (with $1 \le l \le n_v$) denote all possible values of the longitude, the latitude and the vertical coordinate, respectively, of the grid points in our three-dimensional grid. Thus, each grid point will be identified by the triple $(\theta_i, \varphi_j , h_l)$ of its longitude, its latitude and its vertical coordinate, respectively. As a vertical coordinate, we used a synthetic, unitless coordinate, with the values $h_l$ ($1 \le l \le n_v$) being all integers from $h_1=1$ to $h_{64}=64$, each one identifying a different horizontal grid layer.  For more information about the vertical coordinate and a more detailed description of our grid, the interested reader is referred to section \ref{sec:gridDetailed} in the supporting information of this article.\par 

\subsection{Generating the ensemble} \label{sec:GeneratingEnsemble}

After determining our grid, we need to generate the ensemble members $\e_k$ with $1 \le k \le K$. Let us recall that in designing our algorithm to compute the left inverse $(\hat{\P}^s)^{-\frac{1}{2}}$ of the square  root $(\hat{\P}^s)^{\frac{1}{2}}$ of a horizontally localized sample covariance matrix $\hat{\P}^s =\Cc \odot \P^s$, we assumed that the number $K$ of ensemble perturbations equals the number $n_w$ of variables in each vertical column of our grid. Thus, to be able to apply our swift algorithm, in this experiment we will choose $K=n_w=64$. To generate the ensemble perturbations $\e_k$, we first constructed a synthetic $n \times n$ covariance matrix $\P$. We shall describe the procedure to create $\P$ below. For the moment, let us just mention that, among others, we used spherical harmonics techniques to construct $\P$. Once we have $\P$, we could have generated $K$ vectors $\e_k$ ($1 \le k \le K$) as random draws from a normal distribution $\mathcal{N} (\mathbf{0},\P)$ with mean the $n$-dimensional zero vector $\mathbf{0}$ and covariance matrix $\P$.\par 
As it is easy to verify, this procedure would result in vectors $\z_k$ (recall their definition \eqref{eq:zkDefinition}) which sum to $\mathbf{0}$. As a consequence, one can show that the matrix $\A$ would be singular (we will not show this for the sake of concision). Thus, $\A$ would not be invertible and we would not be able to apply our swift algorithm as described in section \ref{sec:AlgorithmDescription} (let us recall that, in designing our algorithm, we assumed $\A$ to be invertible). We were able to circumvent this issue with a simple solution, that we are going to explain in the following.\par
We first generated an ensemble $\{ \e_{k^\prime}\}_{k^\prime =1}^{K^\prime}$ whose number $K^\prime$ of members is greater than $K$ (more specifically, we chose $K^\prime =68$). To do this, we generated $K^\prime$ random vectors $\e_{k^\prime}$ as random draws for the normal distribution $\mathcal{N} (\mathbf{0},\P)$ mentioned above. Then, we computed the differences 
\begin{equation}
    \e_{k^\prime} - \bar{\e} \qquad 1 \le k^\prime \le K^\prime
\end{equation}
where, here, the ensemble mean $\bar{\e}$ is over all $K^\prime$ vectors $\e_{k^\prime}$:
\begin{equation}
    \bar{\e} = \frac{1}{K^\prime} \sum_{k=1}^{K^\prime} \e_{k^\prime}
\end{equation}
After that, we randomly selected a subset of $K$ differences $ \e_{k^\prime} - \bar{\e}$ out of the $K^\prime$ just computed. More specifically, this subset is given by the vector differences $ \e_{k^\prime} - \bar{\e}$ in which the subscript $k^\prime$ takes any integer value from $1$ to $K^\prime$ \emph{except} $K^\prime - K$ values that we randomly chose to discard (we had $K^\prime - K=4$; the $4$ discarded vector differences were no longer used in our experiment). For notational simplicity, let us rename the $K$ selected vector differences $ \e_{k^\prime} - \bar{\e}$ as $\e_{k}- \bar{\e}$ by re-indexing them with the new subscript $k$ taking all integer values from $1$ to $K$. Then, using these (re-indexed) $K$ selected differences $\e_{k}- \bar{\e}$, we computed the vectors $\z_k$ as
\begin{equation} \label{eq:zInTheExperiment}
    \z_k = \frac{1}{K-1} (\e_{k}- \bar{\e}) \qquad 1 \le k \le K
\end{equation}
These vectors $\z_k$ were, in turn, utilized to construct the non-localized sample covariance matrix $\P^s$ considered in our experiment:
\begin{equation} \label{eq:SampleCovarianceMatrixNonZeroSumPerturbations}
    \P^s =  \sum_{k=1}^K \z_k \, (\z_k)\tp
\end{equation}
Then, the localized sample covariance matrix was $\hat{\P}^s =\Cc \odot \P^s$, where the localization matrix $\Cc$ had the simple block form \eqref{eq:blockStructureCEqualBlocks}, that is,

\begin{align} \label{eq:blockStructureCEqualBlocksExperimentSubsection}
    \Cc &= \begin{bmatrix}
        \Cc_{\textbf{b}} & \Cc_{\textbf{b}} & \dots & \Cc_{\textbf{b}} & \dots & \Cc_{\textbf{b}}\\[10pt]
        \Cc_{\textbf{b}} & \Cc_{\textbf{b}} & \dots & \Cc_{\textbf{b}} & \dots & \Cc_{\textbf{b}}\\[10pt]
        \vdots & \vdots & \ddots & \vdots & \iddots & \vdots\\[10pt]
        \Cc_{\textbf{b}} & \Cc_{\textbf{b}} & \dots & \Cc_{\textbf{b}} & \dots & \Cc_{\textbf{b}}\\[10pt]
        \vdots & \vdots & \iddots & \vdots & \ddots & \vdots\\[10pt]
        \Cc_{\textbf{b}} & \Cc_{\textbf{b}} & \dots & \Cc_{\textbf{b}} & \dots & \Cc_{\textbf{b}}\\[10pt]
    \end{bmatrix}
\end{align}

The $n_h \times n_h$ localization matrix block $\Cc_{\textbf{b}}$ was constructed using spherical harmonics techniques, which will be described below.\par 
We also utilized the vectors $\z_k$ ($1 \le k \le K$) computed through \eqref{eq:zInTheExperiment} to construct the matrix $\A$ through equation \eqref{eq:ADef} of section \ref{sec:SquareRoots}.\par
By using the above-described random selection procedure for the vector differences $\e_{k}- \bar{\e}$, we got a slightly less accurate $\P^s$ with respect to the one we would have obtained by using a standard procedure with the full ensemble $\e_{k^\prime}$ ($1 \le k^\prime \le K^\prime$). This is a small price to pay to get a big advantage. Namely, as the vectors $\z_k$ ($1 \le k \le K$), computed through \eqref{eq:zInTheExperiment}, do not exactly sum to $\mathbf{0}$, by using them, we avoided the resulting matrix $\A$ to be singular and we were able to apply our swift algorithm described above in section \eqref{sec:swiftAlgorithm}.\par  
Let us now proceed further in describing our computer experiment setting. For example, we need to describe the covariance matrix $\P$ of the normal distribution $\mathcal{N} (\mathbf{0} , \P )$ used to generate the full ensemble $\e_{k^\prime}$ ($1 \le k^\prime \le K^\prime$).

\subsection{The synthetic covariance matrix $\P$} \label{sec:ConstructingPMainText}

The $n \times n$ synthetic covariance matrix $\P$, utilized in our experiment, had two essential features:
\begin{enumerate}[label=(f\arabic*)]
   \item \label{feature:horizontal} its horizontal covariance structure was homogenous and isotropic;
    \item \label{feature:vertical} its vertical covariance structure approximately mimicked a situation in which the covariance length scale of that vertical structure is the same at low physical height as well as at high physical height.
\end{enumerate}
But, to obtain feature \ref{feature:vertical}, we had to keep in mind that the vertical coordinate used in our experiment is \emph{not} the physical height. As touched on in section \ref{sec:grid} (and explained in more detail in section \ref{sec:gridDetailed}), the values $h_l$ ($1 \le l \le n_v$) of that vertical coordinate are integers from $h_1=1$ to $h_{64}=64$, each one identifying a different horizontal grid layer, from the lowest one to the highest one. Now, typically, in atmospheric models, the vertical resolution is higher at low height and lower at high height. In other words, the lower horizontal layers are closer to each other than the higher horizontal layers. Therefore, moving upwards, the vertical coordinate used in our experiment increases more slowly when the physical height is low, and more swiftly when the physical height is high. Thus, to obtain feature \ref{feature:vertical}, we designed a vertical structure for $\P$ with a covariance scale (measured in terms of our vertical coordinate coordinate) which is larger when that vertical coordinate is low and smaller when that vertical coordinate is high. For a more extensive explanation of this, the interested reader is referred to section \ref{app:constructingP} in the supporting information of this article. More broadly, section \ref{app:constructingP} describes the procedure that we utilized to construct the matrix $\P$, including all technical aspects of that procedure. For example, in that section, we provide an explanation of the techniques (based on spherical harmonics) utilized to obtain feature \ref{feature:horizontal} in $\P$.\par
Besides the essential features \ref{feature:horizontal} and \ref{feature:vertical}, we constructed $\P$ in such a way that all its diagonal elements (i.e. all variances in that covariance matrix) equalled $1$ (an explanation of this is included in section \ref{app:constructingP}). We wanted all variances in $\P$ to equal $1$ solely because this makes it convenient and easy for the reader to interpret the magnitude of many other quantities in our experiment.\par

Now, before moving on to illustrate the covariance matrix $\P$, it will be useful to recall that, as mentioned in section \ref{sec:grid}, in our synthetic experiment, we only have one variable kind (i.e one variable per grid point). To fix ideas, let us call this variable \emph{temperature}, and let us denote it by $T$. We will use this in a moment. Let us now come to illustrate the covariance matrix $\P$. For this, we will depict some relevant aspects of one of its covariance functions, in panel  \subref{fig:CovarianceFunctionOfP28} of figure \ref{fig:CovarianceFunctions28} and panel \subref{fig:CovarianceFunctionOfP27} of figure \ref{fig:CovarianceFunctions27} (regarding, its horizontal structure), and figure \ref{fig:VerticalProfilesOfPAndPs} (regarding its vertical structure). More specifically, as an example, we chose the covariance function with respect to the grid point location $(\theta_{i^\prime},\varphi_{j^\prime}, h_{l^\prime})$ given by $(\ang{180}, \ang{8},28)$\footnote{To be rigorous, let us specify that the longitude $\theta$, the latitude $\varphi$ and their evenly spaced chosen values $\theta_i$ ($1 \le i \le n_\theta$) and $\varphi_i$ ($1 \le i \le n_\varphi$) mentioned above are always in radians in our experiment (and in the computer code for our experiment). In the text of the article, for readability purposes only, the longitude and latitude values are given in degrees.}. That covariance function is defined as
\begin{equation}
    \eta_{\scaleto{\ang{180}, \ang{8},28}{5pt}} (\theta_i, \varphi_j, h_l) \doteq \cov \left[T(\theta_{i^\prime},\varphi_{j^\prime}, h_{l^\prime}), T(\theta_i, \varphi_j, h_l)  \right]
\end{equation}
which is the covariance between the temperature variable $T(\theta_{i^\prime},\varphi_{j^\prime}, h_{l^\prime})$ at the location $(\theta_{i^\prime},\varphi_{j^\prime}, h_{l^\prime})$ given by $(\ang{180}, \ang{8},28)$ and the temperature $T(\theta_i, \varphi_j, h_l)$ at $(\theta_i, \varphi_j, h_l)$, for every location $(\theta_i, \varphi_j, h_l)$. Hence, $\eta_{\scaleto{\ang{180}, \ang{8},28}{5pt}} (\theta_i, \varphi_j, h_l)$ is a function of the grid point location $(\theta_i, \varphi_j, h_l)$. Panel  \subref{fig:CovarianceFunctionOfP28} of figure \ref{fig:CovarianceFunctions28} and panel \subref{fig:CovarianceFunctionOfP27} of figure \ref{fig:CovarianceFunctions27} are two longitude-latitude charts, representing the values of the covariance function $\eta_{\scaleto{\ang{180}, \ang{8},28}{5pt}} (\theta_i, \varphi_j, h_l)$ on the horizontal layer with vertical coordinate $h_{28}=28$ and $h_{27}=27$, respectively. Whereas, in figure \ref{fig:VerticalProfilesOfPAndPs}, the light blue line represents the values of $\eta_{\scaleto{\ang{180}, \ang{8},28}{5pt}} (\theta_i, \varphi_j, h_l)$ as a function of the vertical coordinate $h_l$ ($1 \le l \le n_v$) with the longitude $\theta_i$ and the latitude $\varphi_j$ fixed and given, in degrees, by $\ang{180}$ and $\ang{8}$, respectively. Regarding the green line in figure \ref{fig:VerticalProfilesOfPAndPs}, this refers to the sample covariance matrix $\P^s$ and it will be explained below. 
\newlength{\InterPanelsInverseSqrt}
\setlength{\InterPanelsInverseSqrt}{0.4cm}

\begin{figure}[!]
    \centering
 \includegraphics[width=0.5\textwidth]{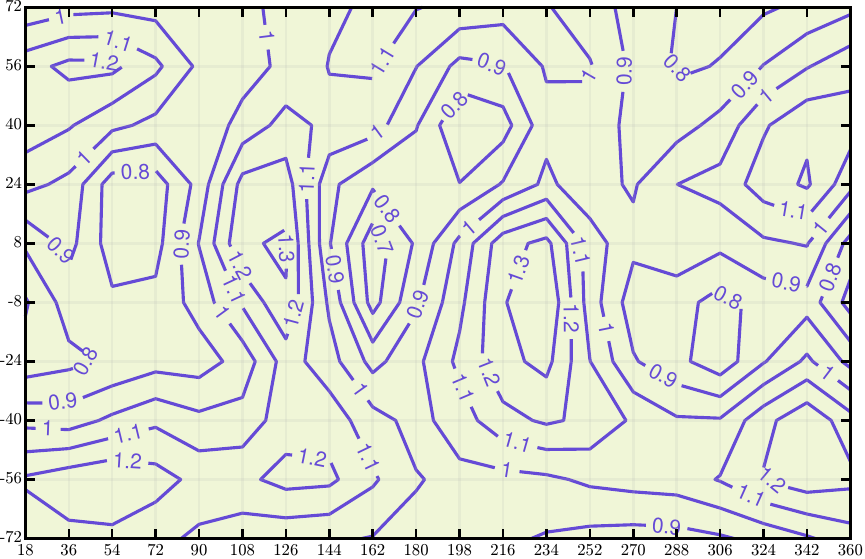} 
\caption{In this longitude-latitude chart, for a fixed vertical coordinate $h_{28}=28$, we represented the values of the variance field (diagonal elements) of the sample covariance matrix $\P^s$ of our synthetic ensemble.}
    \label{fig:VarinceFieldOfPs}
\end{figure}

\subsection{The localization matrix} \label{sec:localizationExperiment}

As mentioned above, the $n \times n$ matrix $\C$ had the simple block form \eqref{eq:blockStructureCEqualBlocksExperimentSubsection}, with each block $\C_{\textbf{b}}$ being a $n_h \times n_h$ matrix. Namely, $\C$ was made of $n_v^2$ blocks $\C_{\textbf{b}}$ in total, arranged in a $n_v \times n_v$ block structure.\par
As explained in section \ref{sec:HorizontalLocalization}, each element of the localization matrix $\Cc$ (and, thus, each element of the matrix block $\Cc_{\textbf{b}}$) corresponds to a pair of model variables. Each model variable is, in turn, identified by its grid point location and its variable kind. As mentioned above, in our synthetic experiment, we have just one variable kind, which we called temperature, to fix ideas. Therefore, in this context, we just need a grid point location $(\theta_i, \varphi_j, h_l)$ to identify a model variable. In principle, each localization matrix element (and, thus, each element of the matrix block $\Cc_{\textbf{b}}$) is a function of the two grid point locations $(\theta_i, \varphi_j, h_l)$ and $(\theta_{i^\prime}, \varphi_{j^\prime}, h_{l^\prime})$ of the pair of variables to which the localization element corresponds. But, as we are dealing with a horizontal localization matrix, its localization elements do \emph{not} depend on the vertical coordinates $ h_l$ and $ h_{l^\prime}$. Hence, each element of $\Cc$ (and, thus,  of $\Cc_{\textbf{b}}$) is just a function of the horizontal coordinates $(\theta_i, \varphi_j)$ and $(\theta_{i^\prime}, \varphi_{j^\prime})$ of the pair of model variables to which that element corresponds. In this sense, we will use the notation $\mathcal{C}_{\scaleto{(\theta_i, \varphi_j), (\theta_{i^\prime}, \varphi_{j^\prime})}{8pt}}$ for the elements of $\Cc$ (and of $\Cc_{\textbf{b}}$).\par
The techniques, used in our experiment, to produce the block $\Cc_{\mathbf{b}}$ used spherical-harmonics and were very similar to the techniques used to create $\P$. However, as the elements of $\Cc_{\mathbf{b}}$ do not depend on the vertical coordinates, a key difference between the techniques for $\Cc_{\mathbf{b}}$ and the ones for $\P$ lies in the absence of vertical structure in the former ones. Regarding the horizontal structure of $\Cc_{\mathbf{b}}$, this is homogenous and isotropic\footnote{To be more rigorous, let us point out that the spherical harmonics techniques that we considered would produce homogenous and isotropic horizontal structures in the case of continuos spacial coordinates $(\theta, \varphi, h)$. But, in our experiment (as in any computer experiment), we used a grid, and, thus, discretized coordinates $(\theta_i, \varphi_j, h_l)$. This led to horizontal structures for $\P$ as well as $\Cc_{\mathbf{b}}$ that are only approximately homogenous and isotropic.}, as the one for $\P$. A detailed description of the procedure to create $\Cc_{\mathbf{b}}$ can be found in section \ref{app:ConstructionOfCb} in the supporting information of this article.\par
\begin{figure}[h]
    \centering
    \includegraphics[width=0.5\textwidth]{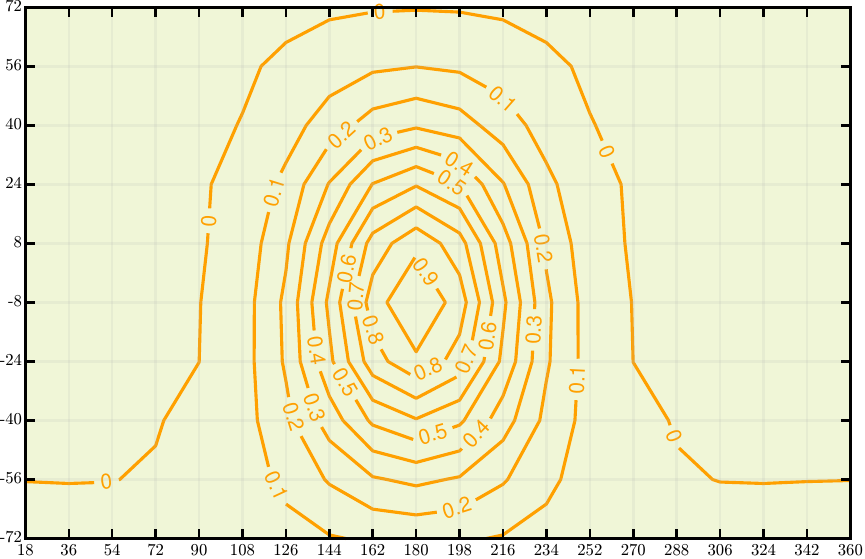}
    \caption[Localization]{The localization matrix $\Cc$ utilized in our experiment has elements which are a function of the horizontal coordinates (i.e. longitude and latitude), but \emph{not} of the vertical coordinates of their corresponding pairs of grid points. In this longitude-latitude chart, we represent the value of the localization elements for which one of the grid points in the pair is fixed and has longitude $\ang{180}$ and latitude $\ang{8}$. }
    \label{fig:RowOfCb}
\end{figure}

Now, let us depict some relevant aspects of the matrix block $\Cc_{\textbf{b}}$.
For this, let us choose a particular pair of longitude-latitude coordinates values $(\theta_{i^\prime}, \varphi_{j^\prime})$, for example, $(\ang{180}, \ang{8})$, where we have expressed those values in degrees instead of radians for the reader's convenience. Then, we can consider all elements $\mathcal{C}_{\scaleto{(\theta_{i^\prime}, \varphi_{j^\prime}) , (\theta_i, \varphi_j)}{7pt}}$ in $\Cc_{\textbf{b}}$ with $(\theta_{i^\prime}, \varphi_{j^\prime})$ given by $(\ang{180}, \ang{8})$. All these elements in $\Cc_{\textbf{b}}$ belong to the same matrix row, and they depend on the latitude-longitude coordinate values $(\theta_{i}, \varphi_{j})$ (the other longitude-latitude coordinate values $(\theta_{i^\prime}, \varphi_{j^\prime})$ are fixed and given by $(\ang{180}, \ang{8})$). Thus, we can represent all these elements in one longitude-latitude chart. We did this in figure \ref{fig:RowOfCb}. In other words, the chart in figure \ref{fig:RowOfCb} depicts a matrix row (the one with $(\theta_{i^\prime}, \varphi_{j^\prime})$ given by $(\ang{180}, \ang{8})$) of the block $\Cc_{\textbf{b}}$. Moreover, let us recall that the elements $\mathcal{C}_{\scaleto{(\theta_i, \varphi_j, h_l), (\theta_{i^\prime}, \varphi_{j^\prime}, h_{l^\prime})}{7pt}}$ of the localization matrix $\Cc$ are related to the elements $\mathcal{C}_{\scaleto{(\theta_i, \varphi_j), (\theta_{i^\prime}, \varphi_{j^\prime})}{8pt}}$ of the block $\Cc_{\textbf{b}}$ by, simply, $\mathcal{C}_{\scaleto{(\theta_i, \varphi_j, h_l), (\theta_{i^\prime}, \varphi_{j^\prime}, h_{l^\prime})}{7pt}}=  \mathcal{C}_{\scaleto{(\theta_i, \varphi_j), (\theta_{i^\prime}, \varphi_{j^\prime})}{8pt}} $. Thus, the elements of $\Cc$ do not depend on the vertical coordinates $h_l$ and $h_{l^\prime}$. Therefore, we can also see the chart of figure \ref{fig:RowOfCb} as representing the values of the localization matrix elements $\mathcal{C}_{\scaleto{(\theta_i, \varphi_j, h_l), (\theta_{i^\prime}, \varphi_{j^\prime}, h_{l^\prime})}{7pt}}$ as a function of the longitude $\theta_i$ and the latitude $\varphi_j$, with $(\theta_{i^\prime}, \varphi_{j^\prime})$ fixed and given by $(\ang{180}, \ang{8})$.\par
This concludes our description of the localization matrix $\Cc$. Let us now move on to illustrating the sample covariance matrix $\P^s$.

\subsection{The sample covariance matrix $\P^s$}

In our experiment, employing $\P$, we generated the ensemble perturbations $\e_k$ ($1 \le k \le K$), as a subset of a bigger set of perturbations as previously explained. Then, to construct a sample covariance matrix $\P^s$, we used equation \eqref{eq:SampleCovarianceMatrixNonZeroSumPerturbations}, i.e.
\begin{equation}
    \P^s = \frac{1}{K-1} \sum_{k=1}^{K} \e_k (\e_k)\tp
\end{equation}
Let us now illustrate some relevant aspects of the matrix $\P^s$. For this, let us firstly consider its diagonal elements $P^s_{ii}$. The quantities $P^s_{ii}$ ($1 \le i \le n$) are the sample variances for the temperature variables $T(\theta_i, \varphi_j, h_l)$ at each grid point location $(\theta_i, \varphi_j, h_l)$. Hence, the $P^s_{ii}$'s are the values of the sample variance field. Figure \ref{fig:VarinceFieldOfPs} represents those values as a function of the longitude $\theta_i$ and latitude $\varphi_j$, for a fixed vertical coordinate given by $h_{28}=28$.\par
\begin{figure}[!]
    \centering
    \captionsetup[subfloat]{position=top}
   \centering
      \subfloat[][]
{ \label{fig:CovarianceFunctionOfP28}\includegraphics[width=0.5\textwidth]{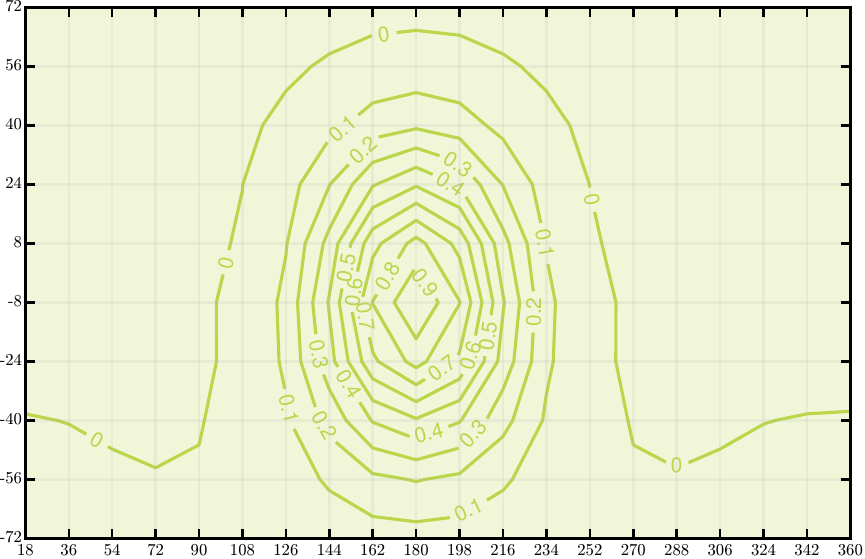}}  \\[\InterPanelsInverseSqrt]
\subfloat[][]
{\label{fig:CovarianceFunctionOfPs28} \includegraphics[width=0.5\textwidth]{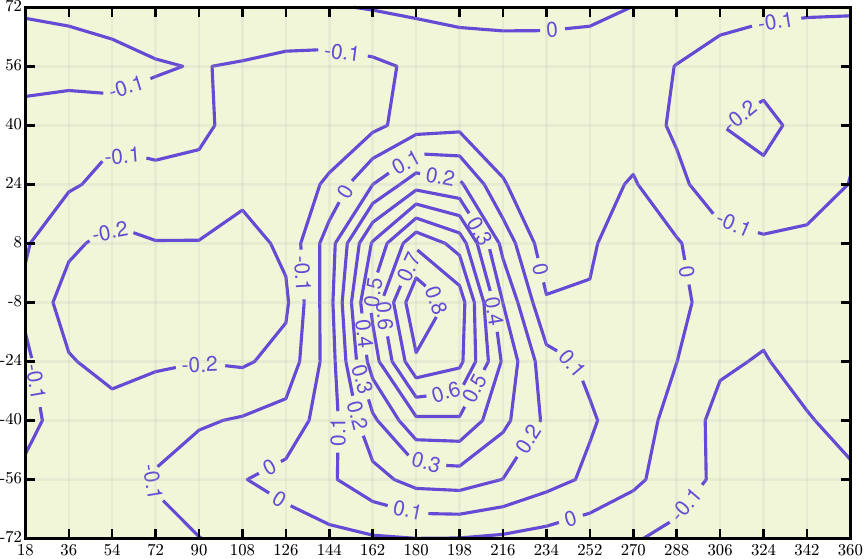}}  \\[\InterPanelsInverseSqrt]
\subfloat[][]
{ \label{fig:CovarianceFunctionOfPsLoc28} \includegraphics[width=0.5\textwidth]{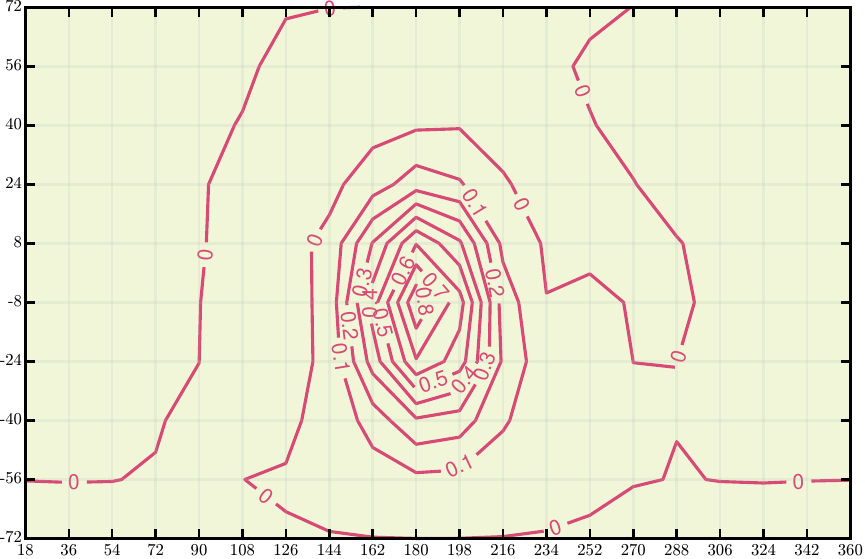}} 
\caption[Covariance function for horizontal layer $h_{28}=28.$]{In these three longitude-latitude charts, for a fixed vertical coordinate $h_{28}=28$, we represented the values the covariance function, with respect to the grid point location $(\ang{180}, \ang{8},28)$,  of:
\begin{enumerate*}[label=\textcolor{gray!60}{\checkmark}]
     \item the synthetic covariance matrix $\P$ (panel \subref{fig:CovarianceFunctionOfP28});
     \item the sample covariance matrix $\P^s$ of a synthetic ensemble generated by sampling a Gaussian with $\P$ as second central moment (panel \subref{fig:CovarianceFunctionOfPs28});
     \item the horizontally localized sample covariance matrix $\hat{\P}^s = \P^s \odot \Cc$ (panel \subref{fig:CovarianceFunctionOfPsLoc28}). 
  \end{enumerate*}
}
    \label{fig:CovarianceFunctions28}
\end{figure}
\begin{figure}[!]
    \centering
    \captionsetup[subfloat]{position=top}
   \centering
      \subfloat[][]
{ \label{fig:CovarianceFunctionOfP27}\includegraphics[width=0.5\textwidth]{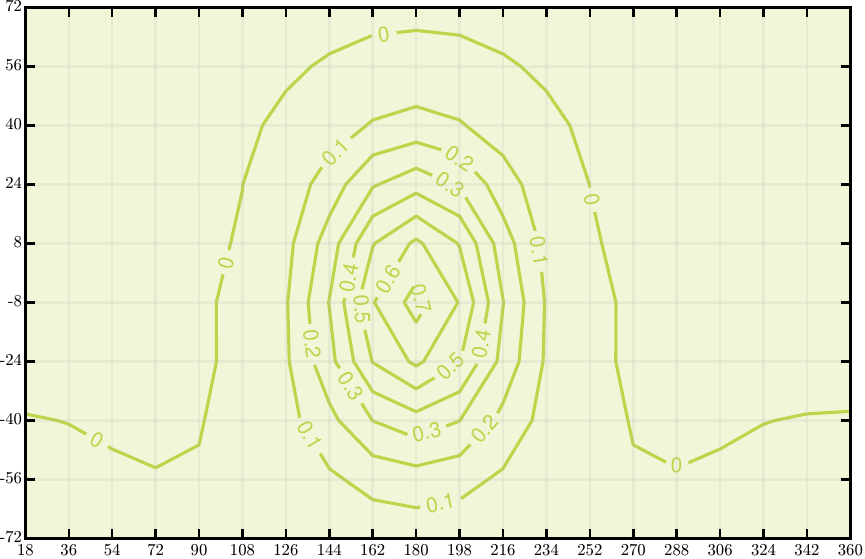}}  \\[\InterPanelsInverseSqrt]
\subfloat[][]
{\label{fig:CovarianceFunctionOfPs27} \includegraphics[width=0.5\textwidth]{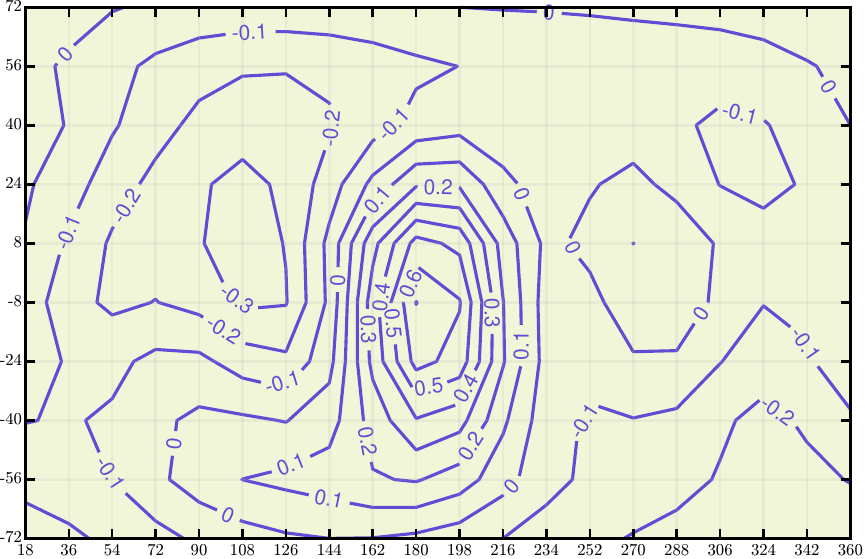}}  \\[\InterPanelsInverseSqrt]
\subfloat[][]
{ \label{fig:CovarianceFunctionOfPsLoc27}\includegraphics[width=0.5\textwidth]{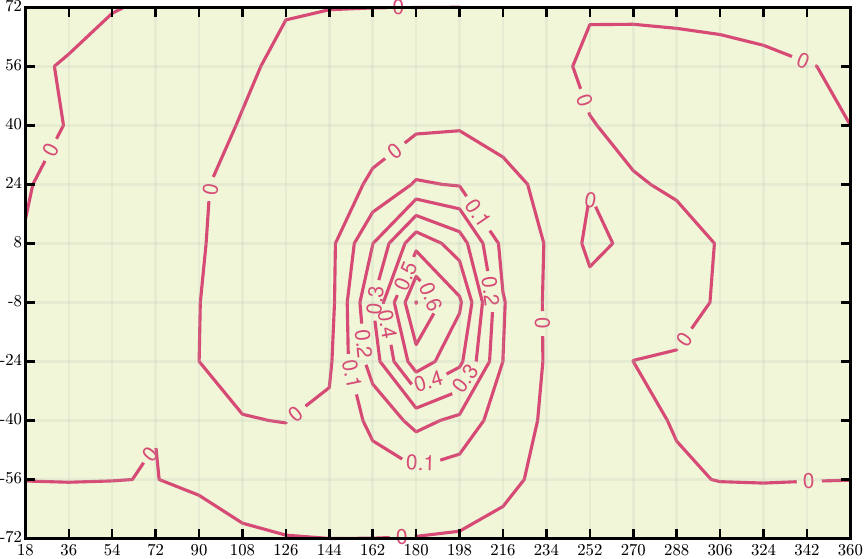}} 
\caption[Covariance function for horizontal layer $h_{27}=27.$]{In these three longitude-latitude charts, for a fixed vertical coordinate $h_{27}=27$, we represented the values the covariance function, with respect to the grid point location $(\ang{180}, \ang{8},28)$,  of:
\begin{enumerate*}[label=\textcolor{gray!60}{\checkmark}]
     \item the synthetic covariance matrix $\P$ (panel \subref{fig:CovarianceFunctionOfP27});
     \item the sample covariance matrix $\P^s$ of a synthetic ensemble generated by sampling a Gaussian with $\P$ as second central moment (panel \subref{fig:CovarianceFunctionOfPs27});
     \item the horizontally localized sample covariance matrix $\hat{\P}^s = \P^s \odot \Cc$ (panel \subref{fig:CovarianceFunctionOfPsLoc27}). 
  \end{enumerate*}
}
    \label{fig:CovarianceFunctions27}
\end{figure}
\begin{figure}[t]
    \centering
     \begin{adjustwidth}{4.8cm}{-1cm}
    \setlength{\unitlength}{1\textwidth}
     \begin{picture}(1,0.66)(0.3,0)
 \put(0.3,0.03){ \includegraphics[width=0.4\textwidth]{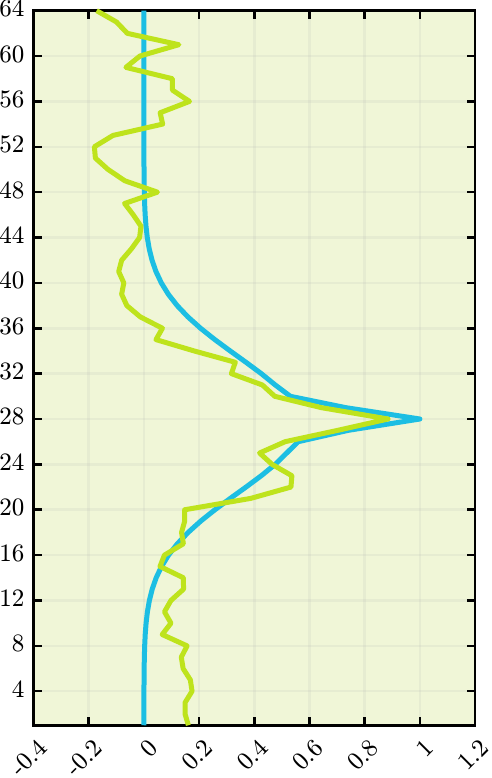}}
  \put(0.27,0.32){\rotatebox{90}{\footnotesize \textit{horizontal layer} }}
 \put(0.45, 0.01) {\footnotesize \textit{covariance value}}
    \put(0.71 , 0.0713){ \colorbox{AxisBackgroundColor}{
\begin{picture}(0.05,0.5832)(0,0)
\put(-0.0048,0.001){\color{gray!50}  \linethickness{0.4mm} \line(1,0){0.055}}
\put(-0.0048,0.00){\color{gray!50}  \linethickness{0.4mm} \line(0,1){0.5833}}
\put(0.05,0.00){\color{gray!50}  \linethickness{0.4mm} \line(0,1){0.5833}}
\put(-0.0048,0.5822){\color{gray!50}  \linethickness{0.4mm} \line(1,0){0.055}}
   \put(0.0226,0.41){\color{PColor}  \linethickness{0.7mm} \line(0,1){0.06}}
    \put(0.0145,0.36){\color{PColor} \Large  $\P$}
    \put(0.0226,0.17){\color{PsColor}  \linethickness{0.7mm} \line(0,1){0.06}}
    \put(0.0115,0.12){\color{PsColor}  \Large $\P^s$}
\end{picture}} }
\end{picture}
\end{adjustwidth}
    \caption[Vertical profiles of the covariance functions]{This figure depicts a vertical profile of: \begin{enumerate*}[label=\textcolor{gray!60}{\checkmark}]
     \item the covariance function $\eta_{\scaleto{\ang{180}, \ang{8},28}{5pt}} (\theta_i, \varphi_j, h_l)$ (\textcolor{PColor}{\rule[0.5ex]{0.4cm}{0.6mm}}) of the synthetic covariance matrix $\P$;
     \item the covariance function $\eta^s_{\scaleto{\ang{180}, \ang{8},28}{5pt}} (\theta_i, \varphi_j, h_l)$ (\textcolor{PsColor}{\rule[0.5ex]{0.4cm}{0.6mm}}) of the sample covariance matrix $\P^s$.
  \end{enumerate*} More specifically, we represented the value (on the \textit{abscissa} axis) of the covariance functions $\eta_{\scaleto{\ang{180}, \ang{8},28}{5pt}} (\theta_i, \varphi_j, h_l)$ and $\eta^s_{\scaleto{\ang{180}, \ang{8},28}{5pt}} (\theta_i, \varphi_j, h_l)$ as a function of the vertical coordinate $h_l$ (on the \textit{ordinate} axis), with fixed longitude $\theta_i$ and latitude $\varphi_j$ given by  $\ang{180}$ and $\ang{8}$, respectively. Here, the subscript $\phantom{i}_{\scaleto{\ang{180}, \ang{8},28}{5pt}} $ specifies the longitude, latitude and vertical coordinate of the grid point with respect to which those covariance functions are defined.}
    \label{fig:VerticalProfilesOfPAndPs}
\end{figure}

After illustrating the sample variance field, let us focus on a covariance function of $\P^s$. More specifically, as an example, we chose the covariance function with respect to the grid point location $(\theta_{i^\prime},\varphi_{j^\prime}, h_{l^\prime})$ given, in degrees, by $(\ang{180}, \ang{8},28)$. Let us denote that covariance function by $\eta^s_{\scaleto{\ang{180}, \ang{8},28}{5pt}} (\theta_i, \varphi_j, h_l)$. Let us remind ourselves that $\eta^s_{\scaleto{\ang{180}, \ang{8},28}{5pt}} (\theta_i, \varphi_j, h_l)$ is, by definition, the sample covariance (that is, a particular off-diagonal element of $\P^s$) between the temperature variable $T(\theta_{i^\prime},\varphi_{j^\prime}, h_{l^\prime})$ at the location $(\theta_{i^\prime},\varphi_{j^\prime}, h_{l^\prime})$ given by $(\ang{180}, \ang{8},28)$ and the temperature $T(\theta_i, \varphi_j, h_l)$ at $(\theta_i, \varphi_j, h_l)$, for every location $(\theta_i, \varphi_j, h_l)$. Hence, $\eta^s_{\scaleto{\ang{180}, \ang{8},28}{5pt}} (\theta_i, \varphi_j, h_l)$ is a function of the grid point location $(\theta_i, \varphi_j, h_l)$. Panel  \subref{fig:CovarianceFunctionOfPs28} of figure \ref{fig:CovarianceFunctions28} and panel \subref{fig:CovarianceFunctionOfPs27} of figure \ref{fig:CovarianceFunctions27} are two longitude-latitude charts, representing the values of the covariance function $\eta^s_{\scaleto{\ang{180}, \ang{8},28}{5pt}} (\theta_i, \varphi_j, h_l)$ on the horizontal layer with vertical coordinate $h_{28}=28$ and $h_{27}=27$, respectively. Whereas, in figure \ref{fig:VerticalProfilesOfPAndPs}, the green line represents the values of $\eta^s_{\scaleto{\ang{180}, \ang{8},28}{5pt}} (\theta_i, \varphi_j, h_l)$ as a function of the vertical coordinate $h_l$ ($1 \le l \le n_v$) with the longitude $\theta_i$ and the latitude $\varphi_j$ fixed and given, in degrees, by $\ang{180}$ and $\ang{8}$, respectively. The sample covariances represented by that green line are affected by sample noise due to the not-very-high number of ensemble members in our synthetic experiment, namely $K=n_w=64$. However, let us recall that, as mentioned in section \ref{sec:AssumptionAboutK}, for today's high resolution atmospheric models,  we have $n_w \sim 10^3$. Consequently, in an operational context, we would have $K=n_w \sim 10^3$, and, thus, in the sample covariances, we would observe a much smaller sample noise than the one in our experiment. 


\subsection{The localized sample covariance matrix $\hat{\P}^s = \Cc \odot \P^s$}

After representing aspects of the non-localized sample covariance matrix $\P^s$, let us focus on the localized sample covariance matrix $\hat{\P}^s = \Cc \odot \P^s$.

As an example, let us consider the covariance function of $\hat{\P}^s = \Cc \odot \P^s$ with respect to the grid point location $(\theta_{i^\prime},\varphi_{j^\prime}, h_{l^\prime})$ given by $(\ang{180}, \ang{8},28)$. Let us denote that covariance function by $\hat{\eta}^s_{\scaleto{\ang{180}, \ang{8},28}{5pt}} (\theta_i, \varphi_j, h_l)$. Let us remind ourselves that $\hat{\eta}^s_{\scaleto{\ang{180}, \ang{8},28}{5pt}} (\theta_i, \varphi_j, h_l)$ is given by:
\begin{equation} \label{eq:LocalizedSampleCovarianceFunction}
   \hat{\eta}^s_{\scaleto{\ang{180}, \ang{8},28}{5pt}} (\theta_i, \varphi_j, h_l) \doteq \eta^s_{\scaleto{\ang{180}, \ang{8},28}{5pt}} (\theta_i, \varphi_j, h_l) \: \, \mathcal{C}_{\scaleto{(\theta_{i^\prime}, \varphi_{j^\prime}), (\theta_i, \varphi_j)}{7pt}}
\end{equation}
where
\begin{itemize}
    \item $\eta^s_{\scaleto{\ang{180}, \ang{8},28}{5pt}} (\theta_i, \varphi_j, h_l)$ is the covariance function of the sample covariance matrix $\P^s$  with respect to the grid point location given by $(\ang{180}, \ang{8},28)$. Aspects of this function were depicted in panel  \subref{fig:CovarianceFunctionOfPs28} of figure \ref{fig:CovarianceFunctions28}, panel \subref{fig:CovarianceFunctionOfPs27} of figure \ref{fig:CovarianceFunctions27} and in figure \ref{fig:VerticalProfilesOfPAndPs};
    \item in the elements $\mathcal{C}_{\scaleto{(\theta_{i^\prime}, \varphi_{j^\prime}) , (\theta_i, \varphi_j)}{7pt}}$ of the localization matrix block $\Cc_{\textbf{b}}$, the pair of longitude-latitude coordinates $(\theta_{i^\prime}, \varphi_{j^\prime})$ is fixed and given by $(\ang{180}, \ang{8})$. Let us recall that, in figure \ref{fig:RowOfCb}, we represented these quantities $\mathcal{C}_{\scaleto{(\theta_{i^\prime}, \varphi_{j^\prime}) , (\theta_i, \varphi_j)}{7pt}}$ as a function of the longitude $\theta_i$ and the latitude $\varphi_j$ (with $(\theta_{i^\prime}, \varphi_{j^\prime})$ is given by $(\ang{180}, \ang{8})$).
\end{itemize}
In panel \subref{fig:CovarianceFunctionOfPsLoc28} of figure \ref{fig:CovarianceFunctions28} and panel \subref{fig:CovarianceFunctionOfPsLoc27} of figure \ref{fig:CovarianceFunctions27}, we illustrated the horizontal structure of $\hat{\P}^s = \Cc \odot \P^s$. More specifically, we represented the values  $\hat{\eta}^s_{\scaleto{\ang{180}, \ang{8},28}{5pt}} (\theta_i, \varphi_j, h_l)$ as a function of the longitude $\theta_i$  and the latitude $\varphi_j$ with a fixed vertical coordinate $h_{28}=28$ (panel \subref{fig:CovarianceFunctionOfPsLoc28} of figure \ref{fig:CovarianceFunctions28}) and $h_{27}=27$ (panel \subref{fig:CovarianceFunctionOfPsLoc27} of figure \ref{fig:CovarianceFunctions27}). As we can see by comparing panels \subref{fig:CovarianceFunctionOfP28}, \subref{fig:CovarianceFunctionOfPs28} and \subref{fig:CovarianceFunctionOfPsLoc28} of figure \ref{fig:CovarianceFunctions28} , and, similarly, by comparing panels \subref{fig:CovarianceFunctionOfP27}, \subref{fig:CovarianceFunctionOfPs27} and \subref{fig:CovarianceFunctionOfPsLoc27} of figure \ref{fig:CovarianceFunctions27}, the values of the $\hat{\P}^s$ covariance function $\hat{\eta}^s_{\scaleto{\ang{180}, \ang{8},28}{5pt}} (\theta_i, \varphi_j, h_l)$ far away from the grid point location $(\ang{180}, \ang{8},28)$ are dampened and closer to the corresponding values of the $\P$ covariance function $\eta_{\scaleto{\ang{180}, \ang{8},28}{5pt}} (\theta_i, \varphi_j, h_l)$ than the corresponding values of the $\P^s$ covariance function $\eta^s_{\scaleto{\ang{180}, \ang{8},28}{5pt}} (\theta_i, \varphi_j, h_l)$. This is the sample-noise-dampening effect of localization.\par
Finally, let us consider a vertical profile of the covariance function $ \hat{\eta}^s_{\scaleto{\ang{180}, \ang{8},28}{5pt}} (\theta_i, \varphi_j, h_l)$. More specifically, let us consider $ \hat{\eta}^s_{\scaleto{\ang{180}, \ang{8},28}{5pt}} (\theta_i, \varphi_j, h_l)$ as a function of the vertical coordinate $h_l$, while keeping the longitude and latitude $(\theta_i, \varphi_j)$ fixed and equal to $(\ang{180},\ang{8})$. From equation \eqref{eq:LocalizedSampleCovarianceFunction}, recalling that, there, $(\theta_{i^\prime}, \varphi_{j^\prime})$ is given by $(\ang{180}, \ang{8})$, we have
\begin{equation}
    \hat{\eta}^s_{\scaleto{\ang{180}, \ang{8},28}{5pt}} (\ang{180},\ang{8}, h_l) = \eta^s_{\scaleto{\ang{180}, \ang{8},28}{5pt}} (\ang{180},\ang{8}, h_l) \: \, \mathcal{C}_{\scaleto{(\ang{180},\ang{8}), (\ang{180},\ang{8})}{5pt}} = \eta^s_{\scaleto{\ang{180}, \ang{8},28}{5pt}} (\ang{180},\ang{8}, h_l)
\end{equation}
because $\mathcal{C}_{\scaleto{(\ang{180},\ang{8}), (\ang{180},\ang{8})}{5pt}}=1$. Thus, because of the use of horizontal localization, the considered covariance function vertical profile of $\hat{\P}^s = \Cc \odot \P^s$ coincides with covariance function vertical profile of $\P^s$ already represented in figure \ref{fig:VerticalProfilesOfPAndPs} as a green line. 

\subsection{The swift algorithm} \label{sec:swiftAlgorithmInTheExperiment}

As explained above, once the perturbations $\e_k$ ($1 \le k \le K$) were produced in our experiment (as a subset of a bigger set of perturbations), we computed the vectors $\z_k \doteq \frac{1}{K-1} \e_k$ ($1 \le k \le K$). These vectors allowed us to build the matrix $\A$ through equation \eqref{eq:ADef}. In the following, we shall outline the remainder of the procedure in our experiment, using the notation defined in the theoretical derivation of our swift algorithm (section \ref{sec:swiftAlgorithm}).\par
In our experiment, after building the matrix $\A$,
\def\RegularTheEnumi{\theenumi}
\renewcommand{\theenumi}{\alph{enumi}}
\begin{enumerate}
    \item We constructed, from $\A$, the matrices $\A_m$ (defined in section \ref{sec:AlgorithmDescription}) for all integers $m$ with $1 \le m \le n_h$;
    \item We computed singular value decompositions of the matrices $\A_m = \Eb_m \Lambdam \F_m\tp$ for all integers $m$ with $1 \le m \le n_h$;
    \item Using the matrices $\Eb_m$, $\Lambdam$ and $\F_m$ ($1 \le m \le n_h$), we constructed the matrices $\Eb$, $\boldsymbol{\Lambda}$ and $\F$ (through equations \eqref{eq:EBlocks}, \eqref{eq:LambdaBlocks} and \eqref{eq:FBlocks});
    \item We performed an economical eigendecomposition of the matrix block $\Cc_{\textbf{b}}$. By this we mean that we wrote $\Cc_{\textbf{b}} = \V_{\textbf{b}} \, \Deltab_{\textbf{b}} \, \V_{\textbf{b}}\tp$, where $\Deltab_{\textbf{b}}$ is a diagonal matrix with the non-zero eigenvalues of $\Cc_{\textbf{b}}$ along its diagonal and $\V_{\textbf{b}}$ is a matrix whose columns are the orthonormal eigenvectors of $\Cc_{\textbf{b}}$ corresponding to the eigenvalues appearing in $\Deltab_{\textbf{b}}$;
    \item We computed $\V$ and $\Deltab$ using $\V_{\textbf{b}}$ and $\Deltab_{\textbf{b}}$, respectively (equations \eqref{eq:VBlocks} and \eqref{eq:DeltaBlocks});
    \item We computed the left inverse $(\hat{\P}^s)^{-\frac{1}{2}}=   \Deltab^{-\frac{1}{2}} \V\tp \Fb \boldsymbol{\Lambda}^{-1} \Eb\tp$ of the square root $(\hat{\P}^s)^{\frac{1}{2}}= \A \V \Deltab^{\frac{1}{2}}$ of the localized sample covariance matrix $\hat{\P}^s = \Cc \odot \P^s$.
\end{enumerate}
\renewcommand{\theenumi}{\RegularTheEnumi}
In what follows, for notational convenience, we shall denote $(\hat{\P}^s)^{\frac{1}{2}}$ by $\hat{\Z}$ and $(\hat{\P}^s)^{-\frac{1}{2}}$ by $\hat{\Z}^{-1}$. Moreover, following the notation utilized in the derivation of our algorithm, we will denote the dimension of the diagonal matrix $\Deltab_{\textbf{b}}$ by $L$, and the dimension of the diagonal matrix $\Deltab$ by $\ns$. Considering \eqref{eq:DeltaBlocks}, it is straightforward to find that, in our experiment, $\ns=L \, n_v$.\par

\subsection{Measuring the algorithm accuracy in computing a left inverse of a square root of $\hat{\P}^s$} \label{sec:Accuracy}
As we have seen, the output of our algorithm is the matrix $\hat{\Z}^{-1}\doteq (\hat{\P}^s)^{-\frac{1}{2}}$, whose size is $\ns \times n$. This matrix is a left inverse of $\hat{\Z}\doteq (\hat{\P}^s)^{\frac{1}{2}}$ (but it is not necessarily its inverse). Hence, if $\hat{\Z}^{-1}$ were computed with perfect numerical accuracy by our algorithm, then $\hat{\Z}^{-1}\, \hat{\Z}$ would be the $\ns \times \ns$ identity matrix $\I_{\ns}$ and, thus, $\hat{\Z}^{-1} \, \hat{\P}^s \, (\hat{\Z}^{-1})\tp$ would also equal $\I_{\ns}$ (because $\hat{\P}^s= \hat{\Z} \, (\hat{\Z})\tp $). In practice, the computed $\hat{\Z}^{-1}$ will not have a perfect numerical accuracy and, as a consequence, $\hat{\Z}^{-1} \, \hat{\P}^s \, (\hat{\Z}^{-1})\tp$ will slightly differ $\I_{\ns}$. It is thus interesting to consider the $\ns \times \ns$ matrix difference $\boldsymbol{\Xi}$ between $\I_n$ and $\hat{\Z}^{-1} \, \hat{\P}^s \, (\hat{\Z}^{-1})\tp$:
\begin{equation} \label{eq:MatrixDifferenceSwiftAlgorithm}
 \boldsymbol{\Xi} \doteq \I_{\ns}  -\hat{\Z}^{-1} \, \hat{\P}^s \, (\hat{\Z}^{-1})\tp 
\end{equation}
A way to assess the accuracy of our algorithm is to measure how close the matrix $\boldsymbol{\Xi}$ is to the zero matrix. To do this, we will define two accuracy indicators based on the elements of $\boldsymbol{\Xi}$. But, before that, we would like to briefly explain in what sense equation \eqref{eq:MatrixDifferenceSwiftAlgorithm} is particularly relevant when applying our algorithm to swiftly perform the standardized variables transform. Let us recall that this important operational application of our algorithm was outlined in section \ref{sec:standardizedVariables}. Let us also recall that, in the context of this application, in an operational centre, $\P^s$ is the sample covariance matrix of a climatological ensemble of forecast error proxies. Moreover, in that context, the matrix $\Cc \odot \P^s$ is utilized as a climatological error covariance matrix in a data assimilation system, i.e. $\Pc=\Cc \odot \P^s$. The left inverse $\hat{\Z}^{-1}=(\Pc)^{-\frac{1}{2}}$ of the square root $\hat{\Z}= (\Pc)^{\frac{1}{2}}$ of $\Pc=\Cc \odot \P^s$ allows us to perform the standardized variables transform, i.e. $\xs = (\Pc)^{-\frac{1}{2}} \x $. A key property of the standardized variables is expressed by \eqref{eq:PcsvIsTheIdentity}, and, in particular, by the last line of \eqref{eq:PcsvIsTheIdentity}, that is,
\begin{equation} \label{eq:PcSVEqualsIForInverseSQRTExperiment}
   (\Pc)^{-\frac{1}{2}}\Pc [(\Pc)^{-\frac{1}{2}}]\tp = \I_{\ns}
\end{equation}
The key property \eqref{eq:PcSVEqualsIForInverseSQRTExperiment} in this future operational application corresponds, in the context of our synthetic experiment, to $\hat{\Z}^{-1} \, \hat{\P}^s \, (\hat{\Z}^{-1})\tp$ being equal to the $\ns \times \ns$ identity matrix $\I_{\ns}$. Hence, it is particularly relevant to determine up to what accuracy, in our experiment, this corresponding property holds true, namely, up to what extent, the matrix difference $\boldsymbol{\Xi}$ defined in \eqref{eq:MatrixDifferenceSwiftAlgorithm} is close to the zero matrix. And, as mentioned above, by determining this, we will assess the accuracy of our algorithm.\par
Thus, to measure how close $\boldsymbol{\Xi}$ is to the zero matrix, let us define two accuracy indicators, which are relevant for our purposes. The first accuracy indicator, which will be denoted as $\text{RMSE}_{\,\boldsymbol{\Xi}}$, is:
\begin{equation}
    \text{RMSE}_{\,\boldsymbol{\Xi}}= \sqrt{ \sum_{i=1}^{\ns} \sum_{j=1}^{\ns} \; \Xi_{ij}^2     } 
\end{equation}
where $\Xi_{ij}$ ($1 \le i \le \ns$ and $1 \le j \le \ns $) are the elements of the matrix $\boldsymbol{\Xi}$.\par
The second accuracy indicator, which we denoted as $\text{MAX}_{\,\boldsymbol{\Xi}}$, was defined as the maximum over the absolute values $ \vert \,\Xi_{ij} \vert$ of all elements $\Xi_{ij}$ of the matrix $\boldsymbol{\Xi}$ ($1 \le i \le \ns$ and $1 \le j \le \ns$).\par
Let us now recall that, in our experiment, a random process is involved, namely the one used to produce the perturbations $\e_k$. Therefore, the values of the accuracy indicators $\text{RMSE}_{\,\boldsymbol{\Xi}}$ and $\text{MAX}_{\,\boldsymbol{\Xi}}$ will depend on that random process. On the other hand, we wanted to assess the accuracy of our algorithm independently of any particular random process realization. For this, we iterated our experiment $10$ times. Then, we computed the mean $\overline{\text{RMSE}}_{\,\boldsymbol{\Xi}}$ of the indicator $\text{RMSE}_{\,\boldsymbol{\Xi}}$ over those $10$ iterations.  Moreover, we computed the maximum $\text{MAX}_{\,\scaleto{\boldsymbol{\Xi}}{4.2pt}}^{\scaleto{\text{MAX}}{3pt}}$ of the indicator $\text{MAX}_{\,\boldsymbol{\Xi}}$ over those $10$ iterations. Thus, we used $\overline{\text{RMSE}}_{\,\boldsymbol{\Xi}}$ and $\text{MAX}_{\,\scaleto{\boldsymbol{\Xi}}{4.2pt}}^{\scaleto{\text{MAX}}{3pt}}$ as our accuracy indicators (instead of particular values of $\text{RMSE}_{\,\boldsymbol{\Xi}}$ and $\text{MAX}_{\,\boldsymbol{\Xi}}$ in any of the $10$ iterations). The values that we obtained for $\overline{\text{RMSE}}_{\,\boldsymbol{\Xi}}$ and $\text{MAX}_{\,\scaleto{\boldsymbol{\Xi}}{4.2pt}}^{\scaleto{\text{MAX}}{3pt}}$ are the following:
\begin{equation}
    \overline{\text{RMSE}}_{\,\boldsymbol{\Xi}}= 2.7 \cdot 10^{-9}     \qquad \text{and} \qquad \text{MAX}_{\,\scaleto{\boldsymbol{\Xi}}{4.2pt}}^{\scaleto{\text{MAX}}{3pt}} = 1.5 \cdot 10^{-6}
\end{equation}
Thus, our algorithm exhibits very good accuracy.\par

\subsection{Measuring the algorithm accuracy when $\hat{\Z}^{-1}$ left-multiplies a vector} \label{sec:AccuracyVector}
 
\begin{figure}[h!]
    \centering
    \setlength{\unitlength}{1\textwidth}
     \begin{picture}(1,0.45)(0,0)
 \put(0.245,0.03){ \includegraphics[width=0.5\textwidth]{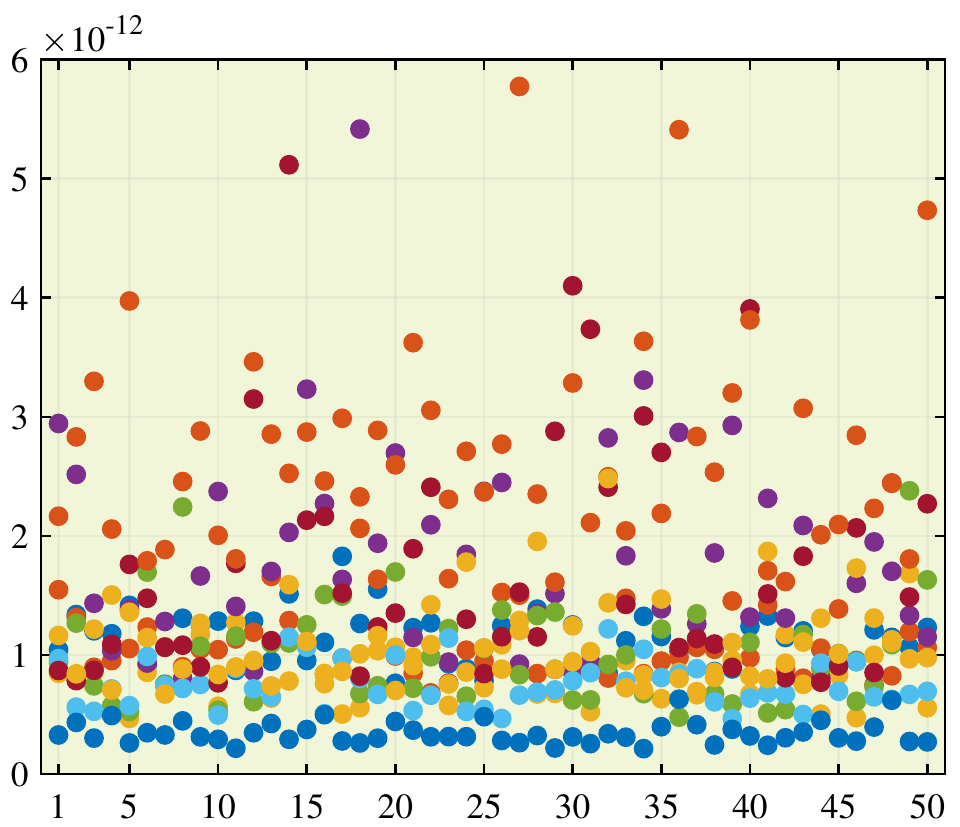}}
  \put(0.22,0.22){\rotatebox{90}{\small \textit{RMSE} }}
 \put(0.49, 0.01) {\small \textit{trial}}
\end{picture}
    \caption[Values of $\text{RMSE}_{\,\boldsymbol{\delta}}$ for different trials and iterations of the synthetic experiment]{In this figure, for $10$ different iterations of our synthetic experiment and $50$ different trials within each iteration, we represent $10 \cdot 50= 500$ values of the indicator $\text{RMSE}_{\,\boldsymbol{\delta}}$, measuring up to what degree of accuracy the matrix $\hat{\Z}^{-1}$ (i.e. the output of our swift algorithm) is a left inverse of $\hat{\Z}$ (which, in turn, is a square root of $\Cc \odot \P^s$), when $\hat{\Z}^{-1}$ left-multiplies a vector $\x$, i.e. in the case of products of the form $\hat{\Z}^{-1} \, \x$. More specifically, we use vectors $\x$ of the form $\hat{\Z}\, \xs$, and the indicator $\text{RMSE}_{\,\boldsymbol{\delta}}$ (defined in \eqref{eq:RMSEdelta}) measures how small the vector difference $\boldsymbol{\delta} \doteq \xs - \hat{\Z}^{-1} \hat{\Z} \, \xs$ is. At each of the $10$ different iterations of our experiment, a different ensemble is generated and used. At each of the $50$ trials within each of those iterations, a different randomly generated vector $\xs$ is used. In the scatter plot shown in this figure, an integer from $1$ to $50$, labelling the trial, is on the abscissa axis, and $\text{RMSE}_{\,\boldsymbol{\delta}}$ is on the ordinate axis. Dots representing values of $\text{RMSE}_{\,\boldsymbol{\delta}}$ in different iterations of the experiment are printed with different colours.
    }
    \label{fig:ScatterAccuracy}
\end{figure}

 As mentioned above, an important operational application of our algorithm is to swiftly perform the standardized variables transform, namely, $\xs = (\Pc)^{-\frac{1}{2}} \x $. As explained in the previous section \ref{sec:Accuracy}, in that context, we have $\hat{\Z}^{-1}=(\Pc)^{-\frac{1}{2}}$. Hence, in our synthetic experiment, we are particularly interested in considering the product of the form $\hat{\Z}^{-1} \x $. If, in this product, we consider a vector $\x$ of the form $\hat{\Z} \,\xs $, where $\xs$ is a generic $\ns$-dimensional vector, we obtain an expression of the form $\hat{\Z}^{-1} \hat{\Z} \, \xs $. If our algorithm were perfectly accurate, the matrix $\hat{\Z}^{-1}$ would be a perfectly accurate left inverse of $\hat{\Z}$, and expression $\hat{\Z}^{-1} \hat{\Z} \,\xs $ would equal $\xs$, for any vector $\ns$-dimensional vector $\xs$. In other words, the vector difference
 \begin{equation} \label{eq:vectorDifferencedelta}
     \boldsymbol{\delta} \doteq \xs - \hat{\Z}^{-1} \hat{\Z} \, \xs
 \end{equation}
would be the zero vector $\boldsymbol{0}$ for all $\xs$. In practice, $\boldsymbol{\delta}$ will not be exactly $\boldsymbol{0}$, and measuring how close $\boldsymbol{\delta}$ is to $\boldsymbol{0}$ is a way to assess and quantify the accuracy of our algorithm concerning the operationally relevant situation when $\hat{\Z}^{-1}$ left-multiplies a vector (given here by $\hat{\Z} \,\xs $). Analogously to how we proceeded in the previous section \ref{sec:Accuracy}, we defined the root mean square error indicator $\text{RMSE}_{\,\boldsymbol{\delta}}$ as:
\begin{equation} \label{eq:RMSEdelta}
     \text{RMSE}_{\,\boldsymbol{\delta}}= \sqrt{ \sum_{i=1}^{\ns}  \; \delta_{i}^2     } 
\end{equation}
where $\delta_{i}$ ($1 \le i \le \ns$) is the $i$-th element of the vector $\boldsymbol{\delta}$. Then, we iterated our synthetic experiment, described in sections \ref{sec:grid} to \ref{sec:swiftAlgorithmInTheExperiment}, $10$ times. In particular, in each of these $10$ iterations, we had a different synthetically generated ensemble, from which the sample covariance matrix $\P^s$ was constructed. For each of these $10$ iterations, we performed $50$ trials, in each of which
\def\RegularTheEnumi{\theenumi}
\renewcommand{\theenumi}{\arabic{enumi}}
\begin{enumerate}
    \item We generated a $\ns$-dimensional vector $\xs$ as a random draw from the normal distribution $\mathcal{N}(\mathbf{0},\I_{\ns})$, where $\I_{\ns}$ is the $\ns \times \ns$ identity matrix;
    \item With that randomly generated vector $\xs$, we computed the vector difference $\boldsymbol{\delta}$ through \eqref{eq:vectorDifferencedelta}, and, thus, the indicator $\text{RMSE}_{\,\boldsymbol{\delta}}$ through \eqref{eq:RMSEdelta}. 
\end{enumerate}
Then, to visually represent the results about the indicator $\text{RMSE}_{\,\boldsymbol{\delta}}$, we constructed a scatter plot with an integer from $1$ to $50$, labelling the trial, on the abscissa axis, and $\text{RMSE}_{\,\boldsymbol{\delta}}$ on the ordinate axis. More specifically, in that scatter plot, which we show in figure \ref{fig:ScatterAccuracy}, for each of those $10$ iterations of our experiment, we represented the $50$ values of $\text{RMSE}_{\,\boldsymbol{\delta}}$ produced as explained above in the $50$ trials. Thus, in that scatter plot, we have $500$ dots in total. And, the dots representing values of $\text{RMSE}_{\,\boldsymbol{\delta}}$ in different experiment iterations are printed with different colours. The $500$ represented values of $\text{RMSE}_{\,\boldsymbol{\delta}}$ range from $2.1 \cdot 10^{-13}$ to $5.8 \cdot 10^{-12}$. This shows that, when it left-multiplies a vector, the matrix $\Z^{-1}$ produced as an output by the swift algorithm is a left inverse of $\Z$ to a very high degree of accuracy.
This concludes the description of our synthetic experiment to test our novel algorithm to compute a left inverse of a square root of a horizontally localized sample covariance matrix.

\section{Conclusions and perspectives} \label{sec:conclusions}

To the authors' knowledge, finding a left inverse $\hat{\Z}^{-1}$ of a square root $\hat{\Z}$ of a localized sample covariance matrix $\Cc \odot \P^s$ was computationally prohibitive in high dimension with known algorithms. In the present article, we tackled this problem and introduced a new, speedy algorithm for that task in the case when a horizontal localization matrix is utilized (for a definition of horizontal localization matrix, see section \ref{sec:HorizontalLocalization}).  This novel algorithm is swift and scalable because it is parallelizable. More specifically, we found that, in this algorithm, calculations involving model variables located at different grid vertical columns can be performed separately, and, thus, in parallel (section \ref{sec:AlgorithmDescription}). Furthermore, we introduced a version of this novel algorithm directly outputting the product $\hat{\Z}^{-1} \x$, for any given $n$-dimensional vector $\x$, without storing the entire matrix $\hat{\Z}^{-1}$ in memory (section \ref{sec:ActionOnVector}). Through a time complexity evaluation (section \ref{sec:ComputationalAspects}), we found that, for operational applications with today's high resolution models, if executed on a single processor, that version of the algorithm is already $9$ orders of magnitude faster than a traditional algorithm, and, with multiple, parallel processors, it  scales until the number of processors  reaches the number $n_h \sim 10^6$ of grid points per horizontal grid layer.\par
We tested the novel algorithm to calculate $\hat{\Z}^{-1}$ in a synthetic experiment (section \ref{sec:swiftAlgorithmTest}) on a three-dimensional grid. In this experiment, using spherical harmonics techniques, we first constructed a synthetic covariance matrix, which we utilized as covariance matrix of a Gaussian distribution with zero mean. We then used random draws from that distribution to create an ensemble. After that, we considered the sample covariance matrix of that ensemble and we localized it with a horizontal localization matrix. We applied our novel algorithm to the resulting localized sample covariance matrix. Through this algorithm, we were able to compute a left inverse of a square root of that matrix. In this experiment, we measured the computational precision of the new algorithm, and we found it to exhibit very good accuracy when calculating $\hat{\Z}^{-1}$ (section \ref{sec:Accuracy}) as well as when computing products of the form $\hat{\Z}^{-1} \, \x$, where $\x$ is a $n$-dimensional vector (section \ref{sec:AccuracyVector}).\par   
In the present article, we have a particular case in mind to which this algorithm can be applied: the case   of a climatological ensemble of forecast error proxies. A horizontally localized sample covariance matrix $\Cc \odot \P^s$ of such an ensemble is a climatological error covariance matrix $\Pc$. This covariance matrix $\Pc=\Cc \odot \P^s$ features two desirable properties: inhomogeneity and anisotropy (namely, geographical dependence), unlike many currently-used methods to construct a climatological error covariance matrix. Our novel algorithm makes it computationally feasible to compute a left inverse $\hat{\Z}^{-1}=(\Pc)^{-\frac{1}{2}}$ of a square root $\hat{\Z}=(\Pc)^{\frac{1}{2}}$ of a climatological error covariance matrix of the form $\Pc=\Cc \odot \P^s$. With that left inverse, one can, in turn, perform the standardized variables transform (section \ref{sec:standardizedVariables}). %
And, standardized variables allow one to construct a new Hybrid forecast covariance model, which two desirable properties: positive definiteness as well as (unlike the current Hybrid) climatologically unbiasedness. The idea of this new Hybrid can be further developed into even more advanced novel Hybrids \parencite{Sardelli2026PhDThesis}, which will be introduced, studied and tested in separate articles.\par
The novel algorithm was developed in the special case when the number of ensemble members equals the number of model variables in each grid vertical column (in current high resolution models, this number is of order $10^3$). Considering this special case played a crucial role in the mathematical derivation of the algorithm. A generalization beyond this special case is left to future research work. Moreover, other prospective work  may concern generalizations to the case when \begin{enumerate*} [label=(\roman*)]
    \item the matrix $\A$ (equation \eqref{eq:ADef}) in the expression of the square root $(\Pc)^{\frac{1}{2}}$  is not invertible or ill-conditioned,
    \item the localization is not necessarily horizontal
    \item the localization is scale dependent.
\end{enumerate*}

\ack Francesco Sardelli gratefully acknowledges financial support from the University of Melbourne’s Faculty of Science [through the \emph{Postgraduate Writing-Up Award}], and also from the University of Melbourne’s School of Geography, Earth and Atmospheric Sciences [through internal funding]. Moreover, Craig. H. Bishop and Francesco Sardelli gratefully acknowledge financial support from the ARC Centre of Excellence for Climate Extremes [CE170100023] and the Bureau of Meteorology. 

\section*{Conflict of interest statement}
The authors declare no conflict of interest.

\section*{Data availability statement}
The computer experiment code utilized in this study is available from the corresponding author upon request.

\section*{Supporting information}
Three appendices are included in the supporting information of this article: appendix \ref{app:SwiftAlgorithmTechnical}, titled \textbf{Technical core of the derivation of the algorithm for $(\hat{\P}^s)^{-\frac{1}{2}}$}; appendix \ref{app:constructingP}, titled  \textbf{Construction of the synthetic covariance matrix $\P$}; and appendix \ref{app:ConstructionOfCb}, titled \textbf{Constructing the localization matrix}.

\printbibliography




\clearpage

\setcounter{page}{1}
\setcounter{section}{0}
\renewcommand{\thesection}{S\arabic{section}}
\setcounter{equation}{0}
\renewcommand{\theequation}{S\arabic{equation}}
\begin{center}
    \textbf{\huge Supporting information}
\end{center}

\section{Technical core of the derivation of the algorithm for $(\hat{\P}^s)^{-\frac{1}{2}}$ }
\label{app:SwiftAlgorithmTechnical}

In this appendix, we will provide the mathematical core of the derivation of the swift algorithm introduced in section \ref{sec:swiftAlgorithm}. More specifically, we will prove equation \eqref{eq:sdvOfA}. Moreover, we will prove that the matrices $\Eb$ and $\Fb$ appearing in that equation are orthogonal. As a consequence, we will show that formula \eqref{eq:sdvOfA} is a singular value decomposition of the matrix $\A$, defined in \eqref{eq:ADef}. This appendix is heavily based on the notation, the concepts and the theory developed in section \ref{sec:swiftAlgorithm}, whose knowledge we will assume as a prerequisite.\par 
Before getting to equation \eqref{eq:sdvOfA}, we will develop a formalism that we allow us to handle some concepts of section \ref{sec:swiftAlgorithm} in a clean and elegant way. To develop this formalism, let us start by considering a segment $\z_{k,l}$ (let us recall that the vectors $\z_{k,l}$ were defined in \eqref{eq:ZSegmentDef} and they are the constituents of the segment structure \eqref{eq:ZSegments} of the vectors $\z_k$). Let us now ask the following question. How can we extract the quantity of the $m$-th grid vertical column $(\z_{k,l})_m$ contained in $\z_{k,l}$? This is a very simple task. Let $\phim$ be a $n_h$-dimensional row vector whose elements are all $0$, but the $m$-th, which equals $1$:
\begin{align} \label{eq:phimDef}
  \phim &\doteq \begin{bmatrix}
                0 & \cdots & \here{1}{fromhere} & \cdots & 0
            \end{bmatrix}
\end{align}
\begin{tikzpicture}[remember picture, overlay]
\node[font=\scriptsize, below right=15pt of fromhere] (tohere) {$m$-th entry};
\draw[Latex-, orcidlogocol, line width=1.5pt] ([yshift=-4pt]fromhere.south) |- (tohere);
\end{tikzpicture}
\newline \vspace{0em}%
Now, to get $(\z_{k,l})_m$, we just need to multiply $\phim$ by the segment $\z_{k,l}$:
\begin{equation} \label{eq:ExtractFromSegment}
    (\z_{k,l})_m= \phim \, \z_{k,l}
\end{equation}
Moreover, if, starting from the quantity $(\z_{k,l})_m$, we would like to get a segment, i.e. a $n_h$-dimensional column vector, with $(\z_{k,l})_m$ in the $m$-th entry and $0$ elsewhere, we can just multiply the transpose $\phim\tp$ of $\phim$ by $(\z_{k,l})_m$:
\begin{equation} \label{eq:ExpandSegment}
   \begin{bmatrix}
                0 \\[10pt] \vdots \\[10pt] \here{$(\z_{k,l})_m$}{fromhere} \\[10pt] \vdots \\[10pt] 0
            \end{bmatrix} =\phim\tp \, (\z_{k,l})_m
\end{equation}
\begin{tikzpicture}[remember picture, overlay]
\node[font=\scriptsize,  left=30pt of fromhere] (tohere) {$m$-th entry};
\draw[Latex-, orcidlogocol, line width=1.5pt] ([xshift=2cm]fromhere) edge (tohere);
\end{tikzpicture}
\newline \vspace{0em}%
Now, let us suppose that we would like to zero out all entries but the $m$-th one of a segment $\z_{k,l}$. From \eqref{eq:ExtractFromSegment} and \eqref{eq:ExpandSegment}, we get that, to do this, we just need to left-multiply $\z_{k,l}$ by $\phim\tp \,\phim$, that is, 
\begin{equation} \label{eq:ZeroOutSegment}
   \begin{bmatrix}
                0 \\[10pt] \vdots \\[10pt] \here{$(\z_{k,l})_m$}{fromhere} \\[10pt] \vdots \\[10pt] 0
            \end{bmatrix} =\phim\tp \,\phim \, \z_{k,l}
\end{equation}
\begin{tikzpicture}[remember picture, overlay]
\node[font=\scriptsize,  left=30pt of fromhere] (tohere) {$m$-th entry};
\draw[Latex-, orcidlogocol, line width=1.5pt] ([xshift=2cm]fromhere) edge (tohere);
\end{tikzpicture}
\newline \vspace{0em}%
The product $\phim\tp \,\phim$ is a $n_h \times n_h$ diagonal matrix whose diagonal entries are all $0$ but the $m$-th one, which equals $1$:
\begin{equation} \label{eq:DiagonalAllZeroButM}
 \phim\tp \,\phim =  \begin{bNiceMatrix}[columns-width = 0.7cm]
                0 &  &  & \Block{2-2}<\Huge>{0} &   \\[10pt] 
                 & \Ddots &  &  & \\[10pt] 
                 &  & \here{$1$}{fromhere} &  & \\[10pt] 
                \Block{2-2}<\Huge>{0} &  &  &\Ddots  & \\[10pt] 
                &  &  &  & 0
            \end{bNiceMatrix} 
\end{equation}
\begin{tikzpicture}[remember picture, overlay]
\node[font=\scriptsize,  right=100pt of fromhere] (tohere) {$m$-th entry};
\draw[Latex-, orcidlogocol, line width=1.5pt] ([xshift=0cm]fromhere) edge (tohere);
\end{tikzpicture}
From \eqref{eq:DiagonalAllZeroButM}, we find the following relation, which will be useful later in this appendix:
\begin{equation}  \label{eq:phiPhiBeforeAfter}
    \phim\tp \,\phim \; \diag (\z_{k,l})=   \phim\tp \,\phim \;\diag (\z_{k,l}) \; \phim\tp \,\phim
\end{equation}
Let us now consider a vector $\z_k$, instead of just a segment of its. If we would like to get a column vector listing all \emph{quantities of $m$-the grid vertical column} in $\z_k$, we can left-multiply each segment $\z_{k,l}$ in $\z_k$ by $\phim$ as follows:
\begin{equation} \label{eq:ContractZ}
\begin{bmatrix}
                (\z_{1,k})_m \\[10pt]
                (\z_{2,k})_m \\[10pt]
                \vdots   \\[10pt]
                (\z_{l,k})_m \\[10pt]
                \vdots   \\[10pt]
                (\z_{n_w,k})_m
            \end{bmatrix} =
  \begin{bmatrix}
               \phim\, \z_{1,k} \\[10pt]
               \phim\, \z_{2,k} \\[10pt]
                \vdots   \\[10pt]
               \phim\, \z_{l,k} \\[10pt]
                \vdots   \\[10pt]
               \phim\, \z_{n_w,k}
            \end{bmatrix} =
            \begin{bNiceMatrix}
               \phim &  &  &  \Block{3-3}<\Huge>{0}& & \\[10pt]
                 & \phim  &  &  & & \\[10pt]
                 &  &\Ddots &   & & \\[10pt]
                \Block{3-3}<\Huge>{0}  & &  & \phim & & \\[10pt]
                 & & &  &\Ddots &  \\[10pt]
                &  &  & & & \phim
             \end{bNiceMatrix} \;
              \begin{bmatrix}
                \z_{1,k} \\[10pt]
              \z_{2,k} \\[10pt]
                \vdots   \\[10pt]
               \z_{l,k} \\[10pt]
                \vdots   \\[10pt]
                \z_{n_w,k}
            \end{bmatrix} 
\end{equation}
For notational simplicity, let us denote the block diagonal matrix in \eqref{eq:ContractZ} as $\Phim$, i.e.
\begin{equation} \label{eq:ContractionMatrix}
\Phim \doteq
            \begin{bNiceMatrix}
               \phim &  &  &  \Block{3-3}<\Huge>{0}& & \\[10pt]
                 & \phim  &  &  & & \\[10pt]
                 &  &\Ddots &   & & \\[10pt]
                \Block{3-3}<\Huge>{0}  & &  & \phim & & \\[10pt]
                 & & &  &\Ddots &  \\[10pt]
                &  &  & & & \phim
             \end{bNiceMatrix} 
\end{equation}
As shown in \eqref{eq:ContractZ}, when left-multiplying $\z_k$ by $\Phim$, we delete all entries in $\z_k$ except the \emph{quantities of the $m$-th vertical column}. For this reason, we will refer to $\Phim$ as the \emph{contraction matrix for the $m$-th grid vertical column} or simply \emph{contraction matrix}.\par
Let us now define the matrices $\Sm$ with $1 \le m \le n_h$ by
\begin{equation}
    \Sm \doteq \Phim\tp \, \Phim
\end{equation}
It is straightforward to check (and we will not do this for the sake of concision) that $\Sm$ is a block diagonal matrix and its blocks along the diagonal are $\boldsymbol{\varphi}_1\tp \,\boldsymbol{\varphi}_1$, $\boldsymbol{\varphi}_2\tp \,\boldsymbol{\varphi}_2$, \dots, $\boldsymbol{\varphi}_{n_h}\tp \,\boldsymbol{\varphi}_{n_h}$. These blocks are the diagonal matrices given by \eqref{eq:DiagonalAllZeroButM}. This means that $\Sm$ is itself diagonal. From \eqref{eq:DiagonalAllZeroButM}, we find the following relations for the blocks $\phim\tp\, \phim$:
\begin{equation} \label{eq:algebraBlocks}
    (\phim\tp \,\phim) \, (\boldsymbol{\varphi}_{m^\prime}\tp \,\boldsymbol{\varphi}_{m^\prime}) = \begin{cases}
        0 & \text{if} \; m \neq m^\prime \\
        (\phim\tp \,\phim)  & \text{if} \; m = m^\prime \qquad \text{for all} \; m 
        \end{cases}
\end{equation}
for all $m$ and $m^\prime$ with $1 \le m \le n_h$ and $1 \le m^\prime \le n_h$, and
\begin{equation} \label{eq:IdentityResolutionBlocks}
    \sum_{m=1}^{n_h} (\phim\tp \,\phim) = \I_{n_h}
\end{equation}
where $\I_{n_h}$ is the $n_h \times n_h$ identity matrix. From \eqref{eq:algebraBlocks} and \eqref{eq:IdentityResolutionBlocks} for the blocks $(\phim\tp \,\phim)$, we find the following corresponding relations for the matrices $\Sm$:
\begin{equation} \label{eq:algebra}
    \Sm\, \mathbf{S}_{m^\prime}  = \begin{cases}
        0 & \text{if} \; m \neq m^\prime \\
        \Sm & \text{if} \; m = m^\prime \qquad \text{for all} \; m 
        \end{cases}
\end{equation}
for all $m$ and $m^\prime$ with $1 \le m \le n_h$ and $1 \le m^\prime \le n_h$, and
\begin{equation} \label{eq:IdentityResolution}
    \sum_{m=1}^{n_h} \Sm = \I_{n}
\end{equation}
where $\I_{n}$ is the $n \times n$ identity matrix.\par
 \colorboxToBeHidden{PaleYellow}{Explain why $k \sim 10^3$ is good enough and no vertical localization needed and so on \dots} \par
From now on in this appendix, let us restrict ourselves to the special case when $n_w=K$, namely, assumption \eqref{eq:nwEqualsK} (see section \ref{sec:AssumptionAboutK}). \par
Using relation \eqref{eq:phiPhiBeforeAfter}, i.e. $ \phim\tp \,\phim \; \diag (\z_{k,l})=   \phim\tp \,\phim \;\diag (\z_{k,l}) \; \phim\tp \,\phim$, and  considering the block structure \eqref{eq:ADef} of the matrix $\A$ and the block structure of the matrix $\Sm$, we find
\begin{equation} \label{eq:SmABeforeAfter}
    \Sm \,\A = \Sm \,\A \,\Sm
\end{equation}
Now, let us consider the matrices $\A_m$, $\Check{\Eb}_m$ and $\Check{\Fb}_m$, defined in section \ref{sec:AlgorithmDescription}. A moment of reflection about the definitions of these matrices will convince the reader that:  
\begin{align}
    \A_m & = \Phim \,\A \, \Phim\tp \label{eq:AmAAndPhiTp}\\
    \Check{\Eb}_m & = \Phim\tp \,\Eb_m \label{eq:EmCheck}\\
    \Check{\Fb}_m & = \Phim\tp \, \Fb_m \label{eq:FmCheck}
\end{align}
for all integers $m$ with $1 \le m \le n_h$. Thus, we have utilized the formalism developed in this appendix to obtain very compact formulas for $\A_m$, $\Check{\Eb}_m$ and $\Check{\Fb}_m$  in terms of the matrices $\A$, $\Eb_m$ and $\Fb_m$, respectively. These formulas will be useful in the next steps of our reasoning, leading us to prove equation \eqref{eq:sdvOfA}, as desired at the beginning of this appendix.\par
To proceed, let us remind ourselves that $\Sm \doteq \Phim\tp \Phim$ and let us use \eqref{eq:IdentityResolutionBlocks}, \eqref{eq:SmABeforeAfter} and \eqref{eq:AmAAndPhiTp} to write $\A$ in terms of the matrices $\A_m$:
\begin{equation}
    \begin{split}
         \A  & = \I_n \: \A =  \sum_{m=1}^{n_h} \Sm \A =  \sum_{m=1}^{n_h} \Sm \A \Sm =  \sum_{m=1}^{n_h} \Phim\tp \A_m \Phim \\
    \end{split}
\end{equation}
Then, let us recall equation \eqref{eq:svdOfAm}, i.e. $ \A_m= \Eb_m \Lambdam \Fb_m\tp$, and let us use \eqref{eq:EmCheck} and \eqref{eq:FmCheck} so that we can write $\A$ as follows
\begin{equation} \label{eq:ProvingSVDOfA}
\begin{split}
    \A & = \sum_{m=1}^{n_h} \Phim\tp \A_m \Phim \\ &= \sum_{m=1}^{n_h} \Phim\tp \Eb_m \:\Lambdam \:\Fb_m\tp \Phim \\ & = \sum_{m=1}^{n_h} \Phim\tp \Eb_m \: \Lambdam \: (\Phim\tp \Fb_m)\tp \\
    & = \sum_{m=1}^{n_h} \Check{\Eb}_m \: \Lambdam \: (\Check{\Fb}_m)\tp
    \\ & =\Eb \:\boldsymbol{\Lambda} \: \Fb\tp
\end{split}
\end{equation}
where $\boldsymbol{\Lambda}$, $\Eb$ and $\Fb$ were defined in \eqref{eq:LambdaBlocks}, \eqref{eq:EBlocks} and \eqref{eq:FBlocks}, respectively. Thus, with the chain of equalities \eqref{eq:ProvingSVDOfA}, we have proven equation \eqref{eq:svdOfAm}, i.e. $\A= \Eb \:\boldsymbol{\Lambda} \: \Fb\tp$, as we wanted.\par
Now, we also would like to show that $\A = \Eb \:\boldsymbol{\Lambda} \: \Fb\tp$ is an actual singular value decomposition. Let us recall that, by its defining equation \eqref{eq:LambdaBlocks}, $\boldsymbol{\Lambda}$ is block diagonal, and its diagonal blocks are $\Lambdam$ ($1 \le m \le n_h$). Moreover, by definition, the $\Lambdam$'s are the diagonal matrices appearing in the singular value decomposition of $\A_m$, i.e. $ \A_m= \Eb_m \Lambdam \Fb_m\tp$. Therefore, all diagonal elements of the $\Lambdam$'s and, consequently, of $\boldsymbol{\Lambda}$ are non-negative. Hence, to show that $\A = \Eb \:\boldsymbol{\Lambda} \: \Fb\tp$ is an actual singular value decomposition, it suffices to prove that the matrices $\Eb$ and $\Fb$ are orthogonal, i.e. 
\begin{equation}
      \Eb\tp \Eb =\Eb \Eb\tp= \I_n  \qquad \text{and} \qquad \Fb\tp \Fb = \Fb \Fb\tp = \I_n
\end{equation}
To prove that $\Eb\tp  \Eb = \I_n$, let us first considers the row vectors $\phim$. For these, we have:
\begin{equation}
    \phim \, \phimp\tp =\begin{cases}
        0 & \text{if} \; m \neq m^\prime \\
        1  & \text{if} \; m = m^\prime 
        \end{cases}
\end{equation}
for all $m$ and $m^\prime$ with $1 \le m \le n_h$ and $1 \le m^\prime \le n_h$. As a consequence, we have the following relation for the block matrices $\Phim$:
\begin{equation} \label{eq:PhimSquared}
    \Phim \, \Phimp\tp =\begin{cases}
         \boldsymbol{0} & \text{if} \; m \neq m^\prime \\
        \I_{n_w}  & \text{if} \; m = m^\prime 
        \end{cases}
\end{equation}
Hence, using \eqref{eq:EmCheck}, we get
\begin{equation} \label{eq:EmTpPhimSquared}
  \Check{\Eb}_m\tp \Check{\Eb}_{m^\prime} = (\Phim\tp \Eb_m)\tp \, (\Phimp\tp \Eb_{m^\prime}) = \Eb_m\tp \Phim \,  \Phimp\tp \Eb_{m^\prime}=\begin{cases}
       \boldsymbol{0} & \text{if} \; m \neq m^\prime \\
        \I_{n_w}  & \text{if} \; m = m^\prime 
        \end{cases}
\end{equation}
where, for the last equality, apart from using \eqref{eq:PhimSquared}, we recalled that $\Eb_m$ is orthogonal and thus $\Eb_m\tp \Eb_m  = \I_{n_w}$, for all $m$ with $1 \le m \le n_h$. From \eqref{eq:EmTpPhimSquared}, by considering the block structure \eqref{eq:EBlocks} of $\Eb$, we get $\Eb\tp  \Eb = \I_n$, as desired.\par
We now need to show that $\Eb \, \Eb\tp = \I_n$. Let us do this through the following chain of equalities and, subsequently, by listing the mathematical facts on which this chain is based:
\begin{equation}
    \begin{split}
    \Eb \, \Eb\tp &=\sum_{m=1}^{n_h}  \Check{\Eb}_m  \: \Check{\Eb}_m\tp \\
    &= \sum_{m=1}^{n_h} \Phim\tp \Eb_m  \: (\Phim\tp \Eb_m)\tp \\
    &= \sum_{m=1}^{n_h} \Phim\tp \Eb_m  \:\Eb_m\tp \Phim \\
    &=\sum_{m=1}^{n_h}\Sm = \I_n
    \end{split}
\end{equation}
where:
\begin{itemize}
    \item In the first line, we used the fact that, by definition, the matrix $\Eb$ has the block structure \eqref{eq:EBlocks};
    \item To go from the first to the second line, we used relation \eqref{eq:EmCheck}, i.e. $\Check{\Eb}_m = \Phim\tp \Eb_m$;
    \item To go from the third to the fourth line, we recalled that, by definition, \mbox{$\Sm= \Phim\tp \Phim $}, and that $\Eb_m$ is orthogonal and thus $\Eb_m \, \Eb_m\tp = \I_{n_w}$;
    \item In the last line, we used relation \eqref{eq:IdentityResolution}.
\end{itemize}
This concludes the proof that $\Eb$ is orthogonal. Moreover, the proof that $\Fb$ is orthogonal is fully analogous to the one for $\Eb$, and it is omitted for the sake of concision. \par
With this, we conclude this appendix, in which:
\begin{itemize}
    \item We have proven equation \eqref{eq:sdvOfA}, i.e. $\A = \Eb \:\boldsymbol{\Lambda} \: \Fb\tp$;
    \item We have shown that $\A = \Eb \:\boldsymbol{\Lambda} \: \Fb\tp$ is an actual singular value decomposition. 
\end{itemize}

\section{The rank of the matrix $\A$} \label{sec:RankOfA}

As explained in section \ref{sec:AlgorithmDescription} and shown in appendix \ref{app:SwiftAlgorithmTechnical}, the singular value decomposition of $\A$ is $\A= \E \:\boldsymbol{\Lambda} \: \Fb\tp$ (equation \eqref{eq:sdvOfA}). Thus, the singular values of $\A$ are the diagonal elements of the matrix $\boldsymbol{\Lambda}$. In turn, $\boldsymbol{\Lambda}$ is a block diagonal matrix, and its diagonal blocks are the matrices $\Lambdam$ ($1 \le m \le n_h$) appearing in the singular value decompositions of $\A_m$, that is, $\A_m= \Eb_m \Lambdam \Fb_m\tp$ ($1 \le m \le n_h$). Therefore, the diagonal elements of $\Lambdam$ are the singular values of $\A_m$, and the set of singular values of $\A$ is given by the collection of the singular values of all the $\A_m$'s with $1 \le m \le n_h$. Hence, in particular, the number of non-zero singular values of $\A$ is given by the sum, over all $m$ with $1 \le m \le n_h$, of the number of non-zero singular values of $\A_m$. Consequently, we have that:
\begin{equation} \label{eq:RankARankAm}
    \rank (\A) = \sum_{m=1}^{n_h} \rank (\A_m)
\end{equation}
Thus, the matrix $\A$ is full rank if and only if all matrices $\A_m$ ($1 \le m \le n_h$) are full rank. Equation \eqref{eq:RankARankAm} suggests that a route to study the rank $\A$ may be to focus on better understanding the matrices $\A_m$. Let us now follow this route.\par
Recalling the definitions of $\A$, i.e. \eqref{eq:ADef}, and of $\Phim$, i.e. \eqref{eq:phimDef}, through straightforward linear algebra calculations, the reader can verify that the $k$-th ($1 \le k \le K$) column of the matrix $\A_m = \Phim \A \Phim\tp$ ($1 \le m \le n_h$) is given by $\Phim \z_k$. In other words, recalling \eqref{eq:ContractZ} which shows the action of the contraction matrix $\Phim$ on the vector $\z_k$, the $k$-th  column of  $\A_m$ is a $n_w$-dimensional vector listing the quantities of the $m$-th grid vertical column contained in the vector $\z_k$. Namely, $\A_m$ is nothing but the matrix $\Z$ after deleting all its (of $\Z$) rows apart from the ones containing quantities of the $m$-th grid vertical column. Let us recall that $\Z$ is a square root of the sample covariance matrix $\P^s$, i.e. $\P^s= \Z \Z\tp$. Therefore, if we compute $\A_m \A_m\tp$, we obtain a submatrix of $\P^s$. The elements of this submatrix are the sample variances and covariances of the quantities of the $m$-th grid vertical column. In other words, by defining the matrix $\P^s_m$ as $\P^s_m \doteq \A_m \A_m\tp$, we have that $\P^s_m$ is the sample covariance matrix of the quantities of the $m$-th vertical grid column. Now, from linear algebra, we know that $\rank(\A_m) = \rank (\A_m \A_m\tp)$. Therefore, we can write
\begin{equation} \label{eq:RankARankPsm}
     \rank (\A) = \sum_{m=1}^{n_h} \rank (\P^s_m)
\end{equation}
Relation \eqref{eq:RankARankPsm} implies that $\A$ is full rank if and only if all matrices $\P^s_m$ ($1 \le m \le n_h$) are full rank. For the purpose of getting full-rank $\P^s_m$'s (and, thus, a full-rank $\A$), a twofold strategy can be adopted. 
Let us conclude this section with an outline of this strategy: \def\RegularTheEnumi{\theenumi}
\renewcommand{\theenumi}{\arabic{enumi}}
\begin{enumerate}
      \item If we carried out ordinary procedures in creating the ensemble $\{\e_k \}_{k=1}^K$ and in defining the vectors $\{\z_k \}_{k=1}^K$, the resulting sample covariance matrix $\P^s$ would not be full rank because the vectors $\z_k$, i.e. the columns of the square root $\Z$ of $\P^s$, would sum to $\mathbf{0}$. Similarly, the columns of $\A_m$ would sum to $\mathbf{0}$, and the matrices $\P^s_m \doteq \A_m \A_m\tp$ would not be full rank, either. Hence, this mechanism is a cause of rank deficiency for the matrix $\A$.  In section \ref{sec:GeneratingEnsemble}, we explained an approach to address this issue by tweaking the procedure to create the ensemble $\{\e_k \}_{k=1}^K$ and to define the vectors $\{\z_k \}_{k=1}^K$. We utilized this approach in our experiment, showing that this allowed us to use our swift algorithm with a resulting very good accuracy (section \ref{sec:Accuracy}). This approach is based on a random selection process of the ensemble members (randomly discarding very few of them). As explained in section \ref{sec:GeneratingEnsemble}, the resulting vectors $\{\z_k \}_{k=1}^K$ do not sum exactly to $\mathbf{0}$. Moreover, with this approach, the columns of each matrix $\A_m$ (let us recall that each of those columns equal $\Phim \z_k$) do not exactly sum to $\mathbf{0}$, either, thus eliminating this cause of rank deficiency for the matrices $\P^s_m \doteq \A_m \A_m\tp$, and, consequently for $\A$.
    \item Another (possible) cause of rank deficiency for the matrices $\P^s_m \doteq \A_m \A_m\tp$ (and, thus, for $\A$) is related with the chosen set of model variables in the atmospheric state vector for a given data assimilation system. To explain this mechanism, let us consider the probability distribution which the ensemble represents, namely, the probability distribution from which the ensemble members can be thought of as being drawn. Let us suppose that, given that probability distribution, there exist a quantity, defined as a linear combination of the chosen model variables, with zero variance. The existence of such a quantity will (almost certainly) cause a zero eigenvalue to be in the spectrum of the sample covariance matrix $\P^s$. This, in turn, may lead to one (or more) matrices $\P^s_m $ (and, consequently, $\A$) to be rank deficient. The existence of the above-mentioned quantity (or more than one such quantity) can originate from linear balance relations. To address this cause of rank deficiency for $\A$, one can simply exclude the balanced part of the variables involved in linear balance relations from the set of model variables in the atmospheric state vector \parencite{Bannister2008}.
\end{enumerate}

\section{Derivation of the swift algorithm for the product between $(\Pc)^{-\frac{1}{2}}$ and a vector} \label{sec:ActionOnVectorTechnical}

Let us consider the algorithm described in section \ref{sec:ActionOnVector}, namely, steps \ref{V:wm} to \ref{V:xs}. As we know, that algorithm takes a generic ($n$-dimensional) state vector $\x$ as an input and it gives a $\ns$-dimensional vector $\xs$ as an output. The aim of this section is to prove that that algorithm is indeed an implementation of the standardized variables transform, namely, that $\xs=(\Pc)^{-\frac{1}{2}} \x$.\par
Let us start our derivation by considering the vectors $\x^{(m)}$, $\w^{(m)}$ and $\w$, defined in section \ref{sec:ActionOnVector}. By recalling equations \eqref{eq:ContractZ} and \eqref{eq:ContractionMatrix}, we realize that, as $\x^{(m)}$ lists the quantities of the $m$-th grid vertical column of $\x$, it can be expressed through the contraction matrix $\Phim$ as follows:
 \begin{align}
   \x^{(m)} &=  \Phim \x \label{eq:xmFromx}
\end{align}
for every $m$ with $1 \le m \le n_h$. Moreover, a moment of reflection about equations \eqref{eq:ExpandSegment} and \eqref{eq:ContractionMatrix} will convince the reader that
 \begin{align}
   \w  &= \sum_{m=1}^{n_h} \Phim\tp \w^{(m)} \label{eq:wFromwm}
\end{align}
Equations \eqref{eq:xmFromx} and \eqref{eq:wFromwm} will be useful below in our derivation. Now, let us consider the inverse $\A^{-1}$ of the matrix $\A$. Recalling the block structures \eqref{eq:FBlocks}, \eqref{eq:LambdaBlocks} and \eqref{eq:EBlocks} of $\F$, $\boldsymbol{\Lambda}$ and $\E$, respectively, and formula \eqref{eq:svdInverseOfA} for the singular value decomposition of $\A^{-1}$, we can write
\begin{equation} \label{eq:InverseOfAAsSum}
\begin{split}
   \A^{-1}&= \Fb \boldsymbol{\Lambda}^{-1} \Eb\tp \\ & = \sum_{m=1}^{n_h} \Check{\Fb}_m \: \Lambdam^{-1} \: (\Check{\Eb}_m)\tp\\  & = \sum_{m=1}^{n_h} \Phim\tp \Fb_m \: \Lambdam^{-1} \: (\Phim\tp \Eb_m)\tp \\
   &= \sum_{m=1}^{n_h} \Phim\tp \Fb_m \:\Lambdam^{-1} \:\Eb_m\tp \Phim \\ 
\end{split}
\end{equation}
where we used relations \eqref{eq:EmCheck} and \eqref{eq:FmCheck}, i.e. $\Check{\Eb}_m = \Phim\tp \Eb_m $ and $\Check{\Fb}_m = \Phim\tp \Fb_m $, respectively. Let us now recall the definition of the vectors $\w^{(m)}$ given at step \ref{V:wm}, namely, $\w^{(m)} \doteq \Eb_m \:\Lambdam^{-1} \:\Fb_m\tp \, \x^{(m)}$ ($1 \le  m \le n_h$). With this definition in mind, and using \eqref{eq:xmFromx}, \eqref{eq:wFromwm} and \eqref{eq:InverseOfAAsSum}, we find:
\begin{equation} \label{eq:InverseOfATimesx}
\begin{split}
   \A^{-1} \, \x &= \sum_{m=1}^{n_h} \Phim\tp \Fb_m \:\Lambdam^{-1} \:\Eb_m\tp \Phim \x\\ 
    &= \sum_{m=1}^{n_h} \Phim\tp \Fb_m \:\Lambdam^{-1} \:\Eb_m\tp \x^{(m)}\\ 
    &= \sum_{m=1}^{n_h} \Phim\tp \w^{(m)} = \w
\end{split}
\end{equation}
Now, let us consider steps \ref{V:wl} to \ref{V:xs} through which the vector $\xs$ is constructed from the vector $\w$. Having those steps in mind and recalling the block structures \eqref{eq:DeltaBlocks} and \eqref{eq:VBlocks} of $\Deltab$ and $\V$, respectively, we realise that 
\begin{equation} \label{eq:xsFromw}
    \xs =  \Deltab^{-\frac{1}{2}} \V\tp \,\w
\end{equation}
Combining \eqref{eq:xsFromw} and \eqref{eq:InverseOfATimesx}, and recalling relation \eqref{eq:InverseSquareRootAlmost}, i.e. $(\Pc)^{-\frac{1}{2}}= \Deltab^{-\frac{1}{2}} \V\tp \A^{-1}$, we find:
\begin{equation}
    \xs = (\Pc)^{-\frac{1}{2}} \x
\end{equation}
as we wanted to show. Thus, the algorithm given by steps \ref{V:wm} to \ref{V:xs} is indeed an implementation of the standardized variables transform.

\section{The grid and the vertical coordinate $h$} \label{sec:gridDetailed}

In this section, let us give a detailed description of the grid used in our synthetic experiment. As mentioned in section \ref{sec:grid}, we considered a three dimensional grid for the atmosphere of a spherical planet. The grid has $n_v= 64$ horizontal layers. There is only $n_{G}=1$ model variable per grid point. Thus, the number of model variables in each vertical column is $n_w = n_G  n_v =64$. Let $\theta$ denote the longitude and let us choose a set of $n_\theta=20$ evenly spaced longitude values $\theta_i$ ($1 \le i \le n_\theta$) starting with $\theta_1=\ang{18}$ and ending with $\theta_{n_{\theta}}= \ang{360}$. Analogously, let $\varphi$ denote the latitude and let us choose a set of $n_\varphi=10$ evenly spaced latitude values $\varphi_i$ ($1 \le i \le n_\varphi$) starting with $\varphi_1=-\ang{72}$ and ending with $\varphi_{n_{\varphi}}=  \ang{72}$. To construct each horizontal layer of our grid, as grid points, let us choose all points with longitude-latitude coordinates given by $(\theta_i,\varphi_j)$, where $\theta_i$ ($1 \le i \le n_\theta$) and $\varphi_j$ ($1 \le j \le n_\varphi$) belong to the chosen sets of values mentioned above for the coordinates $\theta$ and $\varphi$, respectively. With this procedure, we have a grid with $n_h= n_\theta n_\varphi = \num{200}$ grid points on each horizontal layer, with a longitudinal resolution $\Delta\theta =\ang{18}$ and a latitudinal resolution $\Delta\varphi =\ang{16}$. The total number $n$ of grid points in our experiment is $n= n_w n_h=\num{12800}$.\par
To complete our set of coordinates, let us add a vertical coordinate $h$ to the longitude $\theta$ and the latitude $\varphi$. Let $h$ be a synthetic, unitless, positive coordinate, i.e. $ h > 0$. Then, analogously to what we already did for $\theta$ and $\varphi$, let us choose a set $n_v=64$ evenly values $h_l$ ($1 \le l \le n_v$) for $h$ starting with $h_1=1$ and ending with $h_{64}=64$, i.e. all integers from $1$ to $64$. 
Let us set each one of these $n_v=64$ values $h_l$ for the coordinate $h$ to be the vertical coordinate of each horizontal grid layer. Here, we need to make a remark about $h$. This vertical coordinate should \emph{not} be thought of as either equal to the physical height, or directly proportional to it. This is because, typically, in atmospheric models, the vertical resolution is higher at low height and lower at high height. In other words, the lower horizontal layers are closer to each other than the higher horizontal layers. In our synthetic experiment, we would like to take into account and somehow mimic this aspect. Thus, considering how we defined $h$, the same increase in this vertical coordinate should correspond to a relatively small increase in height if we are at the low levels and to a relatively large increase in height at the high levels. Let us keep this point in mind, as it will be useful below when we will construct a synthetic covariance matrix.\par
By adding $h$, we have completed the set of coordinates for our experiment. Thus, the three dimensional coordinates of each grid point are of the form  $(\theta_i, \varphi_j , h_l)$ with $1 \le i \le n_\theta$, $1 \le j \le n_\varphi$ and $1 \le l \le n_v$.

    \section{Construction of the synthetic covariance matrix $\P$} \label{app:constructingP}
    We constructed the $n \times n$ synthetic covariance matrix $\P$ by using a change-of-variables technique. To explain this, let us start by considering the functions of the following product form
\begin{equation} \label{eq:BasisContinuum}
    f_{n,m,r}(\theta, \varphi, h) = Y_{n,m} (\theta, \varphi) \, \xi_r(h)
\end{equation}
where:
\begin{itemize}
    \item $Y_{n,m} (\theta, \varphi)$ is the spherical harmonic of degree $n$ and order $m$, where $n$ and $m$ are  integers with $n \ge 0$ and $-n \le m \le n$;
    \item $\xi_r (h)$ with $1 \le r \le L_{\xi}$ are functions of the vertical coordinate $h$ only. We shall return to these later.
\end{itemize}
Let us restrict ourselves to a finite subset among all functions \eqref{eq:BasisContinuum}. Namely, let us only consider functions \eqref{eq:BasisContinuum} with $n \le n_{\text{max}}$, where $n_{\text{max}}$ is a fixed positive integer. In particular, we chose $n_{\text{max}}= 9$. \colorboxToBeHidden{PaleYellow}{Shall I explain why this choice for $n_{\text{max}}$?}%
\par
Let us recall that, in our synthetic experiment, we only have one variable kind, which, to fix ideas, we called \emph{temperature}. For a moment, let us see the temperature field $T$ as a function of our three continuous coordinates $(\theta, \varphi, h)$. Thus, we will denote it by $T(\theta, \varphi, h)$. We will consider the discretized version of the temperature field (i.e. temperature $T(\theta_i, \varphi_j, h_l)$ as a function of the grid point coordinates $(\theta_i, \varphi_j, h_l)$ only) below. The function $T(\theta, \varphi, h)$ can be expanded as a sum of the functions \eqref{eq:BasisContinuum}, that is,
\begin{equation} \label{eq:3DExpansion}
  T(\theta, \varphi, h)= \sum_{n,m,r} T_{n,m,r} \: f_{n,m,r}(\theta, \varphi, h) = \sum_{n,m,r} T_{n,m,r} \,Y_{n,m} (\theta, \varphi) \, \xi_r(h)
\end{equation}
for a suitable choice of the coefficients $T_{n,m,r}$. To a certain degree of accuracy, we can truncate the sums in \eqref{eq:3DExpansion} by only considering terms with $n \le n_{\text{max}}$. The expansion \eqref{eq:3DExpansion} allows us to represent the temperature field $T(\theta, \varphi, h)$ by simply providing the coefficients $T_{n,m,r}$.\par
In our synthetic experiment, the temperature field is discretized on a grid, that is, we have a temperature variable $T(\theta_i, \varphi_j, h_l)$ at each grid point, where $(\theta_i, \varphi_j, h_l)$ are the coordinates of the grid point itself. We can list all temperature values $T(\theta_i, \varphi_j, h_l)$, one per grid point, in a vector $\x$, which is the state vector of our synthetic experiment. To do this, we need to specify an order according to which we will arrange the values $T(\theta_i, \varphi_j, h_l)$ in the column vector $\x$. We will not go into the details of our particular chosen order because these are essentially irrelevant for the key points of our experiment.\par
After that, we can also discretize the functions $f_{n,m,r}(\theta, \varphi, h)$. For this, we can first evaluate those functions at grid point locations $(\theta_i, \varphi_j, h_l)$, that is, we can consider the quantities $f_{n,m,r}(\theta_i, \varphi_j, h_l)$. Then, we will slightly correct the values of those quantities so that the following orthonormality relations are fulfilled:
\begin{equation} \label{eq:OrthonormalityForFs}
  \sum_{l=1}^{n_h}  \sum_{i=0}^{n_{\theta}} \sum_{j=1}^{n_{\varphi}} \cos (\varphi_j) \: \Check{f}_{n,m,r} (\theta_i, \varphi_j,h_l) \: \Check{f}^\ast_{n^{\prime},m^{\prime},r^{\prime}} (\theta_i, \varphi_j,h_l) = \delta_{n n^{\prime}}\, \delta_{m m^{\prime}} \, \delta_{r r^{\prime}}
\end{equation}
where
\begin{itemize}
    \item $\Check{f}_{n,m,r} (\theta_i, \varphi_j,h_l) $ denote the corrected values just mentioned;
    \item The $\phantom{i}^\ast$ denotes the complex conjugate;
    \item $\delta_{n n^{\prime}}$, $\delta_{m m^{\prime}}$ and $\delta_{r r^{\prime}}$ are Kronecker deltas.
\end{itemize}
The orthonormality relations \eqref{eq:OrthonormalityForFs} are the discretized version of the ones with continuous coordinates, in which the sums are replaced with integrals over $\theta$, $\varphi$ and $h$. More specifically, the sums and the $\cos (\varphi_j)$ on the left-hand side of \eqref{eq:OrthonormalityForFs} are the discrete counterpart of the integral $\int \dd{h} \dd{\theta} \dd{\varphi} \, \cos{\varphi}$ in the spherical coordinates $(\theta, \varphi, h)$.  The reason behind the fact that the corrected values $\Check{f}_{n,m,r} (\theta_i, \varphi_j,h_l)$ instead of the quantities  $f_{n,m,r} (\theta_i, \varphi_j,h_l)$ are needed to fulfill \eqref{eq:OrthonormalityForFs} is related to the use of a grid with evenly spaced values of $\theta$ and $\varphi$. We will not go into the details of this, for the sake of concision. In particular, we will not describe the procedure to compute the corrected values $\Check{f}_{n,m,r} (\theta_i, \varphi_j,h_l)$ because this is a marginal aspect of our experiment.\par
The corrected values $\Check{f}_{n,m,r} (\theta_i, \varphi_j,h_l)$ can then be listed in a column vector $\fmnr$, for each given triple $m,n,r$ with $n \le n_{\text{max}}$, $-n \le m \le +n$ and $1 \le r \le L_\xi$. In doing this, we should arrange the values $\Check{f}_{n,m,r} (\theta_i, \varphi_j,h_l)$ in $\fmnr$ according to the same order with which we listed the quantities $T(\theta_i, \varphi_j, h_l)$ in the state vector $\x$. How many triples  $m,n,r$ are there with $n \le n_{\text{max}}$, $-n \le m \le +n$ and $1 \le r \le L_\xi$? It is possible to show that the answer is $(n_{\text{max}}+1)^2 \, L_\xi$. The proof of this fact is omitted for the sake of concision. We can then fix an order for these $(n_{\text{max}}+1)^2 \, L_\xi$ triples. According to this chosen order, we can concatenate the vectors $\fmnr$ as columns of a $n \times (n_{\text{max}}+1)^2 \, L_\xi$ matrix $\boldsymbol{\mathcal{L}}$.\par
Let us now write the discrete version of equation \eqref{eq:3DExpansion} allowing us to expand $T(\theta_i, \varphi_j, h_l)$ in terms of $\Check{f}_{n,m,r} (\theta_i, \varphi_j,h_l)$ using suitable coefficients $\Check{T}_{n,m,r}$:
\begin{equation} \label{eq:3DExpansionDiscrete}
  T(\theta_i, \varphi_j, h_l) = \sum_{n,m,r} \Check{T}_{n,m,r} \: \Check{f}_{n,m,r} (\theta_i, \varphi_j,h_l)
\end{equation}
where the sum is over all $(n_{\text{max}}+1)^2 \, L_\xi$ triples $n,m,r$ with $n \le n_{\text{max}}$, $-n \le m \le +n$ and $1 \le r \le L_\xi$. If we list the coefficients $\Check{T}_{n,m,r}$ in a column vector $\xr$, arranging them according to the same order for the triples $n,m,r$ used to concatenate the $\fmnr$'s as columns of $\boldsymbol{\mathcal{L}}$, we can express \eqref{eq:3DExpansionDiscrete} in the following much more compact form:
\begin{equation} \label{eq:3DExpansionDiscreteCompact}
    \x = \boldsymbol{\mathcal{L}}\, \xr
\end{equation}
Thus, the state $\x$ in our synthetic experiment can be represented with the $ (n_{\text{max}}+1)^2 \, L_\xi$-dimensional vector $\xr$ listing the coefficients $\Check{T}_{n,m,r}$.\par
Before proceeding further, we should expand a bit on the functions $\xi_r(h)$ of the vertical coordinate $h$ appearing in \eqref{eq:BasisContinuum}. In this section, we introduced the functions $\xi_r(h)$ without mentioning their analytical form. Then, we said that we evaluated the functions $f_{m,n,r}(\theta, \varphi, h) = Y_{n,m} (\theta, \varphi) \, \xi_r(h)$ at each grid point location  $(\theta_i, \varphi_j,h_l)$. In doing this, we evaluated the $\xi_r(h)$ at the values $h_l$, in particular. Thus, we got the quantities $\xi_r(h_l)$. These quantities depend on the $h_l$'s and, thus, on the horizontal layer. Actually, the procedure that we have just described to produce the $\xi_r(h_l)$'s was presented in this section with the only purpose of simplifying and better organizing the description of our experiment. In practice, we proceeded differently to construct the quantities $\xi_r(h_l)$. Let us touch on the technique we employed for this.\par
We first constructed a $n_v \times n_v$ covariance matrix $\P_v$ representing an abstract covariance structure among the different horizontal layers. $\P_v$ has to be seen as a mathematical tool to construct the quantities $\xi_r(h_l)$ and, eventually, the synthetic covariance matrix $\P$. The role played by $\P_v$ in building $\P$ will be described below. For what follows, it will be useful to remind ourselves that the vertical coordinate $h$ should not be thought of as directly proportional to the physical height. As mentioned above, the same increase in $h$ should correspond to a relatively small increase in height if we are at the low levels and to a relatively large increase in height at the high levels. Conversely, the same increase in height corresponds to a large increase in $h$ if we are at the low levels and to a relatively small increase in $h$ at the high levels. Now, let us consider the covariance functions for our synthetic temperature. Each  covariance function is defined with respect to a given grid point. Moreover, each covariance function depends on the spatial coordinates. In particular, it can be expressed as depending on the physical height or the vertical coordinate $h$ (besides latitude and longitude). Let us first express it as depending on the physical height. Then, let us suppose that the \emph{physical height length scale} characterizing the dependence of the covariance function from the physical height does not vary much with the grid point with respect to which the covariance function is defined. If we now express the covariance function as depending on the vertical coordinate $h$, we need to consider a \emph{$h$-coordinate scale} corresponding to the just mentioned physical height length scale. The relation between the two scales will be the same as the relation between physical height increments and $h$-coordinate increments. In other words, as the physical height length scale is (very roughly) the same at low levels and at high levels, the $h$-coordinate scale will be larger at low levels and smaller at high levels. Now, we would like the covariance matrix $\P_v$ to somehow encode the vertical structure of the synthetic covariance matrix $\P$. In particular, we are referring to the vertical structure with respect to the vertical coordinate $h$. Thus, the covariance matrix $\P_v$ should be constructed using a range of $h$-coordinate scales, which are large at the low horizontal layers and small at the high horizontal layers. To crudely represent this aspect, we considered only two $h$-coordinate scales: $\lambda_{\text{small}}=\num{0.75}$ and $\lambda_{\text{big}}=8\,\lambda_{\text{small}}=\num{6.02}$. Then, with each of these two $h$-coordinate scales, we constructed a $n_v \times n_v$ covariance matrix: $\P_{\lambda_{\text{small}}}$ and $\P_{\lambda_{\text{large}}}$, respectively. After that, we combined these two matrices into $\P_v$, weighting $\P_{\lambda_{\text{small}}}$ more at the high levels and $\P_{\lambda_{\text{large}}}$ more at the low levels. In principle, we should carefully describe and clarify how we built $\P_{\lambda_{\text{small}}}$ and $\P_{\lambda_{\text{large}}}$ given the corresponding scales, what formula we used to combine them into $\P$ and in what sense, in this formula, we could give more weight to one or the other at different horizontal layers. However, as these aspects are marginal for the main points of our experiment, we will not go into these details.\par
Once we had the covariance matrix $\P_v$, we performed an eigendecomposition of it, or, more conveniently, an \emph{economical eigendecomposition}, that is, we wrote $\P_v$ as
\begin{equation} \label{eq:PvEigendecomposition}
    \P_v = \boldsymbol{\mathcal{L}}_v \boldsymbol{\Lambda}_v \boldsymbol{\mathcal{L}}_v\tp
\end{equation}
where $\boldsymbol{\Lambda}_v$ is a diagonal matrix of the non-zero eigenvalues $\lambda_r$ only and the columns $\hr$  of the matrix $\boldsymbol{\mathcal{L}}_v$ are the corresponding orthonormal eigenvectors, with $1 \le r \le L_\xi$ (where, now, we have defined $L_\xi $ as the number of non-zero eigenvalues of $\P_v$). We are now ready to build the quantities $\xi_r (h_l)$. \ifBookmarkOn\colorbox{PaleYellow}{proofread from here}\fi As mentioned above, in practice, we did not produce them as explained earlier in this appendix. In particular, we did not use any set of functions $\xi_r(h)$ of the continuous coordinate $h$. Instead, we just considered the $n_v$-dimensional  vectors $\hr$  ($1 \le r \le L_\xi$). Let $(\hr)_l$ be the $l$-th element of the vector $\hr$, where $1 \le l \le n_v$ and $1 \le r \le L_\xi$. Then, we can just set:
\begin{equation} \label{eq:DefineXiAsVectorElements}
   \xi_r(h_l) \doteq  (\hr)_l
\end{equation}
In other words, instead of considering functions $\xi_r(h)$ of the continuous coordinate $h$ and evaluating them at the values $h_l$ to produce the quantities $\xi_r(h_l)$, we simply use the elements $(\hr)_l$ of the vectors $\hr$ define them through \eqref{eq:DefineXiAsVectorElements}. Once we created the $\xi_r(h_l)$'s this way, we got the quantities $ f_{n,m,r}(\theta_i, \varphi_j, h_l) = Y_{n,m} (\theta_i, \varphi_j) \, \xi_r(h_l)$, where $Y_{n,m} (\theta_i, \varphi_j)$ are the spherical harmonics evaluated at the coordinate values $(\theta_i, \varphi_j)$. Having produced the quantities $ f_{n,m,r}(\theta_i, \varphi_j, h_l)$, we then proceeded as explained above in this section, correcting them to fulfill the orthonormality relations \eqref{eq:OrthonormalityForFs}, and thus determining the corrected values $\Check{f}_{n,m,r}(\theta_i, \varphi_j, h_l)$. These, in turn, were used to construct the vectors $\fmnr$, which were concatenated as columns of the matrix $\boldsymbol{\mathcal{L}}$. This matrix allowed us to represent the state of the atmosphere $\x$ with the vector $\xr$ through equation \eqref{eq:3DExpansionDiscreteCompact}, that is, $\x = \boldsymbol{\mathcal{L}} \, \xr$. These steps, that took us from the quantities $ f_{n,m,r}(\theta_i, \varphi_j, h_l)$ all the way to the equation $\x = \boldsymbol{\mathcal{L}} \, \xr$, were already explained above in this section.\par
We are now close to be able to build the synthetic covariance matrix $\P$, which allowed us to generate the ensemble perturbations and, thus, the sample covariance matrix $\P^s$, as described above. The random draws from the normal distribution $\mathcal{N} (\mathbf{0},\P)$, whose covariance matrix is $\P$, are vectors listing temperature field values, i.e. the variables in our state vector $\x$. We can also consider the corresponding covariance matrix $\Pr$, expressed in terms of the variables $\xr$ (that is, the coefficients $\Check{T}_{m,n,r}$), instead of the variables $\x$. The matrices $\P$ and $\Pr$ will then be related through the change-of-variables formula for covariance matrices, that is,
\begin{equation} \label{eq:PFromPTilde}
    \P = \boldsymbol{\mathcal{L}} \, \Pr \, \boldsymbol{\mathcal{L}}\tp
\end{equation}
Thus, in our experiment, we created a synthetic covariance matrix $\Pr$. Then, we produced $\P$ through formula \eqref{eq:PFromPTilde}. Now, we just need to explain how we created $\Pr$. We used a diagonal matrix $\Pr$. To produce the diagonal elements $\Check{P}_{ii}$ of the matrix $\Pr$, we first defined the quantities:
\begin{equation} \label{eq:EigenvaluesOfP}
    \lambda_{n,m,r} \doteq A \,\exp(-\frac{n^2}{2\,n_s^2})\, \lambda_r 
\end{equation}
where
\begin{itemize}
    \item $\lambda_r$ ($1 \le r \le L_\xi$) are the non-zero eigenvalues of the matrix $\P_v$;
    \item $n_s$ is a parameter which controls the width of the Gaussian function appearing in \eqref{eq:EigenvaluesOfP}. In our experiment, we set $n_s=\num{2}$; 
    \item $A$ is a normalization constant, to which we will come back below.
\end{itemize}
For the following, we need to remind ourselves that we chose an order for the triples $n,m,r$ to list the coefficients $\Check{T}_{m,n,r}$ in the vector $\xr$. Using the same order, let us list the quantities $\lambda_{n,m,r}$ as diagonal elements $\Check{P}_{ii}$ ($1 \le i \le (n_{\text{max}+1})^2 \, L_\xi $). This fully determines the diagonal matrix $\Pr$, and, consequently, through \eqref{eq:PFromPTilde}, the synthetic covariance matrix $\P$. As we have seen, we used the non-zero eigenvalues $\lambda_r$ and the corresponding eigenvectors $\hr$ of $\P_v$ to construct $\P$. The eigenvalues $\lambda_r$ appear as a factor in expression \eqref{eq:EigenvaluesOfP} for the eigenvalues of $\P$. This factor encodes the dependence on the index $r$ of the eigenvalues of $\P$. In turn, $r$ indexes the quantities $\xi_r(h_l)$, which equal the elements of the eigenvectors $\hr$ and encode the dependence of the vectors $\fmnr$ on the horizontal layers. The vectors $\fmnr$ appear as columns of $\boldsymbol{\mathcal{L}}$ and, thus, in $\P$ (equation \eqref{eq:PFromPTilde}). In short, $\P_v$, through its eigenvectors and eigenvalues, encodes the vertical aspects of the covariance structure of $\P$.\par
So far, in our description of the procedure to construct $\P$, we have not expanded on the choice of the normalization constant $A$, which appears in \eqref{eq:EigenvaluesOfP}. Let us touch on this point now. For that, let us consider the diagonal elements $P_{ii}$ ($1 \le i \le n$) of $\P$. The quantities $P_{ii}$ are the values of the synthetic variances at each grid point location $(\theta_i, \varphi_j, h_l)$. We wanted all those variances to equal $1$ because this makes it convenient and easy for the scientist to interpret the magnitude of many other quantities in our experiment. With this purpose in mind, we were able to choose a normalization constant $A$ so that all synthetic variances, apart from those at some high values of the latitude, equalled $1$. And, for those high values of the latitude, the variances just slightly differed from $1$. For this, using a different normalization constant $A$ would not have helped because, with the spherical-harmonics-based procedure used to construct $\P$, it is not possible to get all diagonal elements $P_{ii}$ exactly equal to $1$, irrespective of the value of $A$. We will not go into the details of this point, for concision. Let us just touch on how we dealt with this aspect. Given $\P$ constructed with the procedure explained so far, we applied a technique to this matrix to slightly correct it, thus producing a resulting, new covariance matrix $\P_{\scaleto{\text{crt}}{4pt}}$ with all its diagonal elements (variances) equal to $1$, at all latitudes (the subscript $\phantom{l}_{\scaleto{\text{crt}}{4pt}}$ stands for \emph{corrected}). At the same time, this technique ensured that $\P_{\scaleto{\text{crt}}{4pt}}$ was positive semi-definite, as any covariance matrix must be. For the sake of concision and as it is not relevant to our main purposes, we will not describe this technique. In the remainder of our experiment, $\P_{\scaleto{\text{crt}}{4pt}}$ was used as synthetic covariance matrix, whereas $\P$ was completely discarded. In particular, we used $\P_{\scaleto{\text{crt}}{4pt}}$ as covariance matrix of the Gaussian distribution which was, in turn, sampled to create the ensemble $\e_k$. And, $\P_{\scaleto{\text{crt}}{4pt}}$ is the synthetic covariance matrix illustrated in the figures of this article (more specifically, panel  \subref{fig:CovarianceFunctionOfP28} of figure \ref{fig:CovarianceFunctions28}, panel \subref{fig:CovarianceFunctionOfP27} of figure \ref{fig:CovarianceFunctions27}, and the light blue line of figure \ref{fig:VerticalProfilesOfPAndPs}). Having explained that $\P$ was discarded, as no confusion can now arise, for notational simplicity, in the main text of this article (as well as in the next supporting information section \ref{app:ConstructionOfCb}), when referring to $\P_{\scaleto{\text{crt}}{4pt}}$, we will drop the subscript $\phantom{l}_{\scaleto{\text{crt}}{4pt}}$ and we will just write $\P$.

\section{Constructing the localization matrix} \label{app:ConstructionOfCb}

As mentioned in section \ref{sec:localizationExperiment}, the $n \times n$ localization matrix $\Cc$ had the simple block form \eqref{eq:blockStructureCEqualBlocksExperimentSubsection}, with each block $\Cc_{\textbf{b}}$ being a $n_h \times n_h$ matrix. Namely, $\Cc$ was made of $n_v^2$ blocks $\Cc_{\textbf{b}}$ in total, arranged in a $n_v \times n_v$ block structure. Let us explain how we produced the block $\Cc_{\mathbf{b}}$. \par
The procedure for this was similar to the one used to create $\P$, but with no vertical part. Let us briefly explain what we mean by this.\par
Instead of the functions $f_{m,n,r} (\theta, \varphi, h)$ used for $\P$ (see appendix \ref{app:constructingP}), we just considered the spherical harmonics $Y_{n,m} (\theta , \varphi)$ with $ 0 \le n \le n_{\text{max}}^{\text{loc}}$ and $-n \le m \le +n$, where we chose $n_{\text{max}}^{\text{loc}} = 9$. Then, the functions $Y_{n,m} (\theta , \varphi)$ were evaluated at the horizontal grid point coordinate values $(\theta_i, \varphi_j)$, with $1 \le i \le n_\theta$ and $1 \le j \le n_\varphi$, thus obtaining the quantities $Y_{n,m} (\theta_i , \varphi_j)$. Then, we slightly corrected the $Y_{n,m} (\theta_i , \varphi_j)$'s so that the following orthonormality relations are fulfilled:
\begin{equation} \label{eq:OrthonormalityForFs}
  \sum_{i=0}^{n_{\theta}} \sum_{j=1}^{n_{\varphi}} \cos (\varphi_j) \: \Check{Y}_{n,m} (\theta_i, \varphi_j) \: \Check{Y}^\ast_{n^{\prime},m^{\prime}} (\theta_i, \varphi_j) = \delta_{n n^{\prime}}\, \delta_{m m^{\prime}} \, 
\end{equation}
where
\begin{itemize}
    \item $\Check{Y}_{n,m} (\theta_i, \varphi_j) $ denote the corrected values;
    \item The $\phantom{i}^\ast$ denotes the complex conjugate;
    \item $\delta_{n n^{\prime}}$ and $\delta_{m m^{\prime}}$ are Kronecker deltas.
\end{itemize}
For the sake of concision, we will not describe the procedure to compute the corrected values $\Check{Y}_{n,m} (\theta_i, \varphi_j) $. Now, for each given pair of indices $n,m$ (with $ 0 \le n \le n_{\text{max}}^{\text{loc}}$ and $-n \le m \le +n$), we listed the quantities $\Check{Y}_{n,m} (\theta_i, \varphi_j) $ in a vector $\unm$. The order according to which we listed those quantities in $\unm$ needed to be consistent with the order we used to list variables $T(\theta_i, \varphi_j, h_l)$ in the state vector $\x$ of our synthetic experiment. We will not go into the details of this because, concerning our purposes, it is a marginal aspect. Once we obtained the vectors $\unm$, we concatenated them as columns of matrix $\Up$. To arrange the $\unm$'s in $\Up$, we had to choose an order for the pair of indices $n,m$ ($ 0 \le n \le n_{\text{max}}^{\text{loc}}$ and $-n \le m \le +n$). We will not describe our choice for this order for the sake of concision. Then, we defined the quantities:
\begin{equation} \label{eq:EigenvaluesOfCb}
    \lambda_{n,m} \doteq A_{\text{loc}} \,\exp(-\frac{n^2}{2\,(n_s^{\text{loc}})^2})  \qquad  \begin{array}{l}
    \; \, \,0 \le n \le n_{\text{max}}^{\text{loc}}\\
 -n \le m \le +n
 \end{array}
\end{equation}
where
\begin{itemize}
    \item $A_{\text{loc}}$ is a normalization constant that was chosen so that, at the end of the procedure that we are describing, the resulting localization matrix block $\Cc_{\textbf{b}}$ has diagonal elements (approximately) equal to $1$;
    \item We set the parameter $n_s^{\text{loc}}= \num{1.7}$.
\end{itemize}
The quantities $ \lambda_{n,m}$ were arranged as diagonal elements of a diagonal matrix $\Check{\Cc}_{\textbf{b}}$. After this, the localization matrix block $\Cc_{\textbf{b}}$ was produced through the following equation:
\begin{equation} \label{eq:CbThroughSphericalHarmonics}
    \Cc_{\textbf{b}} = \Up \, \Check{\Cc}_{\textbf{b}} \, \Up\tp
\end{equation}
In this matrix $\Cc_{\textbf{b}}$, constructed through \eqref{eq:CbThroughSphericalHarmonics}, let us now consider the diagonal elements $\mathcal{C}_{\scaleto{(\theta_{i}, \varphi_{j}) , (\theta_i, \varphi_j)}{7pt}}$. One finds that, for some high values of the latitude coordinate $\varphi_{j}$, those elements slightly differ from $1$  (whereas, the other diagonal elements equal $1$). As the matrix $\Cc$ has the block form \eqref{eq:blockStructureCEqualBlocksExperimentSubsection}, the diagonal elements of $\Cc$ are given by the diagonal elements of the block $\Cc_{\textbf{b}}$. And, as customary for localization matrices, we would like \emph{all} the diagonal elements of $\Cc$ to equal $1$. This issue is analogous to one that we encountered in the final stage when constructing the synthetic covariance matrix $\P$. Namely, the diagonal elements of the matrix $\P$ produced through our spherical-harmonics-based procedure slightly differed from $1$ for some high values of the latitude (appendix \ref{app:constructingP}). For the diagonal elements of $\Cc_{\textbf{b}}$, we utilized a fully analogous technique to the one used for those of $\P$. Through this technique, we slightly corrected $\Cc_{\textbf{b}}$, thus producing a resulting matrix $\Cc_{\textbf{b},\text{crt}}$ whose diagonal elements are all $1$, at all latitudes ($\phantom{l}_{\text{crt}}$ in the subscript stands for \emph{corrected}; consistently, let the elements of $\Cc_{\textbf{b},\text{crt}}$ be denoted by $\mathcal{C}_{\scaleto{(\theta_{i}, \varphi_{j}) , (\theta_i, \varphi_j)}{7pt}}^{\scaleto{\text{crt}}{3.7pt}}$). For concision, we will not go into the details of the just-mentioned technique utilized to correct $\Cc_{\textbf{b}}$. Then, after producing $\Cc_{\textbf{b},\text{crt}}$, we completely discarded $\Cc_{\textbf{b}}$. The actual localization matrix $\Cc$ that we employed in our experiment was constructed utilizing $\Cc_{\textbf{b},\text{crt}}$ --- instead of the discarded $\Cc_{\textbf{b}}$ --- as a block for its block structure. In this manner, we obtained, for our experiment, a localization matrix $\Cc$ whose diagonal elements are all $1$. Let us also mention and point out that figure \ref{fig:RowOfCb} illustrates $\Cc_{\textbf{b},\text{crt}}$ (and not the discarded $\Cc_{\textbf{b}}$). Having explained that $\Cc_{\textbf{b}}$ was discarded, as no confusion can arise, to simplify our notation, in the main text of this article, when referring to $\Cc_{\textbf{b},\text{crt}}$, we will drop $\phantom{l}_{,\text{crt}}$ from its subscript and we will just write $\Cc_{\textbf{b}}$. Similarly, in the main text of this article, when referring to the elements $\mathcal{C}_{\scaleto{(\theta_{i}, \varphi_{j}) , (\theta_i, \varphi_j)}{7pt}}^{\scaleto{\text{crt}}{3.7pt}}$ of $\Cc_{\textbf{b},\text{crt}}$, as no confusion can arise, to simplify our notation, we will simply write $\mathcal{C}_{\scaleto{(\theta_{i}, \varphi_{j}) , (\theta_i, \varphi_j)}{7pt}}$.\par

\clearpage
\thispagestyle{empty}
\setcounter{page}{1}
\setcounter{section}{0}
\renewcommand{\thesection}{\arabic{section}}
\setcounter{equation}{0}
\renewcommand{\theequation}{\arabic{equation}}
\begin{center}
    \textbf{\huge Graphical abstract}
\end{center}

\vspace{1cm}
\textbf{\Large A tool for Hybrid models: a swift algorithm to calculate the left inverse of a square root of a horizontally localized climatological ensemble covariance matrix}\\
\vspace{0cm}
\begin{center}
    \textbf{\large Francesco~Sardelli\textsuperscript{*} and Craig H.~Bishop}
\end{center}

\vspace{1cm}

\vspace{1cm}
Computing matrix inverses (or, e.g., left inverses) using known algorithms is generally computationally prohibitive in operational weather forecasting. In this article, a swift, parallelizable algorithm to calculate the left inverse of a square root of a horizontally localized ensemble covariance matrix is introduced. The algorithm is tested in a synthetic experiment on a three-dimensional grid, and it is found to have very good accuracy. Potential applications to a new positive definite and climatologically unbiased Hybrid covariance model are pointed out. 

\clearpage
\thispagestyle{empty}

\begin{figure}[h!]
    \centering
    \setlength{\unitlength}{2mm}
    \begin{picture}(60,50)(0,0)
\setlength{\fboxrule}{2pt}%
\setlength{\fboxsep}{3pt}%
    \fcolorbox{gray!50}{GraphicalAbstractBackgroundColor}{
\begin{picture}(60,50)(0,0)
\put(30,40){\begin{tikzpicture}
    \draw[ PhCurrentColor, line width=4pt ] (2, 14.7) -- (2, 14);
\end{tikzpicture}}
\put(25,26){\color{PhCurrentColor} $\scaleto{\vdots}{23mm}$}
\put(42,26){\color{PhCurrentColor} $\scaleto{\vdots}{23mm}$}
\put(6,40){\begin{tikzpicture}
    \draw[ PhCurrentColor, line width=4pt ] (1, 15) -- (10.3, 15);
\end{tikzpicture}}
 \put(5.3,31){\begin{tikzpicture}
     \draw[->, >=Stealth, PhCurrentColor, line width=4pt] (10, 25) -- (10, 23.2);
 \end{tikzpicture}}
 \put(16,31){\begin{tikzpicture}
     \draw[->, >=Stealth, PhCurrentColor, line width=4pt] (10, 25) -- (10, 23.2);
 \end{tikzpicture}}
 \put(33,31){\begin{tikzpicture}
     \draw[->, >=Stealth, PhCurrentColor, line width=4pt] (10, 25) -- (10, 23.2);
 \end{tikzpicture}}
 \put(51.1,31){\begin{tikzpicture}
     \draw[->, >=Stealth, PhCurrentColor, line width=4pt] (10, 25) -- (10, 23.2);
 \end{tikzpicture}}
       \put(4,45){\color{PhTensorColor} \LARGE  $(\Pc)^{\frac{1}{2}}=(\Cc \odot \P^s)^{\frac{1}{2}}= \Z \smalltriangleup  \Cc^{\frac{1}{2}} =\A \diag(\Cc_{\textbf{b}}^{\frac{1}{2}})$}
\put(3,28){\color{PhTensorHIColor} $\A_1 =   \Eb_1 \boldsymbol{\Lambda}_1 \Fb_1\tp $}
\put(14,28){\color{PhTensorHIColor} $\A_2 =   \Eb_2 \boldsymbol{\Lambda}_2 \Fb_2\tp $}
\put(29,28){\color{PhTensorHIColor} $\A_m =   \Eb_m \Lambdam \Fb_m\tp $}
\put(45,28){\color{PhTensorHIColor} $\A_{n_h} =   \Eb_{n_h} \boldsymbol{\Lambda}_{n_h} \Fb_{n_h}\tp $}
\put(29,15){\begin{tikzpicture}
    \draw[->, >=Stealth, PhCurrentColor, line width=4pt, ] (2, 14) -- (2, 13);
\end{tikzpicture}}
\put(6,20){\begin{tikzpicture}
    \draw[ PhCurrentColor, line width=4pt, ] (1, 15) -- (10.3, 15);
\end{tikzpicture}}
 \put(6.3,20){\begin{tikzpicture}
     \draw[ PhCurrentColor, line width=4pt]  (10, 24) -- (10, 23.1);
 \end{tikzpicture}}
 \put(17,20){\begin{tikzpicture}
     \draw[ PhCurrentColor, line width=4pt]  (10, 24) -- (10, 23.1);
 \end{tikzpicture}}
 \put(34,20){\begin{tikzpicture}
     \draw[ PhCurrentColor, line width=4pt]  (10, 24) -- (10, 23.1);
 \end{tikzpicture}}
 \put(52.2,20){\begin{tikzpicture}
     \draw[ PhCurrentColor, line width=4pt] (10, 24) -- (10, 23.1);
 \end{tikzpicture}}
  \put(22,12){\color{PhTensorColor} \LARGE  $\A = \mathbf{E} \boldsymbol{\Lambda} \F\tp$}
 \put(29,5.7){\begin{tikzpicture}
    \draw[->, >=Stealth, PhCurrentColor, line width=4pt, ] (2, 14) -- (2, 13);
\end{tikzpicture}}
 \put(14,3){\color{PcColor} \LARGE  $(\Pc)^{-\frac{1}{2}} = \diag(\Cc_{\textbf{b}}^{-\frac{1}{2}}) \, \! \F \boldsymbol{\Lambda}^{-1} \mathbf{E}\tp$}
\end{picture}} 
\end{picture}
       \label{fig:GraphicalAbstract}
\end{figure}

\end{document}